\documentclass[12pt]{iopart}

\usepackage{graphicx}
\usepackage{amssymb}
\usepackage[mathscr]{eucal}

\newcommand{\dfrac}[2]{\frac{\displaystyle #1}{\displaystyle #2}}
\newcommand{\email}[1]{\ead{#1}}
\newcommand{\affiliation}[1]{\address{#1}}
\newcommand{\acknowledgments}{\ack}

\newcommand{\sss}[1]{{\scriptscriptstyle{#1}}}
\newcommand{\mean}[1]{\left\langle #1 \right\rangle}
\newcommand{\order}[1]{\mathscr{O}\!\left(#1\right)}
\newcommand{\diag}{\mathrm{diag}}
\newcommand{\vol}[1]{\mathrm{Vol}\!\left(#1\right)}
\newcommand{\hypergauss}[4]{\mathrm{F}\!\left(#1,#2;#3;#4\right)}

\newcommand{\Mpc}{\mathrm{Mpc}}
\newcommand{\MeV}{\mathrm{MeV}}

\newcommand{\GeV}{\mathrm{GeV}}
\newcommand{\km}{\mathrm{km}}
\newcommand{\scnd}{\mathrm{s}}
\newcommand{\Gy}{\mathrm{Gyrs}}

\newcommand{\dd}{\mathrm{d}}
\newcommand{\uPl}{\mathrm{Pl}}
\newcommand{\uin}{\mathrm{in}}
\newcommand{\udm}{\mathrm{dm}}
\newcommand{\ure}{\mathrm{re}}
\newcommand{\uend}{\mathrm{end}}

\newcommand{\ureh}{\mathrm{reh}}
\newcommand{\urad}{\mathrm{rad}}

\newcommand{\unuc}{\mathrm{nuc}}

\newcommand{\ue}{\mathrm{e}}

\newcommand{\ud}{\mathrm{d}}
\newcommand{\ub}{\mathrm{b}}
\newcommand{\us}{\mathrm{s}}

\newcommand{\ui}{\mathrm{i}}

\newcommand{\uS}{\mathrm{S}}
\newcommand{\uM}{\mathrm{M}}

\newcommand{\uT}{\mathrm{T}}

\newcommand{\uDBI}{\sss{\mathrm{DBI}}}
\newcommand{\uUV}{\sss{\mathrm{UV}}}
\newcommand{\ugrav}{\mathrm{grav}}
\newcommand{\ustrg}{\mathrm{strg}}

\newcommand{\utotal}{\mathrm{tot}}
\newcommand{\ubulk}{\mathrm{bulk}}
\newcommand{\uthroat}{\mathrm{throat}}
\newcommand{\usssS}{\sss{\uS}}
\newcommand{\usssT}{\sss{\uT}}
\newcommand{\usssPl}{\sss{\uPl}}

\newcommand{\calG}{\mathcal{G}}
\newcommand{\calC}{\mathcal{C}}

\newcommand{\calD}{\mathcal{D}}
\newcommand{\calH}{\mathcal{H}}
\newcommand{\calP}{\mathcal{P}}

\newcommand{\nS}{n_\usssS}
\newcommand{\nT}{n_\usssT}
\newcommand{\muS}{\mu_\usssS}
\newcommand{\muT}{\mu_\usssT}
\newcommand{\muST}{\mu_{\usssS,\usssT}}
\newcommand{\omegaT}{\omega_\usssT}
\newcommand{\omegaS}{\omega_\usssS}
\newcommand{\omegaST}{\omega_{\usssS, \usssT}}
\newcommand{\QQoverTT}{\frac{Q^{2}_{\mathrm{rms-PS}}}{T^{2}}}

\newcommand{\CAMB}{\texttt{CAMB} }
\newcommand{\COSMOMC}{\texttt{COSMOMC} }
\newcommand{\ini}{\uin}
\newcommand{\ie}{\textrm{i.e.}~}
\newcommand{\eg}{\textrm{e.g.}~}

\newcommand{\gs}{g_\us}
\newcommand{\ells}{\ell_\us}
\newcommand{\alphas}{\alpha'}
\newcommand{\mpl}{m_\usssPl}

\newcommand{\muophi}{x}
\newcommand{\trans}{1}
\newcommand{\Htrans}{H_\trans}
\newcommand{\Ntrans}{N_\trans}
\newcommand{\muophitrans}{\muophi_\trans}
\newcommand{\epsone}{{\epsilon_1}}
\newcommand{\epstwo}{{\epsilon_2}}
\newcommand{\epsthree}{{\epsilon_3}}
\newcommand{\epstwoobs}{{\epstwo_{\mathrm{obs}}}}

\newcommand{\phione}{\phi_{\epsone}}
\newcommand{\phitwo}{\phi_{\epstwo}}
\newcommand{\phidotDBI}{\dot{\phi}_\uDBI}
\newcommand{\phiDBI}{\phi_\uDBI}
\newcommand{\phitrans}{\phi_\trans}
\newcommand{\phistrg}{\phi_\ustrg}
\newcommand{\phiuv}{\phi_\uUV}
\newcommand{\phizero}{\phi_0}
\newcommand{\phiini}{\phi_{\mathrm{in}}}
\newcommand{\phiend}{\phi_{\mathrm{end}}}
\newcommand{\phistar}{\phi_*}
\newcommand{\phifluct}{\phi_{\mathrm{fluct}}}
\newcommand{\phiin}{\phi_\uin}
\newcommand{\phiinstrg}{\phi_{\uin,\ustrg}}
\newcommand{\phiintwo}{\phi_{\uin,\epstwo}}

\newcommand{\rzero}{r_0}
\newcommand{\ruv}{r_\uUV}

\newcommand{\zstrg}{z_\ustrg}
\newcommand{\Nstar}{N_*}
\newcommand{\Nflux}{\mathcal{N}}
\newcommand{\Mflux}{\mathcal{M}}
\newcommand{\Kflux}{\mathcal{K}}
\newcommand{\Ntot}{N_\usssT}
\newcommand{\Vthroat}{V_6^\uthroat}
\newcommand{\Vbulk}{V_6^\ubulk}
\newcommand{\Vtotal}{V_6^\utotal}

\newcommand{\tens}{h}
\newcommand{\kstar}{k_*}

\newcommand{\usssPMoneP}{\sss{(\uM=1)}}
\newcommand{\calPnum}{\calP^{\usssPMoneP}}
\newcommand{\OmegaCDM}{\Omega_\udm}

\newcommand{\OmegaR}{{\Omega_{\urad}}}
\newcommand{\OmegaB}{\Omega_\ub}
\newcommand{\Rrad}{\mathsf{R}_\urad}
\newcommand{\Rreh}{\mathsf{R}}
\newcommand{\wstate}{w}
\newcommand{\zre}{z_\ure}
\newcommand{\optdepth}{\tau}

\newcommand{\xib}{x}
\newcommand{\metric}{g}
\newcommand{\Ricci}{R}

\newcommand{\cl}{\mathrm{cl}}

\begin{document}

\title[Brane inflation and the WMAP data]{Brane inflation and the WMAP
data: a Bayesian analysis}

\author{Larissa Lorenz}
\email{lorenz@iap.fr}
\affiliation{Institut d'Astrophysique de Paris, UMR
7095-CNRS, Universit\'e Pierre et Marie Curie, 98bis boulevard Arago,
75014 Paris, France}

\author{J\'er\^ome Martin}
\email{jmartin@iap.fr}
\affiliation{Institut d'Astrophysique de Paris, UMR
7095-CNRS, Universit\'e Pierre et Marie Curie, 98bis boulevard Arago,
75014 Paris, France}

\author{Christophe Ringeval}
\email{ringeval@fyma.ucl.ac.be}
\affiliation{Theoretical and Mathematical Physics Group, Center for
  Particle Physics and Phenomenology, Louvain University, 2 Chemin du
  Cyclotron, 1348 Louvain-la-Neuve, Belgium}
\date{today}

\begin{abstract}
The Wilkinson Microwave Anisotropy Probe (WMAP) constraints on string
inspired ``brane inflation'' are investigated. Here, the inflaton field
is interpreted as the distance between two branes placed in a
flux-enriched background geometry and has a Dirac-Born-Infeld (DBI)
kinetic term. Our method relies on an exact numerical integration of the
inflationary power spectra coupled to a Markov-Chain Monte-Carlo
exploration of the parameter space. This analysis is valid for any
perturbative value of the string coupling constant and of the string
length, and includes a phenomenological modelling of the reheating era
to describe the post-inflationary evolution. It is found that the data
favour a scenario where inflation stops by violation of the slow-roll
conditions well before brane annihilation, rather than by tachyonic
instability. Concerning the background geometry, it is established that
$\log v > -10$ at $95\%$ confidence level (CL), where $v$ is the
dimensionless ratio of the five-dimensional sub-manifold at the base of
the six-dimensional warped conifold geometry to the volume of the unit
five-sphere. The reheating energy scale remains poorly constrained, $T
_\ureh > 20 \, \GeV$ at $95\%$ CL, for an extreme equation of state
($\wstate_\ureh \gtrsim -1/3$) only. Assuming the string length is
known, the favoured values of the string coupling and of the
Ramond-Ramond total background charge appear to be correlated. Finally,
the stochastic regime (without and with volume effects) is studied using
a perturbative treatment of the Langevin equation. The validity of such
an approximate scheme is discussed and shown to be too limited for a
full characterisation of the quantum effects.

\end{abstract}




\section{Introduction}
\label{sec:introduction}

Inflation is presently considered our best description of the early
Universe, since it solves the puzzles of the Standard Hot Big Bang
Model~\cite{Starobinsky:1980te,
  Guth:1980zm,Linde:1981mu,Starobinsky:1982ee, Albrecht:1982wi,
  Linde:2005ht} and, in addition, provides a very convincing mechanism
for structure formation leading to an almost scale-invariant power
spectrum~\cite{Mukhanov:1981xt, Hawking:1982cz,
  Guth:1982ec,Bardeen:1983qw}. The small deviations from
scale-invariance are connected to the micro-physics of inflation and
are, therefore, of utmost importance when investigating the physical
conditions in the very early
Universe~\cite{Martin:2003bt,Martin:2004um,Martin:2007bw}. In
practice, inflation is usually driven by a scalar field (or possibly
several fields) and for this type of matter, the effective pressure
can become negative, if the scalar field's potential is sufficiently
flat. This condition is necessary to produce a phase of accelerated
expansion (\ie inflation) in General Relativity. In this context, one
of the main open issues is the physical nature of the inflaton
field. Since it is clear that this question should be addressed in the
framework of extensions of the Standard Model of particle physics,
String Theory seems to be a promising place to look for a physically
well-motivated candidate field. In recent years, this issue has given
rise to several studies, for reviews see
references~\cite{Dvali:2001fw,Burgess:2001fx,
  Cline:2006hu,HenryTye:2006uv,Kallosh:2007ig,Burgess:2007pz}. In
particular, string inspired brane world scenarios have turned out to
be especially fruitful. In this context, the inflaton field is usually
interpreted as the distance between two branes moving relative to each
other in the extra
dimensions~\cite{Dvali:1998pa,Alexander:2001ks,Kachru:2003sx}. The
corresponding potential generically contains a Coulomb term of the
form $\sim C-\phi ^{-4}$, where $C$ is a constant and $\phi $ the
inflaton field~\cite{Kachru:2003sx, Firouzjahi:2005dh,
  Baumann:2006cd,Baumann:2006th,Baumann:2007ah,
  Baumann:2007np,Krause:2007jk, Panda:2007ie}. The end of inflation
can then either occur by violation of the slow-roll conditions or by a
mechanism of tachyonic instability, as in hybrid inflation. In either
case, when the distance between the branes becomes of the order of the
string length, the simple single field description breaks down and a
more refined (fully stringy) treatment is necessary. The main goal of
the present article is to carry out a detailed comparison of the
inflationary predictions derived from one particular class of brane
inflation scenarios, namely the
Kachru-Kallosh-Linde-Maldacena-McAllister-Trivedi (KKLMMT) models,
with the Cosmic Microwave Background (CMB) data presently available.

\par

Analysing this class of potentials is further motivated by two rather
different reasons. Firstly, although it is possible (for instance by
using the slow-roll approximation) to span the space of (single field)
inflationary models without relying on a specific form of the inflaton
potential~\cite{Peiris:2006sj,Alabidi:2006qa,
deVega:2006hb,Kinney:2006qm}, it is also interesting to not use any such
approximation except, of course, the linear theory for the cosmological
perturbations~\cite{Martin:2006rs, Ringeval:2007am, Lesgourgues:2007gp}.
Let us also mention that with the incoming flow of more accurate data
from the Planck satellite~\cite{Lamarre:2003zh, Mennella:2003qp} or from
the future arc-minute resolution CMB experiments~\cite{Barker:2005gx,
Kosowsky:2006na, Ruhl:2004kv}, the errors linked with the current
approximation schemes may become a problem. The no-approximation
approach often requires numerical computations. As a consequence, one
has to pay the price of choosing a particular form for the inflaton
potential, which implies some loss of generality. A possible way out and
a systematic strategy to span the space of inflationary models is to
choose potentials that are representative examples of a whole class of
scenarios. For instance, a scenario where inflation takes place in the
regime of large (in Planck mass units) vacuum expectation value (vev)
for the field, and where the potential goes to infinity as the vev of
the field increases towards infinity, is illustrated by chaotic and/or
hybrid inflation. The difference between these two models consists in
the way inflation stops. On the contrary, a model where inflation occurs
for small values of the inflaton vev is well represented by models in
which the potential has a symmetry breaking shape. These classes of
models were studied in great detail and compared to the WMAP third-year
data in reference~\cite{Martin:2006rs}. From this general perspective,
the KKLMMT models with their above-mentioned Coulomb term generate
inflation when the vev of the field is large, but with a potential
bounded at infinity.

\par

Another motivation, different in spirit from the previous one, lies in
model building issues. The Coulomb type potential naturally arises in
string inspired scenarios of inflation. Moreover, this type of potential
has been known for a long time, even without relying on stringy
inspiration, since it can occur in models of hybrid inflation from
dynamical supersymmetry breaking~\cite{Lyth:1998xn}. Recently, in the
context of String Theory, corrections to the Coulomb potential
associated with the problem of moduli stabilisation have been
explored~\cite{HenryTye:2006uv,Baumann:2006cd,Baumann:2006th,
Baumann:2007ah,Baumann:2007np, Panda:2007ie}. Although interesting from
the theoretical point of view, they are not considered in the
following. Including these effects in the analysis would greatly
increase the number of free parameters relevant for the description of
the inflationary part of the model and their degeneracy by opening the
parameter space. As a result, given the current data, no constraint
could be obtained (as it is the case, for instance, for running mass
inflation~\cite{Martin:2006rs}). We have therefore chosen to perform an
exhaustive Bayesian analysis of the CMB data by considering only the
generic Coulomb contribution to the potential in the context of string
inspired brane inflation.

\par

Compared to the existing literature~\cite{Huang:2006ra, Zhang:2006cc,
  Bean:2007hc, Peiris:2007gz}, our study uses a complete Markov-Chain
  Monte-Carlo (MCMC) analysis based on an exact numerical integration of
  this type of potential. In this respect, we do not consider particular
  values of the string length $\sqrt{\alpha'}$ and the string coupling
  $\gs$ but, on the contrary, develop a parameter scanning strategy
  which allows to leave these parameters free\footnote{Some conditions
  on these parameters must be nevertheless fulfilled as, for instance,
  $\gs <1$ in order to be in the perturbative regime.}. However, we will
  see in the following that, already with only the Coulomb part of the
  potential, degeneracies in the parameter space do not permit the WMAP
  data to fully constrain these stringy quantities. Nevertheless, the
  throat geometry and some combinations of the KKLMMT model parameters
  are constrained. Finally, we present new approximate solutions for the
  field evolution in the DBI regime and refine the analysis of
  inflationary quantum effects. In particular, it is shown that the
  approximate solutions usually considered are not sufficient to fully
  characterise this regime.

\par

This article is organised as follows. In
section~\ref{sec:braneinfmodels}, we briefly recall some basic facts
about the KKLMMT scenario of brane inflation, deriving the governing
equations from the effective four-dimensional action based on type IIB
String Theory. In particular, we discuss the intrinsic string features
of the model, which among others manifest themselves through an unusual
kinetic behaviour as well as a restricted domain of validity for the
model. In section~\ref{sec:stochastic}, we briefly consider the impact
of quantum fluctuations that may affect the classical trajectory in
certain regions. Section~\ref{sec:slowroll} is devoted to the slow-roll
phase of the scenario, which can be treated analogously to small field
models. However, in the case of the KKLMMT scenario, the slow-roll
discussion has to be complemented by a study of the stringy aspects:
this is presented in section~\ref{sec:stringeffects}. After deriving the
restrictions on the parameter space from model-intrinsic arguments in
section~\ref{sec:restrictions}, we present our results from the WMAP
third-year data in sections~\ref{sec:srwmap}
and~\ref{sec:mcmc}. Conclusions are drawn in the final section
\ref{sec:conclusions}. Finally, in~\ref{app:stocha}, we give details on
the stochastic regime.

\section{Brane inflation}
\label{sec:braneinfmodels}

In the following, we discuss the KKLMMT model of brane inflation, first
proposed in reference~\cite{Kachru:2003sx} as a realisation of inflation
within String Theory. Our goal here is to consider all aspects (string
related or not) at play during the inflationary phase. We begin with a
intuitive description of how the KKLMMT model arises in String Theory.

\subsection{Qualitative description}
\label{subsec:qualitative}

The model considers a D3 and an anti-D3 brane in a ten-dimensional
supergravity background whose world-volume (time-axis included) is
aligned with the ${\xib^{\mu}}$ coordinate axes, $\mu$ varying from $0$
to $3$. The six extra dimensions $y^A$ with $A=4,\dots,9$, are
compactified such that the ``radial'' distance separating brane and
anti-brane is $y^{4}=r$, and their distance in the other coordinates
vanishes. While inflationary models involving D3 (or higher dimensional)
branes have been considered in the literature before, the KKLMMT model
assumes that the six-dimensional section forms a so-called
Klebanov-Strassler (KS) throat~\cite{Candelas:1989js,Klebanov:1999rd,
Klebanov:2000nc, Klebanov:2000hb,PandoZayas:2000sq}, i.e. a warped
deformed conifold with background fluxes for certain background
fields.

\par

Intuitively, this background can be thought of as populated by $\Nflux$
heavy D3 branes with Ramond-Ramond (RR) charge. The anti-D3 brane is
embedded at a fixed position \(r_{0}\) (the bottom of the KS throat),
representing an additional source of RR field strength that can be
described as a small perturbation. The D3 brane is inserted into this
perturbed background at \(r_{1}\gg r_{0}\) (far from the bottom of the
throat) and is considered as a probe: it does not affect the geometry
itself, but experiences the forces due to gravity and RR interaction,
brought about by the exchange of light closed string modes between the
branes. The distance \(r=r_{1}-r_{0}\) between the test D3 and the fixed
anti-D3 brane corresponds (up to normalisation) to the inflaton field
\(\phi\) whose potential \(V(\phi)\) can be calculated from the forces
experienced by the test D3 brane in the limit $r \gg \ells$,
$\ells=\sqrt{\alphas}$ being the string length scale. Inflation takes
place while \(\phi\) ``rolls down'' along a flat interaction potential
in the direction of decreasing \(\phi\). When the branes become too
close, in the sense that their proper distance approaches the string
scale, a tachyon, the lightest open string mode stretching from one
brane to the other, appears, and the long distance potential $V(\phi)$
is no longer valid. When the D3 brane reaches the bottom of the throat
\(r_{0}\), the two branes annihilate in a complex process whose details
are beyond the scope of the present paper. We will, however, introduce a
phenomenological model which assumes that the brane annihilation
triggers a reheating era.

\subsection{The inflaton effective action}
\label{subsec:effective}

In this section, we use an effective field theory representation of the
KKLMMT model, and our starting point will be the effective action of the
inflaton field \(\phi\). Considering that the inflaton $\phi$ is itself
an open string mode, it is related to the distance $r$ between the brane
and the anti-brane by
\begin{equation}
\label{eq:phir}
\phi=\sqrt{T_{3}}\, r\,,
\end{equation}
and its effective four-dimensional action reads
\begin{eqnarray}
\label{eq:generalaction}
S&=&-\frac{1}{2\kappa}\int  \Ricci \sqrt{-\metric} \, \ud^4 \xib  -
T_{3}\int \dfrac{1}{\tilde{h}(\phi)} \sqrt{-\metric} \, \ud^4 \xib 
\nonumber\\ &- & T_{3}\int 
\dfrac{1}{\tilde{h}(\phi)} \sqrt{1+\dfrac{\tilde{h}(\phi)}{T_{3}}
  \metric^{\mu \nu} \partial_{\mu}\phi
  \partial_{\nu}\phi } \, \sqrt{-\metric} \, \ud^4 \xib  ,
\end{eqnarray}
where the quantity $\kappa $ is defined by 
\begin{equation}
\label{eq:kappadef}
\kappa \equiv \dfrac{8\pi}{\mpl^2}\,,
\end{equation}
and
\begin{equation}
\label{eq:deftension}
T_{3}= \dfrac{1}{(2\pi)^{3} \gs \alphas^{2}}\,,
\end{equation}
is the D3 brane tension, $\gs$ being the string coupling constant. The
quantity \(\tilde{h}(\phi)\) is the (perturbed) warp factor of the
ten-dimensional metric whose form can be calculated from the full
Einstein equations~\cite{PandoZayas:2000sq}. To simplify the notation,
one can introduce the effective brane tension \(T(\phi)\)
\begin{equation}
T(\phi) \equiv\frac{T_{3}}{\tilde{h}(\phi)},
\end{equation}
allowing us to re-write equation~(\ref{eq:generalaction})
as~\cite{Alishahiha:2004eh}
\begin{eqnarray}
\label{eq:generalaction2}
S & =&  S_\ugrav + S_\uDBI \nonumber \\
&=& -\frac{1}{2\kappa} \int \Ricci
\sqrt{-\metric} \, \ud^4 \xib \nonumber \\ & & - \int \left[T(\phi)\,
  \sqrt{1+\frac{1}{T(\phi)} \metric^{\mu\nu} \partial_\mu \phi
    \partial_\nu \phi } +T(\phi)\right]\sqrt{-\metric}\, \ud^4 \xib .
\end{eqnarray}
It is clear that, for a slowly varying scalar field, the terms in
$(\partial \phi)^2$ are small and the square root in $S_\uDBI$ can be
Taylor expanded in the field derivatives. The resulting action at
leading order identifies with the one of a canonically normalised scalar
field evolving in a self-interaction potential.

\subsection{The kinetic term}
\label{subsec:kinetic}

On a flat Friedmann--Lema\^{\i}tre--Robertson--Walker (FLRW) brane, the
metric reads
\begin{equation}
\label{eq:flrwmetric}
\ud s^2 = a(\eta)^2\left(-\ud \eta^2 + \delta_{ij} 
\ud x^i \ud x^j \right),
\end{equation}
where $\eta$ and $x^i$ are the conformal time and spatial
coordinates. In terms of the cosmic time $\ud t=a \ud \eta$, the metric
tensor reduces to $\diag
\left[-1,a^{2}(t),a^{2}(t),a^{2}(t)\right]$. For such a homogeneous
metric, $\phi$ can only depend on time and keeping the first non-trivial
term in the square root expansion reproduces a standard kinetic term,
namely
\begin{equation}
S_\uDBI = \int \left[\frac{1}{2} \dot{\phi}^{2} -2 T(\phi)
  \right]a^{3}(t) \ud^4 \xib ,
\end{equation}
where a dot denotes a derivative with respect to the cosmic
time. However, this remains true as long as higher order terms in the
expansion can be neglected, \ie for
\begin{equation}
\label{eq:DBIdominance}
\dot{\phi}^{2}\lesssim T(\phi).
\end{equation}
Let us denote by $\phiDBI$ the field value from which onwards the field
evolution would be dominated by the string corrections to the standard
kinetic term. Notice that equation~(\ref{eq:DBIdominance}) is also the
condition required to have a positive square root argument in
equation~(\ref{eq:generalaction2}). In this sense, \(\sqrt{T(\phi)}\)
can be understood as the field maximum speed in the warped throat, and
one can, by relativistic analogy, define a Lorentz factor $\gamma$ such
that~\cite{Silverstein:2003hf}
\begin{equation}
\label{eq:gamma}
\gamma \left(\dot{\phi },\phi \right)\equiv
\frac{1}{\sqrt{1-\dot{\phi}^2/T\left(\phi \right)}}\, .
\end{equation}
When \(\gamma\sim\order{1}\), the field \(\phi\) follows the dynamics of
a canonically normalised scalar field which can slow-roll (and hence,
produce inflation) on a sufficiently flat potential. However, there also
exists an ``ultra-relativistic'' regime, \(\gamma\gg 1\), where the
string-intrinsic DBI form of the kinetic term is important. In this
regime, inflation may be possible even if the scalar field potential is
not flat.

\subsection{The inflaton potential}
\label{subsec:potential}

\begin{figure}
\begin{center}
\includegraphics[width=7.7cm]{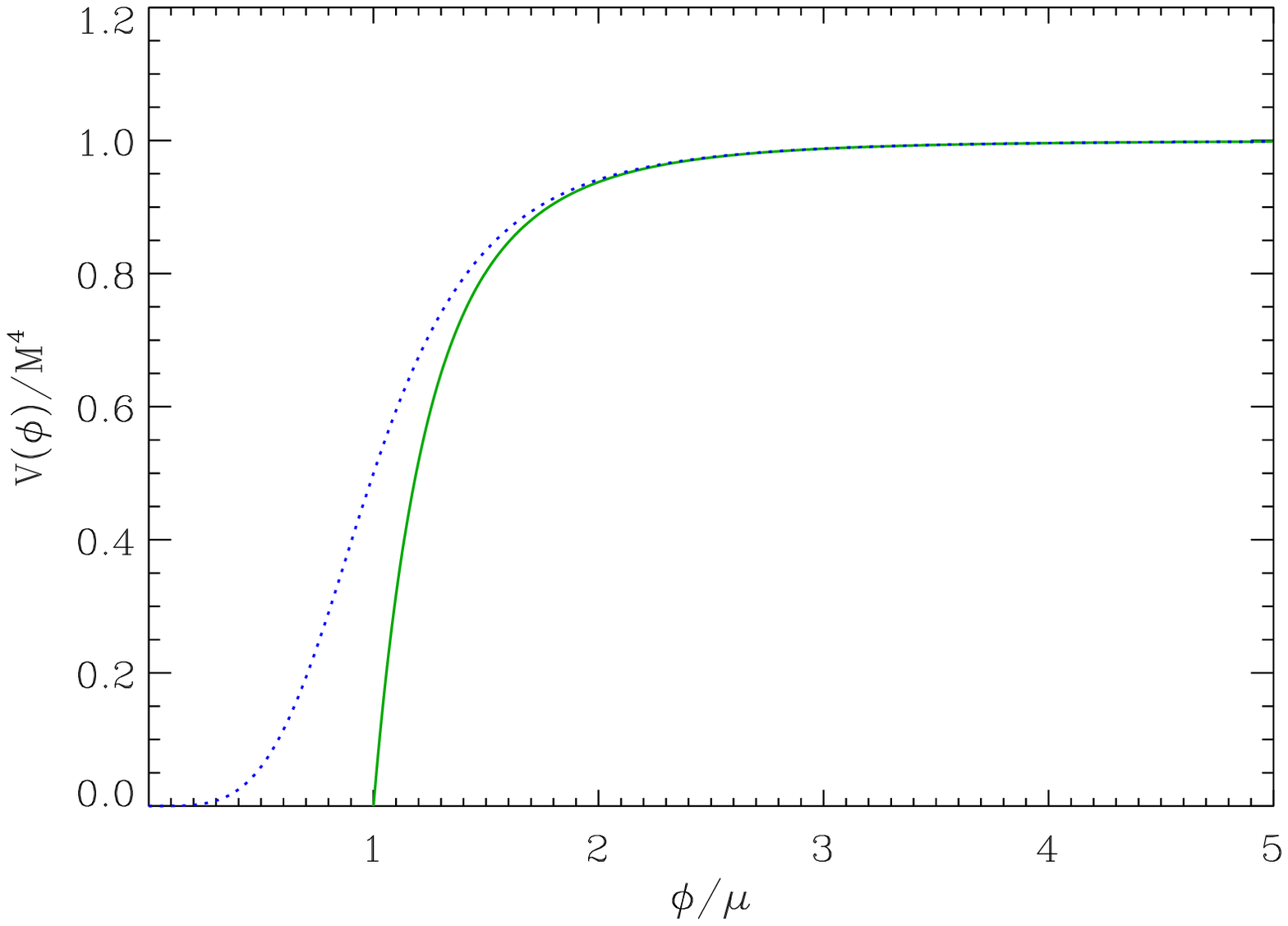}
\includegraphics[width=7.7cm]{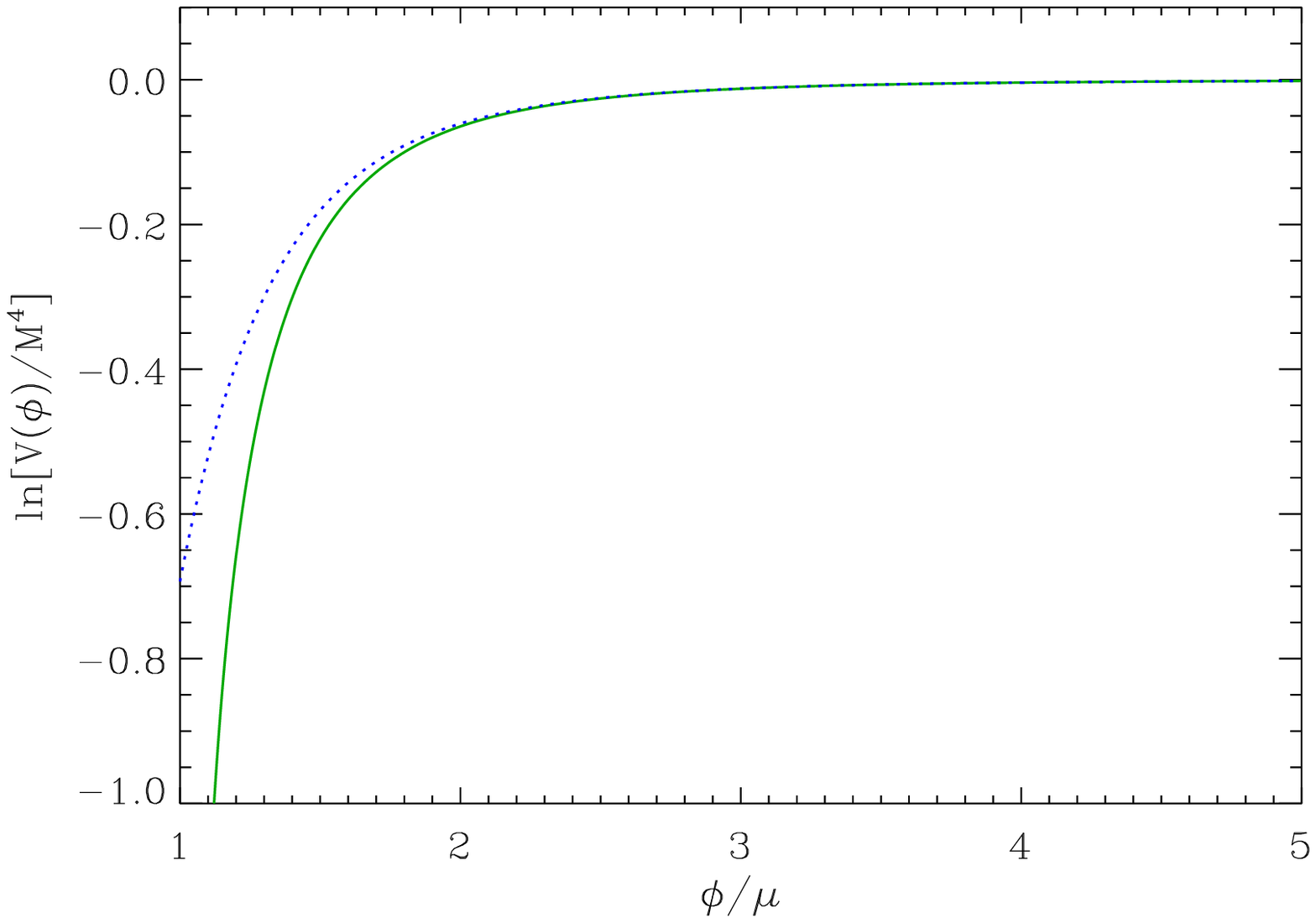} \caption{Left panel: The
dotted blue line represents the potential given by
equation~(\ref{eq:Vofphifull}). The solid green line shows the
approximate potential for \(\phi\gg\mu\), see
equation~(\ref{eq:Vofphiapprox}). The conventional slow-roll phase
occurs while the field is rolling on the extremely flat region to the
right of the plot. Right panel: Same as the left panel, but in
logarithmic unit for the potential.}  \label{fig:potkklt}
\end{center}
\end{figure}

Having established the inflaton dynamics, we now identify the potential
\(V(\phi)\). There is only one term in the expansion of
equation~(\ref{eq:generalaction2}) which does not contain field
derivatives and hence we conclude that
\begin{equation}
\label{eq:Vofphigeneral}
V(\phi)=2T(\phi)=\frac{2T_{3}}{\tilde{h}(\phi)}\,.
\end{equation}
So far we have not specified the form of the warp factor
\(\tilde{h}(\phi)\). In the case of a KS background perturbed by one
anti-D3 brane at the bottom of the throat, we
have~\cite{Kachru:2003sx, PandoZayas:2000sq}
\begin{equation}
\tilde{h}(\phi)=\frac{2T_{3}}{M^{4}}
\left[1+\left(\frac{\mu}{\phi}\right)^{4}\right],
\end{equation}
leading to the potential
\begin{equation}
\label{eq:Vofphifull}
V(\phi)=\frac{M^{4}}{1+\left(\mu/\phi\right)^{4}}\, .
\end{equation}
In the case where $\phi\gg\mu$, the first order expansion in $\mu/\phi$
reads
\begin{equation}
\label{eq:Vofphiapprox}
V(\phi) \simeq M^{4}\left[1-\left(\frac{\mu}{\phi}\right)^{4}\right],
\end{equation}
which is the most commonly used form of the KKLMMT Coulomb-type
potential. The full expression of the potential as well as its
approximate expression are represented in figure~\ref{fig:potkklt}.

\subsection{Parameter sets}
\label{sec:params}

{}From the String Theory point of view, a complete description of the
KKLMMT model is given by the parameter set
\begin{equation}\label{eq:stringset}
\left(g_{\rm s},\,\alpha',\,v,\,{\cal M},\,{\cal K}\right) .
\end{equation}
where we remind that \(g_{\rm s}\) and \(\sqrt{\alpha'}\) are the string
coupling and length, respectively. The quantity \(v\) is the
(dimensionless) volume ratio of the five-dimensional sub-manifold
forming the basis of the six-dimensional conifold geometry to the volume
of the five-sphere
\begin{equation}
\label{eq:defv}
v=\dfrac{\vol{X_{5}}}{\vol{\mathscr{S}_{5}}}\,.
\end{equation}
For instance, the base of the KS throat is the Einstein space $T_{1,1}$,
hence in this case one has $v=16/27$~\cite{Klebanov:1998hh}. The
quantity $\Mflux$ and $\Kflux$ are two positive integers associated with
the separately quantised background fluxes. They satisfy
\begin{equation}
\Nflux=\Kflux \Mflux,
\end{equation}
where $\Nflux$ is a positive integer representing the total background
RR charge.

\par

Another set of parameters equivalent to (\ref{eq:stringset}) is given in
terms of the extra dimensions' geometry. The two natural length scales
are the location of the bottom of the throat $r_{0}$, which corresponds
to the field value
\begin{equation}\label{eq:phi0}
\phizero=\sqrt{T_{3}}\,r_{0},
\end{equation}
and the throat edge $\ruv$. The position \(r=\ruv\) where the throat is
smoothly glued into the rest of the extra dimensional bulk. Provided the
depth of the throat is comparable to its width, the edge field value can
be approximated by~\cite{Baumann:2006cd}
\begin{equation}
\label{eq:phiuv}
\phiuv=\sqrt{T_{3}} \, \ruv ,
\end{equation}
with
\begin{equation}\label{eq:ruv}
\ruv^4 = 4\pi \gs \alphas^2 \dfrac{\Nflux}{v}\,.
\end{equation}
Consequently, one can deduce the maximum warp factor between the edge
and the bottom of the throat~\cite{Giddings:2001yu}
\begin{equation}
\left(\frac{\phiuv}{\phizero}\right)^{4} = \left(\frac{\ruv}{\rzero}
\right)^{4} \simeq \exp\left(\frac{8\pi \Kflux}{3 \gs \Mflux}
\right) ,
\end{equation} 
where the anti-D3's small perturbation to the KS warp factor has been
neglected. The set of geometrical parameters is therefore given by
\begin{equation}\label{eq:geometryset}
\left(\gs,\alphas,\phizero,\phiuv,\Nflux \right).
\end{equation}
\par

The potential expression in equation~(\ref{eq:Vofphifull}) contains
the parameters \(M\) and \(\mu\), which are defined from the string
parameters as
\begin{equation}
\label{eq:mandmu}
M^{4}=\dfrac{4\pi^{2} v \phizero^{4}}{\Nflux},\qquad
\mu^{4}=\dfrac{\phizero^{4}}{\Nflux} = \dfrac{M^{4}}{4\pi^{2}v}\, .
\end{equation}
Note that \(v\) can be expressed through \(M\) and \(\mu\),
\begin{equation}
\label{eq:vdef}
v=\frac{M^{4}}{4\pi^{2}\mu^{4}}\,.
\end{equation}
Since both \(M\) and \(\mu\) are proportional to the ratio
\(\phizero^{4}/\Nflux\), one has to provide another parameter to
maintain equivalence with the set of equations~(\ref{eq:stringset}) and
(\ref{eq:geometryset}), for instance $\Nflux$. In that case,
equation~(\ref{eq:mandmu}) can be inverted for $\phizero$:
\begin{equation}\label{eq:phi0ofM}
\phizero = \mu \Nflux^{1/4} = \dfrac{M}{\sqrt{2\pi}} \left(
\dfrac{\Nflux}{v} \right)^{1/4} .
\end{equation}
The model is thus equally well described by the ``cosmological''
parameter set
\begin{equation}\label{eq:cosmoset}
\left(\gs,\alphas,M,\mu,\Nflux  \right).
\end{equation}

\subsection{Domain of validity}
\label{subsec:validity}

We move on to discuss the domain of validity of the effective field
theory description. It turns out that in order for
equation~(\ref{eq:generalaction}) to be justified, one has to impose
some constraints on the model parameters and the accessible field values
\(\phi\).

\subsubsection{Throat within the overall six-dimensional volume.} 
\label{subsubsec:throatincluded}

In equation~(\ref{eq:kappadef}), the four-dimensional gravitational
coupling constant $\kappa$ is expressed as a function of the
four-dimensional Planck mass, which, in a ten-dimensional geometry,
depends on the volume of the compactified extra dimensions,
\begin{equation}
\label{eq:4dmpl}
\mpl^{2} = 8 \pi \frac{\Vtotal}{\kappa_{10}},
\end{equation}
where $\kappa_{10}$ is related to the string parameters by
\begin{equation}
\label{kappaten}
\kappa_{10} = \dfrac{1}{2} (2\pi)^{7} \gs^{2} \alphas ^{4},
\end{equation}
and $\Vtotal$ is the total volume of the six compactified extra
dimensions~\cite{Polchinski:1998rq, Polchinski:1998rr}. Now remembering
that the total six-dimensional volume consists in the volume $\Vthroat$
of the KS throat, where inflation takes place, plus the bulk volume
$\Vbulk$, into which this throat is glued at \(r=\ruv\), one has
\begin{equation}
\Vtotal = \Vbulk + \Vthroat,
\end{equation}
implying $\Vtotal/\Vthroat>1$. Although we ignore the precise form of
the six-dimensional bulk, the throat volume can be calculated
from~\cite{Baumann:2006cd}
\begin{equation}
\Vthroat =2\pi^{4} \gs \Nflux \alphas^{2} \ruv^{2},
\end{equation}
with \(\ruv\) given by equation~(\ref{eq:phiuv}). Then, re-writing
equation~(\ref{eq:4dmpl}) as
\begin{equation}
\mpl^{2}=8\pi \frac{\Vthroat}{\kappa_{10}} \frac{\Vtotal}{\Vthroat}\,,
\end{equation}
gives the total to throat volume ratio
\begin{equation}
\label{eq:volratio}
\frac{\Vtotal}{\Vthroat} = \alphas \mpl^{2} \sqrt{4 \pi^3 \gs
  \dfrac{v}{\Nflux^3}} \,.
\end{equation}
Since this ratio must exceed one, we obtain a condition on the possible
combinations of $\Nflux$ and $v$, depending on the choice of $\gs$ and
$\alphas$:
\begin{equation}
\label{eq:throatsize}
\Nflux^{3/2} v^{-1/2} < 2\pi^{3/2} \alphas \mpl^{2} \gs^{1/2} .
\end{equation}
Note that this condition heavily depends on the choice of $\gs$ and
$\alphas$, the value of which can vary significantly according to the
specific String Theory considered. It is also convenient to recast
equation~(\ref{eq:throatsize}) in terms of $\phiuv$. From
equations~(\ref{eq:phiuv}) and (\ref{eq:volratio}), one has
\begin{equation}
\kappa \phiuv^2 = \dfrac{4}{\Nflux} \dfrac{\Vthroat}{\Vtotal}\,,
\end{equation}
and the volume ratio limit (\ref{eq:throatsize}) implies
\begin{equation}
\label{eq:phiuvmax}
\phiuv < \dfrac{\mpl}{\sqrt{2\pi \Nflux}}\,.
\end{equation}
Let us also remark that, for the supergravity effective description to
be valid, the volume of the extra dimensions, and therefore of the
throat, should be larger than $\alpha '^3$, even when the volume of the
five-dimensional conifold basis is very small. However, one can show
that this condition does not play a role in the following analysis.

\subsubsection{Inflation within a single throat.}
\label{subsubsec:inflationthroat}

We will restrict our attention to the case where the D3 brane begins its
journey towards the anti-D3 brane inside the KS throat, \ie for
\begin{equation}
\label{eq:onethroat}
\phi<\phiuv,
\end{equation}
where $\phiuv$ is the field value on the throat edge [see
  equation~(\ref{eq:phiuv})]. In cosmological terms, this means that
all the e-folds required to solve the Standard Big Bang Model problems
occur inside one and the same throat. We will discuss in
section~\ref{sec:restrictions} in which sense
equation~(\ref{eq:onethroat}) allows us to impose restrictions on the
parameter values.

\subsubsection{End of brane motion by an instability-like mechanism.}
\label{subsubsec:instability}

The derivation of ~(\ref{eq:Vofphifull}) assumes that brane interaction
is due to the exchange of the lightest closed string modes only. This is
only approximately true: Firstly, one could include the contribution of
heavier modes (whose propagation through the bulk is Yukawa suppressed)
as well as consider more general effects such as the K\"ahler potential
necessary to stabilise the moduli. In this case, deriving the full
brane--anti-brane interaction potential is highly non-trivial. A
phenomenological approach is to assume that these effects can be
summarised in an additional potential term [\eg of the form
$V(\phi)=m^{2}\phi^{2}$] which can be added to
equation~(\ref{eq:Vofphifull}). This route has been explored in the
literature, see e.g. \cite{Bean:2007hc}. Very recently, the theoretical
underpinning of form and origin of these terms has also received much
attention \cite{Baumann:2006cd,Baumann:2006th,Baumann:2007ah,
Baumann:2007np,Krause:2007jk}. For our cosmological purposes, these
effects have not been considered, following the argumentation outlined
earlier based on the accuracy of the present data. Secondly, in addition
to the above-mentioned corrections to the potential, the effective field
model intrinsically has a restricted domain of validity.  When the two
branes come too close in terms of their proper distance $s$ (with
respect to the string scale $\ells$), an open string can stretch between
them and a tachyon, being the lightest excitation of this open string,
appears. Let us determine the value \(\phistrg\) for which the proper
distance between the branes $s_\uend \simeq \ells$, where
$\ells=\sqrt{\alphas}$. It is solution of the following equation
\cite{Bean:2007hc}
\begin{equation}
\int_{s_{0}}^{s_\uend} \ud s = \int_{r_{0}}^{r_\ustrg} h^{1/4}(r) \ud r ,
\end{equation}
where the origin \(s_{0}\) can be chosen at the brane annihilation
location $r_0$, \ie \(s_{0}=0\), with \(r_{0}\) the radial coordinate
of the bottom of the throat. Using the warp factor of the unperturbed
background, $h(r)=\ruv^{4}/r^{4}$, one gets $s_\uend = \ruv
\ln(r_\ustrg/r_{0})$. Since $s_\uend = \sqrt{\alphas}$, we have
$r_\ustrg = r_{0} \ue^{\sqrt{\alphas}/\ruv}$, which after
normalisation gives
\begin{equation}
\label{eq:phistrg}
\phistrg = \phi_{0} \ue^{\sqrt{\alphas}/\ruv} = \mu \Nflux^{1/4} \exp 
\left[\left(4\pi \gs \frac{\Nflux}{v} \right)^{-1/4} \right],
\end{equation}
where use has been made of equations~(\ref{eq:phiuv}) and
(\ref{eq:phi0ofM}). Notice that \(\phistrg\) explicitly depends on the
background flux $\Nflux$. As a consequence, another equivalent
formulation of the parameter set (\ref{eq:cosmoset}) is therefore
given by
\begin{equation}\label{eq:cosmoset2}
(\gs,\alphas,M,\mu,\phistrg).
\end{equation} 
{}From equation~(\ref{eq:phistrg}), one immediately gets
\begin{equation}
\label{eq:phigtmu}
\phistrg>\mu ,
\end{equation}
since the background flux number $\Nflux \ge 1$. Although the String
Theory details of the tachyon appearance and brane annihilation are out
of the scope of this work, we will assume that these processes trigger a
reheating era which precedes the cosmological radiation-dominated era
(see section~\ref{subsec:mcmcmethod}).

\subsection{Equations of motion}
\label{subsec:eom}

The equations of motion for the inflaton in a FLRW metric on the brane
can be obtained from the total action~(\ref{eq:generalaction2}). From
the stress-tensor associated with the scalar field \(\phi\), the energy
density and pressure read
\begin{eqnarray}
\rho =(\gamma-1) T(\phi)+V(\phi),\qquad P =\dfrac{\gamma-1}{\gamma}
T(\phi) - V(\phi),
\end{eqnarray}
from which one obtains the Friedmann--Lema\^{\i}tre equations
\begin{eqnarray}
\label{eq:friedman}
H^2 = \frac{\kappa }{3}\left[T\left(\phi \right)\left(\gamma -1\right)
  +V\left(\phi \right)\right],\qquad \dot{H} =
\dfrac{\kappa}{2}T(\phi)\left(\dfrac{1}{\gamma} - \gamma \right).
\end{eqnarray}
When the velocity of the field is small, the Lorentz factor \(\gamma\)
defined in (\ref{eq:gamma}) is close to one and its square root can be
expanded. Likewise, varying the action with respect to the inflaton
field leads to the following ``Klein-Gordon like'' equation of motion
\begin{eqnarray}
\label{eq:kglike}
\dfrac{\ud^2 \phi}{\ud N^2} + \left(\dfrac{3}{\gamma^2} + \dfrac{\ud
  \ln H}{\ud N} \right) \dfrac{\ud \phi}{\ud N} +
  \left(\dfrac{3}{\gamma^2} - 1\right) \dfrac{T'(\phi)}{2 H^2} +
  \dfrac{V'(\phi) - T'(\phi)}{H^2 \gamma^3}=0\,,
\end{eqnarray}
where, in the present context, a prime denotes a derivative with respect
to $\phi $. Note that $N$ here is the number of e-folds (not to be
confused with the background flux $\Nflux$), which will be used as a
convenient measure of time. In the limit $\gamma \rightarrow 1$,
equations~(\ref{eq:friedman}) and (\ref{eq:kglike}) reduce to the
standard Friedmann--Lema\^{\i}tre and Klein-Gordon equations. It will be
important to establish in which regime the DBI corrections to the
kinetic term (and therefore, deviation from the standard slow-roll
dynamics) may be important.

\subsection{Field evolution overview}
\label{subsec:overview}

Brane inflation according to the KKLMMT scenario incorporates both
features of usual slow-roll inflation as well as intrinsic String Theory
effects. Let us sketch the successive phases of the field evolution as
$\phi$ rolls down the potential \(V(\phi)\) towards its small values.

\par

\begin{figure}
\begin{center}
\includegraphics[width=7.7cm]{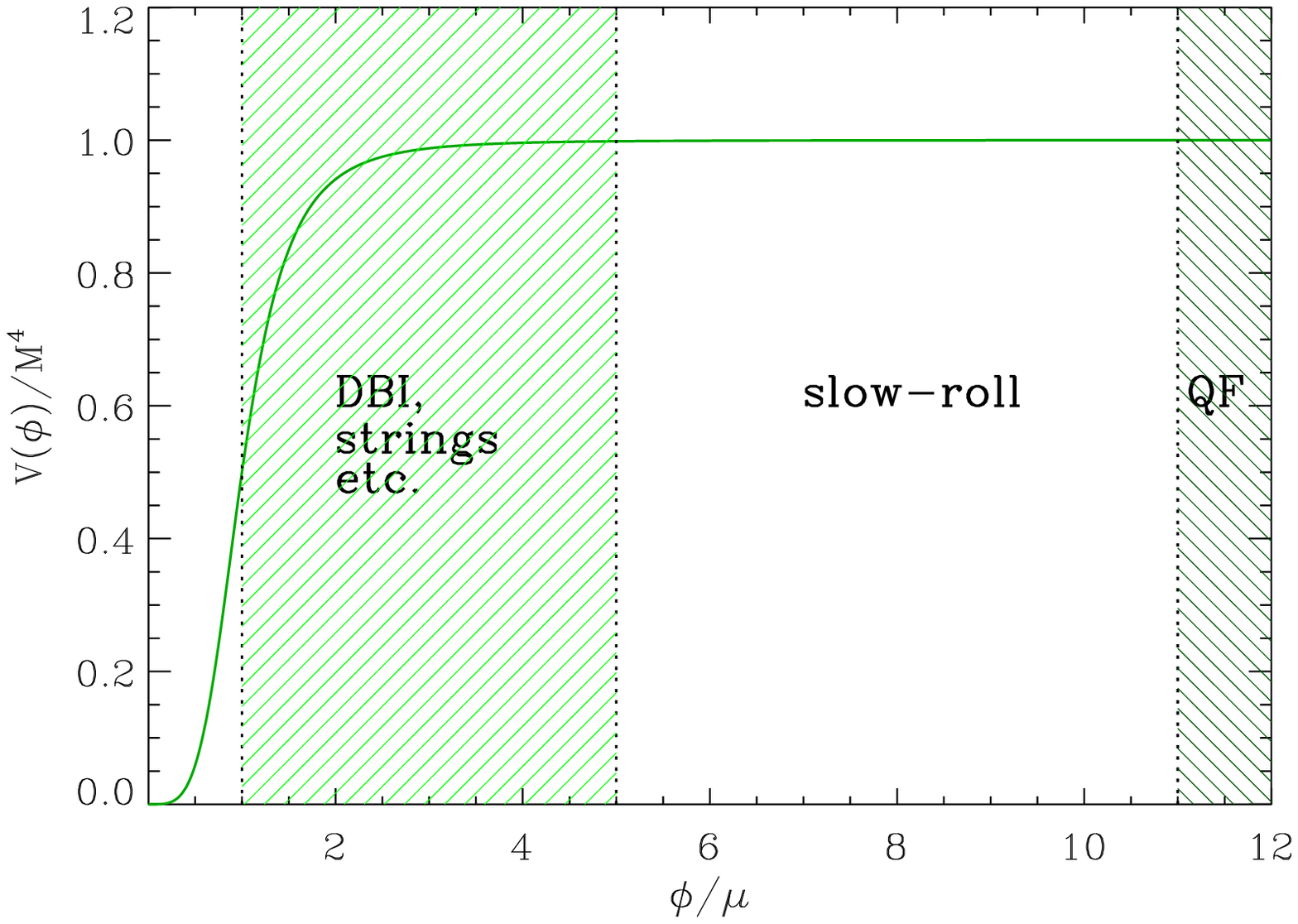}
\includegraphics[width=7.7cm]{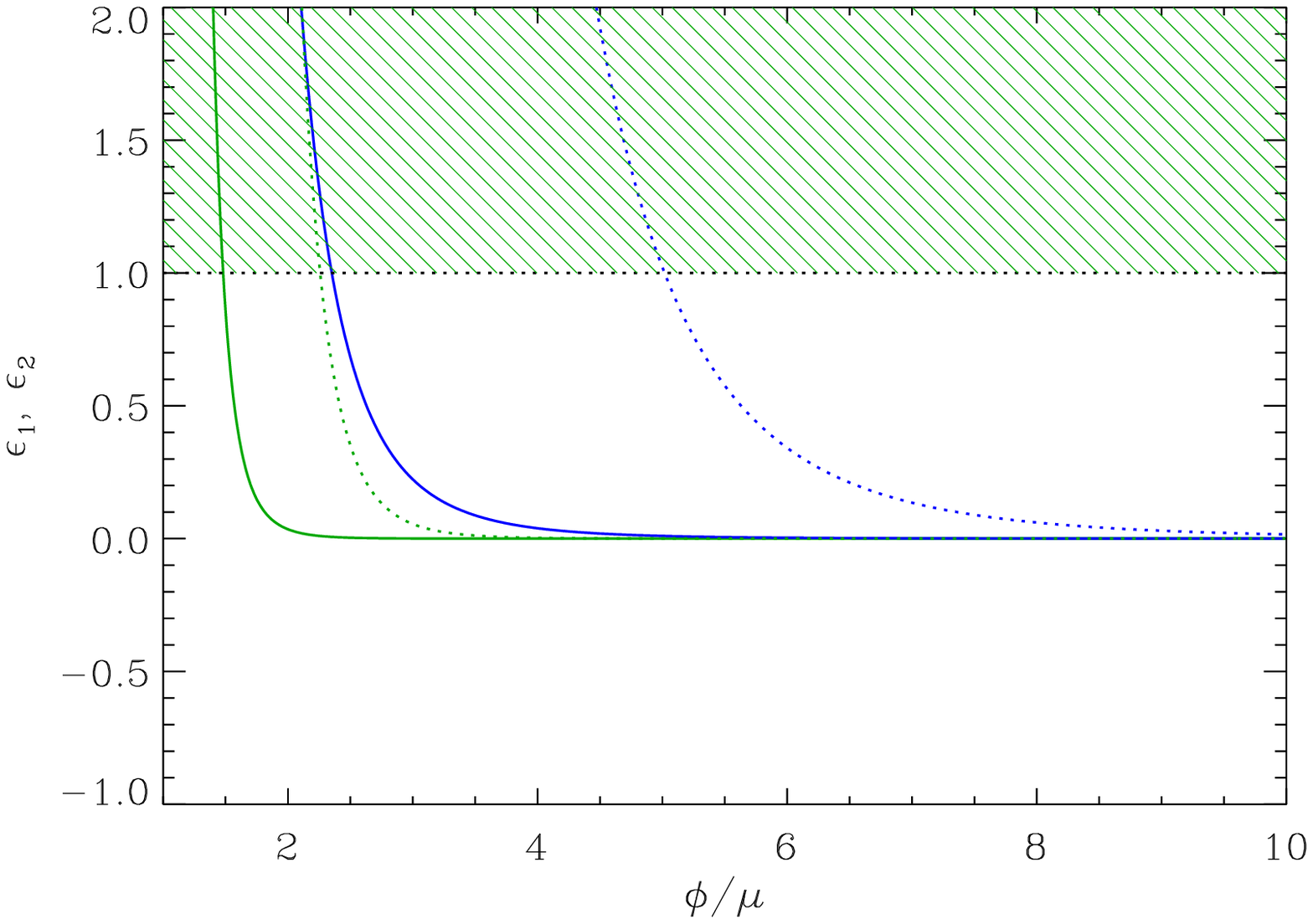} \caption{Left panel:
Sketch of the expected dynamical regimes according to the vev of the
inflaton field for the potential~(\ref{eq:Vofphifull}). The field starts
out in the flat (no hatched) region of the potential and rolls towards
smaller field values. Eventually, it will reach the light green hatched
region where the slope of the potential becomes noticeable and where the
DBI effects can no longer be neglected. This regime may or may not be
reached according to the value of \(\phistrg\) for which the derivation
of equation~(\ref{eq:Vofphifull}) breaks down. On the far right (dark
green hatched region), the potential is extremely flat and quantum
fluctuations are expected to dominate over the field classical
evolution. Right panel: The slow-roll parameters $\epsilon _1$ (green
line) and $\epsilon _2$ (blue line) for the KKLMMT potential. The solid
curves have been obtained for $\mu/\mpl=0.1$ whereas the dotted ones for
$\mu/\mpl=0.01$. Decreasing \(\mu/\mpl\) increases the differences
between $\phione$ and $\phitwo$, the field values for which $\epsone=1$
and $\epstwo=1$, respectively. Notice that we always have
$\phitwo>\phione$.}  \label{fig:potkklt_s}
\end{center}
\end{figure}

In figure~\ref{fig:potkklt_s} (left panel), one can distinguish a region
(for very large field values, dark green shaded) in which the quantum
character of the inflaton field cannot be neglected: Quantum
fluctuations are of the same order as the classical ones where the
potential is extremely flat, \ie for \(\phi\) exceeding a certain
\(\phi_{\rm fluct}\) . This is notably of interest if the brane starts
its journey above this limit. The peculiar properties of this regime
will be discussed in section~\ref{sec:stochastic} and \ref{app:stocha}.

\par

For the moment, let us assume that the brane motion starts at
\(\phi_{\rm in}<\phi_{\rm fluct}\). In this region, the potential
(\ref{eq:Vofphifull}) is still very flat since $\phi \gg \mu$, allowing
for most of the inflationary expansion to take place. One may therefore
expect the usual slow-roll approximation to be valid and it will be used
in section~\ref{sec:slowroll} to derive the shape of the induced
primordial power spectra.

\par

Rolling further to the left, the field \(\phi\) eventually enters a
region where the slope of the potential becomes noticeable (light green
shaded region). Therefore, the conventional slow-roll approximations are
certainly no longer sufficient to describe this evolution, and, since
the field velocity should increase, one may expect the DBI dynamics to
be the driving force for $\phi<\phiDBI$. In addition, $\phistrg$ is also
located around this potential domain. The precise order of events in
this region will be discussed in section~\ref{sec:stringeffects}.

\section{Stochastic inflation}
\label{sec:stochastic}

In this section, we will denote the \emph{classically} predicted field
value by \(\phi_{\rm cl}\), and consider the effect of the quantum
fluctuations on the classical
trajectory~\cite{Starobinsky:1986fx,Chen:2006hs,Martin:2005ir,
Martin:2005hb}. The
field values for which quantum fluctuations are comparable to the
classical ones can be obtained by first estimating the size of the
classical fluctuations in the slow-roll regime. Denoting \(V_{\phi_{\rm
cl}}\) the potential derivative with respect to the field, one gets
\begin{equation}
\label{eq:class}
\Delta\phi_{\rm cl}\simeq -\frac{V_{\phi_{\rm cl}}}{3H(\phi_{\rm
cl})}\,\Delta t \,,
\end{equation}
which comes from the Klein-Gordon equation by neglecting
\(\ddot{\phi}\). For \(\Delta t\) it is convenient to take the typical
time scale at work during expansion, namely one Hubble time, \(\Delta
t=1/H\). Concerning the quantum fluctuations, one has for a test field
in de-Sitter space-time,
\begin{equation}
\label{eq:quant}
\Delta\phi_{\rm qu}\simeq \frac{H(\phi_{\rm cl})}{2\pi}\, .
\end{equation}
The field value \(\phi_{\rm fluct}\), for which classical and quantum
fluctuations are of equal amplitude, verifies $\Delta\phi_{\rm
  cl}(\phifluct)=\Delta\phi_{\rm qu}(\phifluct)$. From
equations~(\ref{eq:class}) and~(\ref{eq:quant}) it follows that
\begin{equation}
\label{eq:phifluct}
\frac{\phi_{\rm
fluct}}{\mu}\simeq\left[\frac{3}{8\pi}\left(\frac{\mpl}{\mu}\right)^{2}
\left(\frac{\mpl}{M}\right)^{4}\right]^{1/10}.
\end{equation}
Using equation~(\ref{eq:mandmu}), we can express (\ref{eq:phifluct}) in
terms of \(v\) and the overall scale of inflation \(M/\mpl\). According
to their value, \(\phi_{\rm fluct}\) can generically be much larger than
the minimal initial field value needed to produce \(\Ntot \simeq
10^{2}\) e-folds of inflation. To ensure the consistency of the
semi-classical approach, we require
\begin{equation}
\label{eq:nofluctuations}
\phi_{\rm in}<\phi_{\rm fluct}\, ,
\end{equation}
in the following. Quantum effects can be nevertheless described using
the stochastic approach~\cite{Martin:2005ir, Martin:2005hb} which is
presented in detail in~\ref{app:stocha}. Since the initial value of the
field is also limited by the requirement of having the D3 brane inside
the throat, the effective semi-classical model remains valid only if
$\phiin<\min(\phi_{\rm fluct},\phiuv)$.

\section{The slow-roll regime}
\label{sec:slowroll}

In this section we analyse the KKLMMT potential according to its
slow-roll characteristics. When denoted as in~(\ref{eq:Vofphifull}), the
analogy between $V(\phi)$ and small field models (investigated in detail
in reference~\cite{Martin:2006rs}) is evident. Notice, however, that for
small field inflation, $\phi$ is increasing while inflation is under
way, whereas KKLMMT inflation proceeds from large to small field values
(see figure~\ref{fig:potkklt}).

\subsection{Slow-roll parameters}
\label{subsec:srparameters}

Among the several definitions of the slow-roll parameters, we here use
the Hubble-flow functions $\epsilon_n $ defined
by~\cite{Schwarz:2001vv,Schwarz:2004tz,Leach:2002ar}
\begin{equation}
\label{eq:flow}
\epsilon_{n+1} \equiv \frac{\dd \ln |\epsilon_n|} {\dd N}\,, \qquad
n\geq 0\, ,
\end{equation} 
where $N$ is the number of e-folds since some initial time
$\eta_\uin$. The above hierarchy starts from
$\epsilon_0=H_\ini/H$. Slow-roll inflation proceeds as long as
$|\epsilon_n| \ll 1$, for all $n>0$, while inflation takes place if the
scale factor is accelerating, \ie $\epsilon _1<1$.

\par

{}From the potential~(\ref{eq:Vofphifull}), the first two Hubble flow
parameters in their slow-roll formulation read
\begin{eqnarray}
\label{eq:eps1kklt} \epsilon_1 &=& \frac{1}{\pi} \left(\frac{\mpl}{\mu
}\right)^2 \frac{\left( \dfrac{\phi}{\mu} \right)^{10}} {\left[ 1+
    \left( \dfrac{\phi}{\mu} \right)^{-4}\right]^2}\, , \\ \epsilon_2
& = & \frac{1}{\pi} \left( \frac{\mpl}{\mu } \right)^2 \left(
\dfrac{\phi }{\mu} \right)^{-6} \frac{5+ \left( \dfrac{\phi}{\mu}
  \right)^{-4}} {\left[ 1+ \left( \dfrac{\phi}{\mu} \right)^{-4}
    \right]^2} \,,
\label{eq:eps2kklt}
\end{eqnarray}
and they are represented in figure~\ref{fig:potkklt_s} (right
panel). They depend on the ratio \(\phi/\mu\) and the vev energy scale
\(\mu/\mpl\) [or \(M/\mpl\), by virtue of equation~(\ref{eq:mandmu})]
only. As above-mentioned, inflation ends when \(\epsilon_{1}=1\), while
the slow-roll approximation itself breaks down when \(\epsilon_{2}=1\)
(and/or $\epsilon_{1}=1$). Let us stress again that the field theory
description of the brane motion fails at $\phi=\phistrg$.

\subsection{Classical field trajectory}
\label{subsec:cltrajec}

The next step is to obtain the classical field trajectory; this is
possible while the slow-roll approximation is valid but even in this
case, the trajectory is found only implicitly. The total number of
e-folds $N(\phi)$ reads~\cite{Liddle:1994dx}
\begin{eqnarray}
\label{eq:trajectory} N(\phi) &=& {2\pi }\frac{\mu ^2}{\mpl
^2}\left[\frac{1}{6}\left(\frac{\phi_{\rm in} }{\mu}\right)^{6}
  +\frac{1}{2}\left(\frac{\phi _{\rm in}}{\mu }\right)^2
- \frac{1}{6}\left(\frac{\phi}{\mu }\right)^{6} -
  \frac{1}{2}\left(\frac{\phi }{\mu}\right)^2\right] .
\end{eqnarray} 
For the generic situation where $\phi/\mu \gg 1$, the two quadratic
terms are sub-dominant. Neglecting them yields the following explicit
solution 
\begin{equation}
\label{eq:trajecapprox} 
\frac{\phi (N)}{\mu}\simeq
\left[\left(\frac{\phi _\ini}{\mu
}\right)^{6}-\frac{3}{\pi}\left(\frac{\mpl} {\mu}\right)^2N\right]^{1/6}
.
\end{equation} 
As seen in figure~\ref{fig:potkklt_s}, given that the field evolution
starts on the very flat part of the potential (but outside the domain of
quantum fluctuations), there will surely be an initial phase of
evolution during which the slow-roll approximation, and hence the
trajectory (\ref{eq:trajecapprox}), is valid. However, moving towards
smaller field values, this trajectory will eventually become inaccurate
and one has to determine at which field value the slow-roll
approximation breaks down and how this field value compares to the end
of inflation and/or to $\phistrg$.

\subsection{When does the slow-roll regime break down?}
\label{subsec:srend}

The slow-roll approximation breaks down when one of the slow-roll
parameters becomes of order one, which need not coincide with the end of
accelerated expansion ($\epsilon_1=1$). If slow-roll breaks down first,
inflation can still proceed (typically during a very small number of
e-folds), but the evolution of the field is no longer described by the
slow-roll trajectory (\ref{eq:trajecapprox}). Let us define the field
values $\phione$ and $\phitwo$ by
\begin{equation}
\epsone(\phione) = 1\,, \qquad \epstwo(\phitwo) = 1. 
\end{equation}
{}From equations~(\ref{eq:eps1kklt}) and (\ref{eq:eps2kklt}), one
obtains the algebraic equations
\begin{eqnarray}
1+2\left(\frac{\mu}{\phi_{\epsilon_{1}}}\right)^{4}+
\left(\frac{\mu}{\phi_{\epsilon_{1}}}\right)^{8}-\frac{1}{\pi}\,
\left(\frac{\mpl}{\mu}\right)^{2}
\left(\frac{\mu}{\phi_{\epsilon_{1}}}\right)^{10}=0 \, ,
\label{eq:defphieps1}
\end{eqnarray}
and 
\begin{eqnarray}
  1 &+ & 2\left(\frac{\mu}{\phi_{\epsilon_{2}}}\right)^{4}
  +\left(\frac{\mu}{\phi_{\epsilon_{2}}}\right)^{8}
  -\frac{5}{\pi} \left(\frac{\mpl}{\mu}\right)^{2}
  \left(\frac{\mu}{\phi_{\epsilon_{2}}}\right)^{6} \nonumber \\  & - &
\frac{1}{\pi}\,\left(\frac{\mpl}{\mu}\right)^{2}
  \left(\frac{\mu}{\phi_{\epsilon_{2}}}\right)^{10}=0 \,.
\label{eq:defphieps2}
\end{eqnarray}
These equations cannot be solved explicitly (except numerically). Some
approximate analytical solutions are derived in the following.

\subsubsection{Case $\mu > \mpl$ with $\phi_{\epsilon_{1},
\epsilon_{2}}<\mu$.}
\label{subsubsec:mularge}

In this case, keeping only the two dominant terms in
equation~(\ref{eq:defphieps1}) gives
\begin{equation}
\left(\frac{\mu}{\phi_{\epsilon_{1}}}\right)^{8}-\frac{1}{\pi} 
\left(\frac{\mpl}{\mu}\right)^{2}\,
\left(\frac{\mu}{\phi_{\epsilon_{1}}}\right)^{10}\simeq 0 \,,
\end{equation}
whose solution is
\begin{equation}
\label{eq:phieps1large}
\frac{\phi_{\epsilon_{1}}}{\mu}\simeq\frac{\mpl}{\mu}
\frac{1}{\sqrt{\pi}}\,.
\end{equation}
This expression is consistent with our assumptions: $\mu > \mpl$ indeed
implies $\phi_{\epsilon_{1}} <\mu$. Still in the same limit, the three
remaining terms in equation~(\ref{eq:defphieps2}) are
\begin{equation}
\left(\frac{\mu}{\phi_{\epsilon_{2}}}\right)^{8}-\frac{5}{\pi}
\left(\frac{\mpl}{\mu}\right)^{2}
\left(\frac{\mu}{\phi_{\epsilon_{2}}}\right)^{6}-\frac{1}{\pi}
\left(\frac{\mpl}{\mu}\right)^{2}
\left(\frac{\mu}{\phi_{\epsilon_{2}}}\right)^{10}\simeq 0 \, .
\end{equation}
This equation has an exact solution with two acceptable roots
\begin{equation}
\label{eq:phieps2large}
\left(\frac{\phi_{\epsilon_{2}}}{\mu}\right)_{\pm}=\frac{\mpl}{\mu}\,
\sqrt{\frac{2}{\pi}}\left[1\pm\sqrt{1-\frac{20}{\pi^{2}}
    \left(\frac{\mpl}{\mu}\right)^{4}}\right]^{-1/2} .
\end{equation}
For $\mu > \mpl$, slow-roll inflation can proceed till the field $\phi$
reaches $\max(\phione,\phitwo) <\mu$. By comparing
equation~(\ref{eq:phieps1large}) and (\ref{eq:phieps2large}), we further
see that $\phitwo >\phione$, \ie the slow-roll approximation breaks down
before the end of inflation.

\par

However, as already shown in section~\ref{sec:stringeffects}, the KKLMMT
model itself is no longer well-defined for \(\phi<\phistrg\) due to the
stringy origin of the potential (\ref{eq:Vofphifull}). The value
\(\phistrg\) given in equation~(\ref{eq:phistrg}) is always greater than
\(\mu\). Hence, in the case where $\mu > \mpl$, inflation will
definitely come to an end at $\phistrg > \max(\phione,\phitwo)$. As a
result, the entire field evolution occurs in the slow-roll regime and,
as it will be shown in the following, we do not need to worry about DBI
effects which remain negligible in that case.

\subsubsection{Case $\mu <\mpl$ with
$\phi_{\epsilon_{1},\epsilon_{2}}>\mu$.}
\label{subsubsec:musmall}

In this limit, the second and third term of
equation~(\ref{eq:defphieps1}) are small compared to one, which leads to
\begin{equation}
\label{eq:phieps1mu}
\frac{\phi _{\epsilon _1}}{\mu}\simeq \left(\frac{1}{\sqrt{\pi
}}\frac{\mpl}{\mu}\right)^{1/5} .
\end{equation} 
Similarly, in equation~(\ref{eq:defphieps2}), the second and third term
are small compared to one, while the last term is small compared to the
fourth, so by keeping only the two dominant terms we obtain
\begin{equation}
\label{eq:phieps2mu}
\frac{\phi _{\epsilon _2}}{\mu}\simeq \left[\frac{5}{\pi
}\left(\frac{\mpl}{\mu}\right)^2\right]^{1/6} .
\end{equation} 
Comparing equation~(\ref{eq:phieps2mu}) and (\ref{eq:phieps1mu}) shows
that one still has $\phitwo>\phione$, and the slow-roll approximation
breaks down before inflation stops. This is confirmed in
figure~\ref{fig:potkklt_s}. However, this time, the field does not
necessarily reach $\phistrg$ first and non-slow-roll inflation may
continue between $\phitwo$ and $\phione$. Moreover, from $\phione$ to
$\phistrg$, the field encounters a transitory regime interpolating
between the conventional slow-roll and DBI dynamics.  As discussed
above, all the slow-roll formulae are strictly speaking no longer valid
after $\epsilon_{2}=1$. As a result, the region $\phi< \phitwo$ is only
accessible numerically and we will show in this way that the number of
e-folds occurring between $\phitwo$ and $\phione$ (or $\phistrg$) is
actually small. The key lesson to be learned from
\(\phi_{\epsilon_{2}}>\phi_{\epsilon_{1}}\) is that whenever we want to
rely on analytic results of the slow-roll calculus, we have to confine
the field to values \(\phi>\phi_{\epsilon_{2}}\) for consistency and
keep in mind that the results will be applicable only if the number of
e-folds occurring after $\phitwo$ remains negligible.

\par

To proceed further, let us summarise the possible field evolution given
that $\mu<\mpl$: The first option is that, also in this case, the field
reaches the value \(\phistrg\) first, \ie \(\phistrg > \phitwo\); this
is possible depending on the values of $\Nflux$ and $v$ for the chosen
background geometry [see equation~(\ref{eq:phistrg})]. In that case, the
entire field evolution corresponds to slow-roll inflation and is
described by the analytical trajectory~(\ref{eq:trajecapprox}).

\par

If $\phistrg<\phitwo$, then the situation is more complex and depends on
the exact order of $\phione$, $\phiDBI$ and \(\phistrg\). For instance,
if $\phione$ is the next field value to be reached after $\phitwo$, the
accelerated expansion would come to a halt here for a field \(\phi\)
with standard dynamics, rendering the number of e-folds between
$\phitwo$ and $\phione$ extremely small. However, the subsequent
evolution for $\phi<\phione$ could bring the field in the regime of DBI
dominance and inflation may re-start. At the same time, $\phistrg$ could
be located between $\phione$ and $\phiDBI$ and, as a consequence,
$\phiDBI$ would never be reached. All other combinations are of course a
priori possible. Which one is realised in practice depends on the value
of the free parameters characterising the model.

\section{String-intrinsic aspects}
\label{sec:stringeffects}

We now move on to investigating those aspects of the KKLMMT model due to
its stringy origin. In particular, two questions remain to be properly
addressed: the unusual DBI dynamics, and the appearance of a tachyon
once the proper brane distance approaches the string scale.

\subsection{When does inflation end?}
\label{subsec:infend}

The potential (\ref{eq:Vofphifull}) results from String Theory under the
assumption that only the lightest closed string modes (gravitons and the
RR particles) contribute to the brane interaction. When the proper
distance \(s\) between the branes becomes comparable to the string
length scale $\ells$, the exchange of heavier closed modes become
relevant, and an open string can stretch from one brane to the
other. The appearance of a tachyon triggers the brane annihilation
process, or reheating from a cosmological point of view.

\par

We already calculated the field value \(\phistrg\) at which the tachyon
appears in equation~(\ref{eq:phistrg}). The crucial question is, does
the field first reach the value \(\phistrg\) or \(\phi_{\epsilon_{2}}\)?
If \(\phi_{\epsilon_{2}}<\phistrg\), inflation ends at brane
annihilation for \(\phi_{\rm end}=\phistrg\) and the field \(\phi\)
spends its entire ``lifetime'' in the usual slow-roll regime. Such is
the course of events when $\mu>\mpl$. For $\mu<\mpl$, one has to compare
$\phi_{\epsilon_{2}}$ given in equation~(\ref{eq:phieps2mu}), with
$\phistrg$ given by equation~(\ref{eq:phistrg}). Their ratio reads
\begin{equation}
\frac{\phitwo}{\phistrg}=\left(\dfrac{M}{\mpl}\right)^{-1/3}
\Nflux^{-1/4} v^{1/12} 10^{1/6} \exp\left[-\left(4\pi \gs
  \dfrac{\Nflux}{v}\right)^{-1/4}\right] .
\label{eq:phi2eqstrg}
\end{equation}
This expression involves the energy scale $M/\mpl$ of the potential
because \(\phitwo\) and \(\phistrg\) both depend on $M/\mpl$, but not in
the same way. However, from an observational point of view, this scale
is involved in the amplitude of the cosmological perturbations and
certainly fixed by the CMB normalisation. Therefore, we may go further
and access the above ratio using the current WMAP measurement of the CMB
quadrupole.

\par

Let us call \(\phi_{*}\) the field value at the time when the wavelength
of the cosmological perturbations of physical interest today left the
Hubble radius during inflation. The slow-roll field trajectory allows us
to express \(\phi_{*}\) in terms of \(N_{*}\), the number of e-folds
between the time of Hubble exit and the end of inflation at \(\phi_{\rm
end}\). Using the approximated classical trajectory
(\ref{eq:trajecapprox}), this leads to
\begin{eqnarray}
\label{eq:phistarapprox}
\left(\frac{\phi _*}{\mu }\right)^{6}\simeq \left(\frac{\phi
_\uend}{\mu
}\right)^{6}+\frac{3}{\pi}\left(\frac{\mpl}{\mu}\right)^2
N_*\,.
\end{eqnarray} 
To proceed further, we have to specify when exactly inflation ends. But,
at the same time, this is also the question we try to address. Changing
the mechanism which stops inflation will change the scale $M/\mpl$ and,
hence, will affect the ratio~(\ref{eq:phi2eqstrg}).

\par

\begin{figure}
\begin{center}
\includegraphics[width=7.7cm]{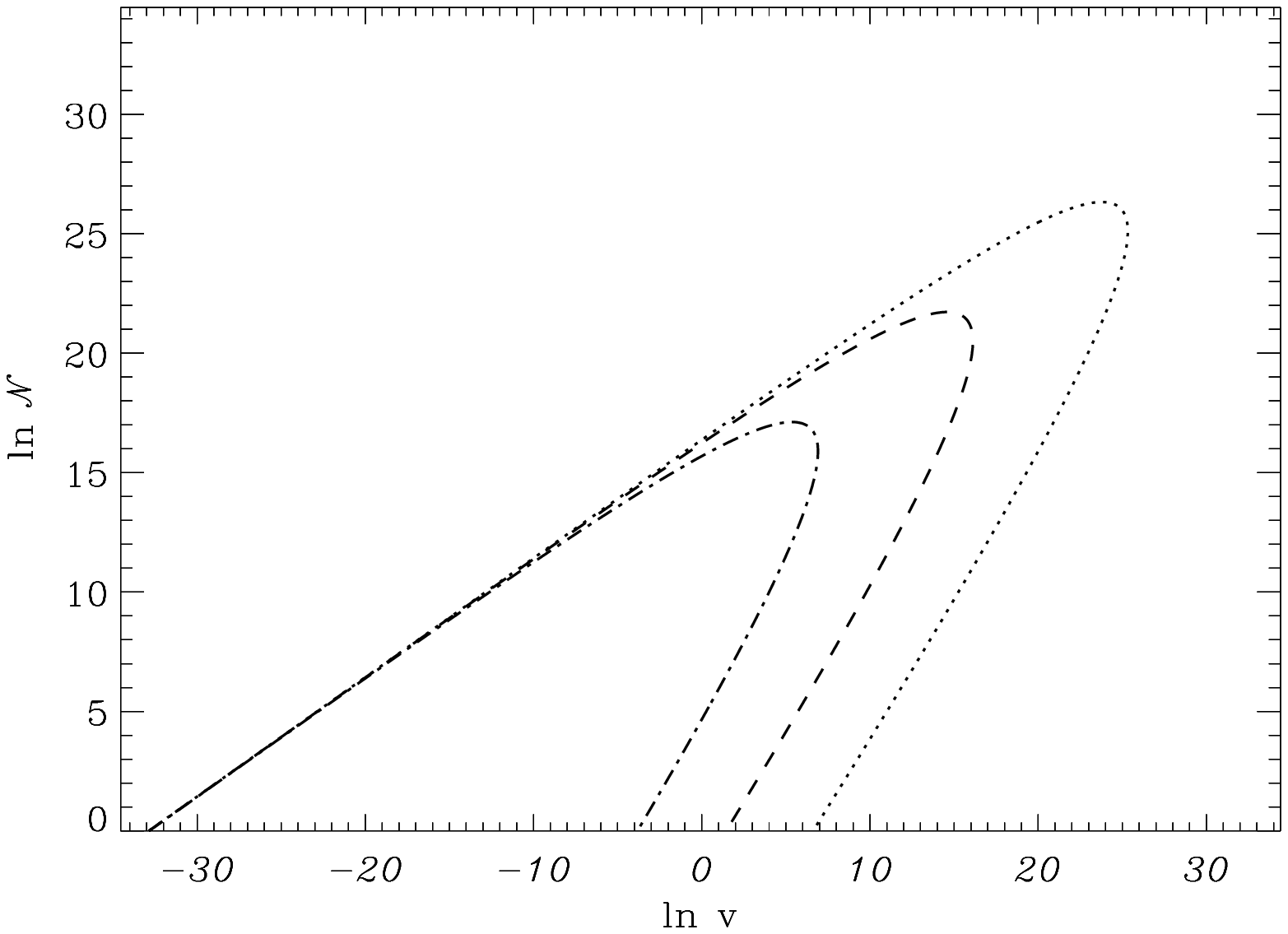}
\includegraphics[width=7.7cm]{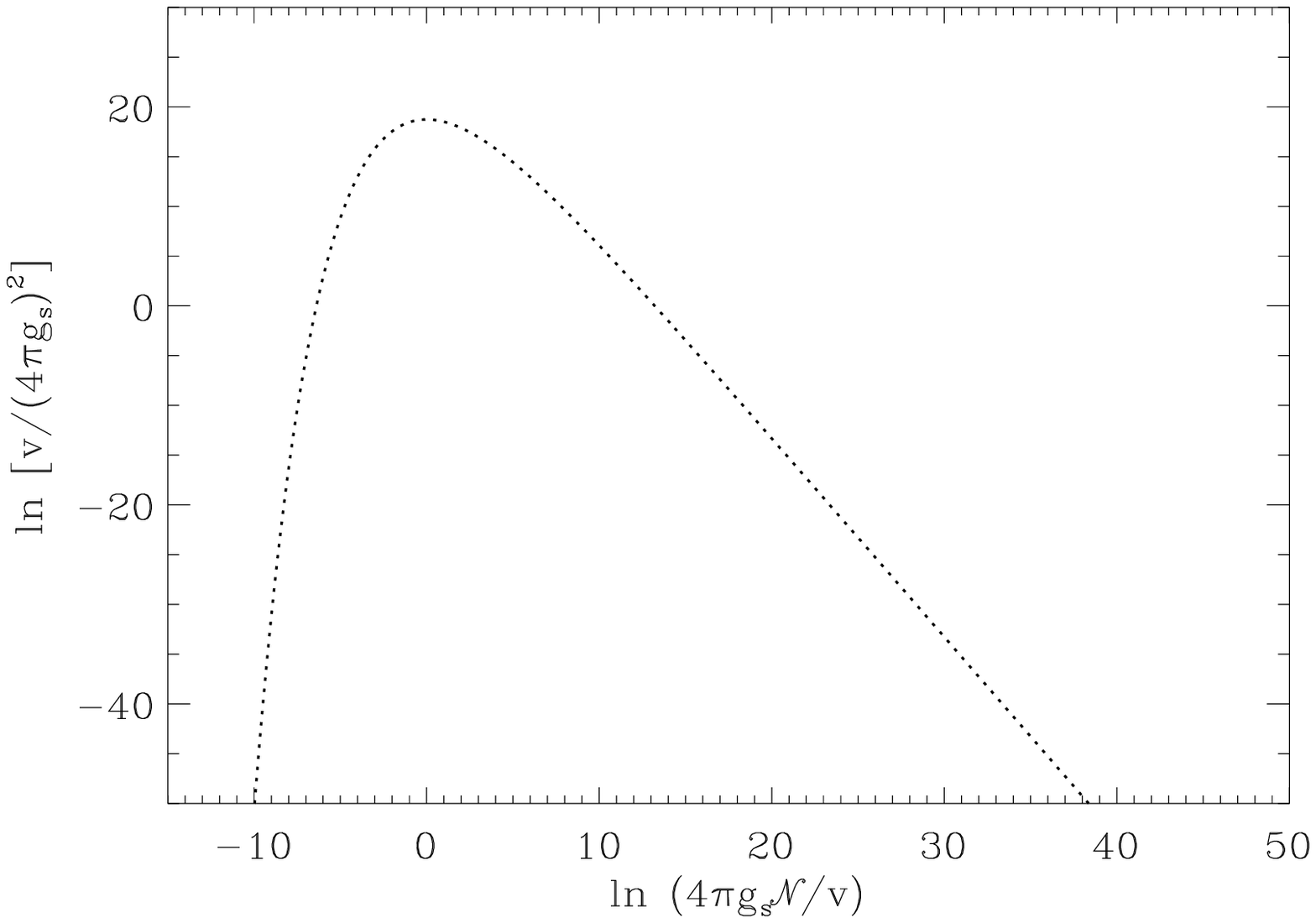} \caption{Left panel:
The contour $\phitwo=\phistrg$ in the plane $(\ln v,\ln\Nflux)$,
obtained from equations~(\ref{eq:phi2eqstrg}) and
(\ref{eq:phi2eqstrgnorm}) using the normalisation given by the CMB
quadrupole with $N_*=50$ [see equation~(\ref{eq:scaleMclas})]. The
dotted line corresponds to $\gs=0.1$, the dashed line to $\gs=10^{-3}$
and the dotted-dashed one to $\gs=10^{-5}$. In each case, the area
enclosed by the contour is the part of the parameter space where the
slow-roll conditions are violated before brane annihilation and
$\phistrg$ does not play an important role for the end of inflation. It
is clear from this plot that the contour sensitively depends on the
value of $\gs$. However, such a dependence can be absorbed by an
appropriate rescaling of the parameters as shown in the right panel. The
same contour $\phitwo=\phistrg$ is represented in the plane $(\ln
x,\ln\bar{v})$, these parameters being defined in
equation~(\ref{eq:rescaledparam}). It is universal for all values of the
string coupling $\gs$.}  \label{fig:rapp}
\end{center}
\end{figure}

It is convenient to determine the frontier $\phitwo=\phistrg$ in the
parameter space. Setting for convenience $\phi_\uend=\phitwo$ in
equation~(\ref{eq:phistarapprox}) gives
\begin{equation}
\label{eq:phistarclas}
\frac{\phi _*}{\mu }\simeq \left[\frac{3}{\pi }
\left(\frac{\mpl}{\mu
}\right)^2\left(N_*+\frac{5}{3}\right)\right]^{1/6}\, .
\end{equation}
Note that, had we used \(\phi_{\rm end}=\phi_{\epsilon_{1}}\) instead,
the result would be (\ref{eq:phistarclas}) up to the replacement
$N_*+5/3 \rightarrow N_{*}$. Therefore, ending inflation at
\(\phi_{\epsilon_{2}}\) instead of \(\phi_{\epsilon_{1}}\) only causes a
small shift in $N_*$. This is expected since, as soon as $\epstwo>1$,
slow-roll is violated and the field starts to evolve rapidly: even if
the precise evolution can only be probed numerically (see
section~\ref{sec:dbi}), the number of e-folds spent in this regime is
generically small. Inserting the result~(\ref{eq:phistarclas}) into the
expressions~(\ref{eq:eps1kklt}) and (\ref{eq:eps2kklt}) for the
slow-roll parameters, one arrives at 
\begin{eqnarray}
\label{eq:sr1approxclas}
\epsilon _1 & \simeq & \frac{1}{\pi }\left(\frac{\mpl }{\mu
}\right)^2 \left[\frac{3}{\pi }\left(\frac{\mpl }{\mu
}\right)^2\left(N_*+\frac{5}{3}\right)
\right]^{-5/3},\\
\label{eq:sr2approxclas}
\epsilon _2 & \simeq & \frac{5}{3 N_*}
\left(1+\frac{5}{3N_*}\right)^{-1}\simeq
\frac{5}{3 N_*}\, .
\end{eqnarray} 
As can be seen in these equations, the value of $\epsilon _1$ is
generically much smaller than $\epsilon _2$. Moreover, the quantity
\(\epsilon_{2}\) does not depend on the scale $\mu/\mpl $, at first
order. The CMB quadrupole normalisation gives the additional
relation~\cite{Martin:2006rs}
\begin{equation}
\label{eq:WMAPnormalization}
\frac{V_*}{\mpl ^4}\simeq \frac{45 \epsilon
_1}{2}\QQoverTT\, ,
\end{equation}
where, for $\phistar \gg \mu$, one may approximate $V_*\simeq
M^4$. Using the above expression (\ref{eq:sr1approxclas}) for
\(\epsilon_{1}\), we find the relations
\begin{eqnarray}
\frac{M}{\mpl}&\simeq&\left(45 \QQoverTT \right)^{3/8}
\left(6N_{*}+10\right)^{-5/8} v^{-1/8},
\label{eq:scaleMclas}\\
\frac{\mu}{\mpl}&\simeq&\left(45 \QQoverTT \right)^{3/8}
\left(6N_{*}+10\right)^{-5/8} (2\pi)^{-1/2} v^{-3/8}.
\label{eq:scalemuclas}
\end{eqnarray}
The numerical value of the WMAP quadrupole is~\cite{Hinshaw:2006ia}
\begin{equation}
\dfrac{Q_\mathrm{rms-PS}}{T} \simeq 6\times 10^{-6}.
\end{equation}
Inserting this expression into equation~(\ref{eq:phi2eqstrg}), the ratio
$\phitwo/\phistrg$ becomes a function of $\Nflux$ and $v$ only, at fixed
string coupling $\gs$. Explicitly, one obtains
\begin{equation}
\frac{\phitwo}{\phistrg} = \calC \Nflux ^{-1/4} v^{1/8} 10^{1/6}
\exp\left[-\left(4\pi \gs \frac{\Nflux}{v} \right)^{-1/4} \right] ,
\label{eq:phi2eqstrgnorm}
\end{equation}
where the constant $\calC$ reads
\begin{equation}
\calC = 10^{1/6} (6N_{*}+10)^{5/24} \left(45 \QQoverTT
\right)^{-1/8} .
\end{equation}
In the plane $(v,\Nflux)$, the condition $\phitwo=\phistrg$ is a curve
separating the plane into the two domains $\phistrg<\phitwo$ and
$\phistrg>\phitwo$. Notice that the shape and position of the frontier
depend on the value of $\gs$; this is illustrated in the left panel of
figure~\ref{fig:rapp}.

It is convenient to re-scale the parameters $\gs$, $\Nflux$ and $v$ to
minimise the dependence on $\gs$ of the contour
$\phi_{\epsilon_{2}}=\phistrg$. For this purpose, we define the new
variables
\begin{equation}
\label{eq:rescaledparam}
x \equiv 4\pi \gs \dfrac{\Nflux}{v}\, ,\qquad 
\bar{v}\equiv \frac{v}{(4\pi
  \gs)^{2}}\, .
\end{equation}
Usually, the analysis is performed with some specific values of
$\gs$. In this paper, our strategy is different and more general. With
the help of the above rescaling, our results will be valid for any
values of the coupling constant. In terms of these new parameters, the
condition $\phistrg=\phitwo$ from equation~(\ref{eq:phi2eqstrgnorm})
now reads, in logarithmic units
\begin{equation}
\label{eq:ratiocontour}
\ln\bar{v}=8\ln \calC-2\ln x-8x^{-1/4}.
\end{equation}
This contour is represented in figure~\ref{fig:rapp} (right panel). All
dependence of this line on \(\gs\) has been absorbed in the rescaling,
so one can now state universally that inside the contour, slow-roll
breaks down before the model-intrinsic instability is reached
($\phistrg<\phitwo$), whereas outside, slow-roll inflation proceeds all
the way until brane annihilation ($\phistrg>\phitwo$). In fact, this
result is more general and we demonstrate in the following that the
rescaling permits to absorb the $\gs$ dependence in all the equations
expressing a physically relevant condition for the model.

\par

Note that in the above discussion, we have used $\phiend=\phitwo$ to
derive the normalisation (\ref{eq:scaleMclas}). In doing so, we have in
fact ignored the other ``stringy'' characteristic of the KKLMMT model,
namely the DBI kinetic term. While brane evolution definitely ends at
$\phistrg$, it is possible that, due to the unusual dynamics, a phase of
``DBI inflation'' occurs even after violation of the slow-roll
conditions. If a considerable number of e-folds could be produced in the
DBI regime, the value of $\Nstar$ used in equation~(\ref{eq:scaleMclas})
would no longer be correct. In the following, we discuss for which field
value $\phiDBI$ those effects will be important and how much expansion
may occur outside the slow-roll regime.

\subsection{When is the DBI regime important?}
\label{sec:dbi}

An order of magnitude of the field value $\phiDBI$ for which the DBI
regime is relevant can be obtained from the potential
(\ref{eq:Vofphigeneral}) by use of the condition
(\ref{eq:DBIdominance}):
\begin{equation}
\phidotDBI^2 \simeq V(\phiDBI)\, .
\end{equation}
For a standard kinetic term ($\gamma=1$), this value would precisely
coincide with $\epsone=1$. This suggest that the DBI regime appears when
standard inflation ends. Of course, given the increasingly non-canonical
dynamics as the field is approaching $\phiDBI$, the standard slow-roll
formula is no longer valid and should be replaced by its generalised
form
\begin{equation}
\label{eq:epsonedef}
\epsone \equiv -\dfrac{\ud \ln H}{\ud N} = \dfrac{1}{2} \kappa \gamma 
\left(\dfrac{\ud \phi}{\ud N} \right)^2 = \dfrac{2}{\kappa
\gamma} \left(\dfrac{\ud \ln H}{\ud \phi} \right)^2,
\end{equation}
where use has been made of equation~(\ref{eq:gamma}) and of the exact
expression~(\ref{eq:friedman}). Therefore, in the ultra-relativistic
limit $\gamma \gg1$, equation~(\ref{eq:epsonedef}) suggests that the
expansion of the universe may accelerate even with a steep potential. If
such a second phase of inflation occurs for more than typically $60$
e-folds, then the slow-roll phase becomes observationally irrelevant and
the DBI regime crucial. We address this issue by analysing the DBI
regime using both analytical and numerical methods.

\par

The above expression for $\epsone$ can be further simplified by using
equation~(\ref{eq:gamma}) to express the field derivatives in terms of
$\gamma$, $H$ and $T$ solely. The Friedmann--Lema\^{\i}tre equation
(\ref{eq:friedman}) can then be used to remove any explicit dependency
in the Hubble parameter in favour of $\gamma$:
\begin{equation}
\label{eq:hubbledbi}
H^2 = \kappa \dfrac{V}{3 - \dfrac{2\epsone}{1 + \gamma^{-1}}}\,.
\end{equation}
As a result, the first slow-roll parameter reads, in terms of
$\gamma$, $V$ and $T$
\begin{equation}
\label{eq:epsonedbi}
\epsone = \dfrac{3}{2} \dfrac{1 - \gamma^{-2}} {1 +
  \gamma^{-1} \left(\dfrac{V}{T} - 1 \right)}\,.
\end{equation}
Equations~(\ref{eq:hubbledbi}) and~(\ref{eq:epsonedbi}) are exact and
clearly enhance the effect of $\gamma$. In the limit $\gamma \rightarrow
1$, one recovers the usual slow-roll formulae obtained with a standard
kinetic term, whereas an approximate analytic solution can be derived in
the ultra-relativistic limit $\gamma \gg 1$. Under the assumption
$V/T=\order{1}$, at leading order in $\gamma^{-1}$, one gets
\begin{equation}
\label{eq:epsonedbilo}
\epsone = \dfrac{3}{2} + \order{\gamma^{-1}},
\end{equation}
showing that deep in the DBI regime the universe is not inflating but
expands as in a matter dominated era. In our case $V/T=2$ which
ensures the validity of the above expansion. Notice, however, that
this relationship does no longer hold if one considers an additional
term in the potential, \eg $m^2 \phi^2$. In that case the $V/T$ term
in equation~(\ref{eq:epsonedbi}) can no longer be neglected and
inflation may indeed proceed in the DBI regime~\cite{Bean:2007hc,
  Peiris:2007gz}.

\par

Although the universe is not inflating for $\gamma \gg 1$, we still have
to check that the number of e-folds the field spends in the DBI ``matter
dominated'' era remains small. From equations~(\ref{eq:epsonedef}) and
(\ref{eq:epsonedbilo}), the Hubble trajectory at leading order in
$\gamma^{-1}$ reads
\begin{equation}
\label{eq:hubbledbilo}
H(N) = \Htrans \exp\left[-\dfrac{3}{2}\left(N-\Ntrans\right)\right],
\end{equation}
where $\Htrans$ and $\Ntrans$ respectively denote the Hubble
parameter and the e-fold at which the limit $\gamma \gg 1$ becomes
relevant. They could, for instance, be defined as the ones
corresponding to the end of the slow-rolling phase provided that
the transitory regime occurring between the conventional slow-roll and
the region of DBI dominance is short. The field evolution at
leading order is given in terms of $\gamma$ by
equation~(\ref{eq:epsonedef}) and reads
\begin{equation}
\label{eq:fieldimpdbilo}
\kappa \left(\dfrac{\ud \phi}{\ud N}\right)^2 = 3 \gamma^{-1} +
\order{\gamma^{-2}}.
\end{equation}
The system of equations is closed by using expression
(\ref{eq:friedman}) for the Hubble parameter
\begin{equation}
\label{eq:friedmandbilo}
H^2 = \dfrac{\kappa T}{3} \left(\gamma + 1 \right) ,
\end{equation}
which is exact since, in our case, one has $V/T=2$. As a consequence,
one gets the field trajectory (neglecting one compared to $\gamma\gg1$)
\begin{equation}
\dfrac{\ud \phi}{\ud N} \simeq - \dfrac{\sqrt{T(\phi)}}{\Htrans}
\exp\left[\dfrac{3}{2}\left(N-\Ntrans\right) \right].
\end{equation}
This expression can be implicitly integrated in terms of the Gauss
hyper-geometric function through use of
equations~(\ref{eq:Vofphigeneral}) and (\ref{eq:Vofphifull}), yielding
\begin{eqnarray}
\label{eq:efolddbilo}
N & = \Ntrans + \dfrac{2}{3} \ln\left(1 + \dfrac{3}{\sqrt{2}}
\dfrac{\mu}{M} \dfrac{\Htrans}{M} \left\{
\dfrac{\sqrt{1+\muophi_{\phantom{1}}^4}}{\muophi} -
\dfrac{\sqrt{1+\muophitrans^4}}{\muophitrans}
 \right. \right. \nonumber \\ 
& -  \left. \left. \dfrac{2}{\muophi}
\hypergauss{-\dfrac{1}{4}}{\dfrac{1}{2}}
{\dfrac{3}{4}}{-\muophi^4} +
\dfrac{2}{\muophitrans} \hypergauss{-\dfrac{1}{4}}{\dfrac{1}{2}}
{\dfrac{3}{4}}{-\muophitrans^4} \right\} \right),
\end{eqnarray} 
where $\muophi\equiv \mu/\phi$. A more illuminating form can be obtained
in both limits $\muophi \gg 1$ and $\muophi \ll 1$. In the limit
$\muophi \ll 1$, that is to say $\phi \gg \mu$, using the Taylor series
defining the hyper-geometric function, equation~(\ref{eq:efolddbilo})
becomes
\begin{equation} 
N \simeq \Ntrans
+ \dfrac{2}{3} \ln\left[1+\dfrac{3}{\sqrt{2}} \dfrac{\Htrans}{M}
\dfrac{\mu}{M} \left(\dfrac{1}{\muophitrans} - \dfrac{1}{\muophi}
\right) \right],
\end{equation}
which gives for the field
\begin{equation}
\label{eq:largefielddbilo}
\dfrac{\phi}{\mu} \simeq \dfrac{\phitrans}{\mu} - \dfrac{\exp
\left[\dfrac{3}{2}
\left(N-\Ntrans\right)\right]-1}{\dfrac{3}{\sqrt{2}} 
\dfrac{\Htrans}{M} \dfrac{\mu}{M}}\,.
\end{equation}
Similarly, in the limit $\muophi \gg 1$ or $\phi \ll \mu $, using the
linear transformation formulae associated with the Gauss hyper-geometric
function~\cite{Abramovitz:1970aa}, equation~(\ref{eq:efolddbilo}) yields
\begin{equation}
N \simeq \Ntrans + \dfrac{2}{3} \ln\left[1+\dfrac{3\sqrt{2}}{2}
\dfrac{\Htrans}{M} \dfrac{\mu}{M} \left(\muophi -
\muophitrans \right) \right],
\end{equation}
and, therefore, inverting the previous relation, one obtains
\begin{equation}
\label{eq:smallfielddbilo}
\dfrac{\phi}{\mu} \simeq \left\{
\dfrac{\mu}{\phitrans} + \dfrac{\exp
\left[\dfrac{3}{2}\left(N-\Ntrans\right)\right]-1}
{\dfrac{3\sqrt{2}}{2} \dfrac{\Htrans}{M} \dfrac{\mu}{M}}
\right\}^{-1}.
\end{equation} 

\begin{figure}
\begin{center}
\includegraphics[width=12.5cm]{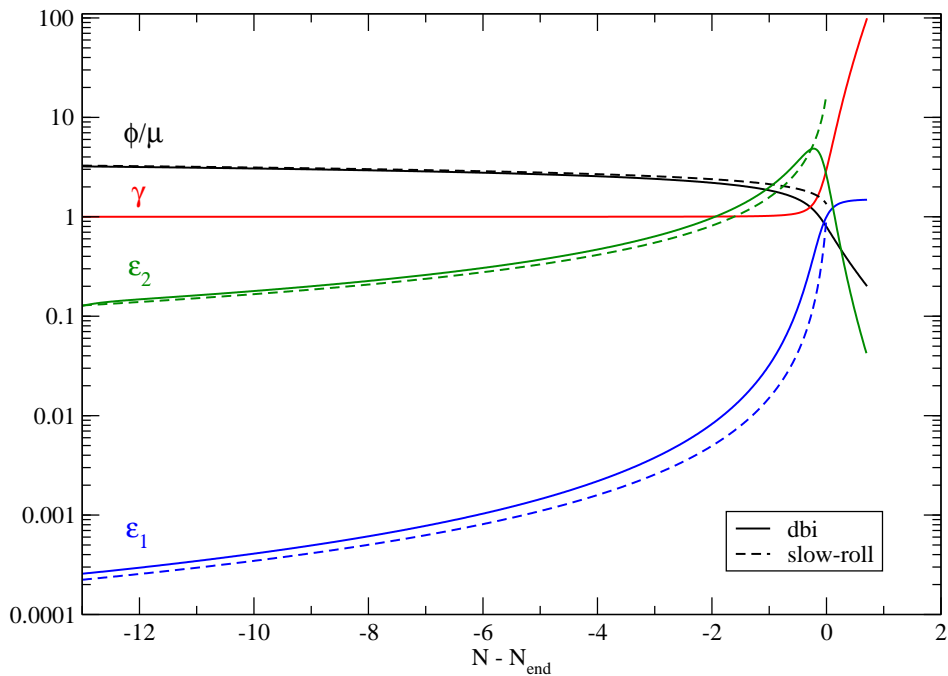} \caption{Evolution of the
field $\phi$, the Hubble-flow functions $\epsone$, $\epstwo$ and the DBI
parameter $\gamma$ in the last e-foldings of an extreme model living
close to the throat edge ($\mu=\mpl/\sqrt{32 \pi}$). This model has been
chosen to emphasise the DBI effects: the solid lines correspond to an
exact numerical integration of the full action (\ref{eq:generalaction2})
whereas the dashed lines are obtained by using the slow-roll
approximations (\ref{eq:eps1kklt}), (\ref{eq:eps2kklt}) and
(\ref{eq:trajectory}). The DBI regime smoothly connects the slow-roll
evolution ($\gamma\simeq 1$) to the ultra-relativistic matter-like
expansion ($\gamma \gg 1$). Note that in the model at hand, brane
annihilation occurs at $\phistrg>\mu$ preventing any observable effects
coming from the DBI evolution.}  \label{fig:numdbi}
\end{center}
\end{figure}
In fact, the limit $\muophi \gg 1$ ($\phi \ll \mu$) cannot be reached
in the model under scrutiny since the tachyonic instability occurs for
$\phi \simeq \phistrg \ge \mu$. As a result, the
ultra-relativistic DBI regime could only occur for $\phi>\mu$. As can
be seen in equation~(\ref{eq:largefielddbilo}), the corresponding
field evolution is exponentially fast implying that the number of
e-folds during which $\phi>\phistrg$ and $\gamma \gg 1$ is negligible,
provided $\phitrans/\mu$ and the denominator are not too large. This
is indeed the case for the following reasons. Firstly, as already
mentioned, equation~(\ref{eq:epsonedbi}) shows that $\gamma$ can
deviate from unity only if $\epsone$ is also of order unity, \ie the
field should be in the non-flat region of the potential and therefore
$\phitrans \gtrsim \mu$. Concerning the denominator, as can be seen
from equation~(\ref{eq:hubbledbi}), $\Htrans/M^2 \lesssim 1/\mpl$ and
this term is at most of order $\mu/\mpl$. The
constraint~(\ref{eq:phiuvmax}) coming from the size of the throat,
together with the fact that brane annihilation occurs at
$\phistrg>\mu$, require that
\begin{equation}
\label{eq:mumax}
\mu < \dfrac{\mpl}{\sqrt{2\pi}}\,,
\end{equation}
ensuring that $\Htrans \mu/M^2$ cannot be large. 

\par

To end this section, we have numerically checked that the previous
analysis was still qualitatively valid during the intermediate regime in
which neither the slow-roll nor the ultra-relativistic approximations
can be used. Figure~\ref{fig:numdbi} shows the last e-foldings of
evolution for an extreme model, close to the limit~(\ref{eq:mumax}), for
which the effects of the DBI regime can be seen. The differences between
the slow-roll approximation (under the standard kinetic term hypothesis)
and the exact DBI integration appear only during less than an e-fold and
close to the region where $\epsone=1$. Remembering that the effective
field description of the brane motion physically ends at $\phistrg$, we
conclude that for all practical purposes, the DBI regime has no
cosmological observable effects for the model based on the potential
(\ref{eq:Vofphifull}). The situation, however, can be entirely different
in models that include additional terms (e.g. $m^{2}\phi^{2}$) in the
potential, as shown in reference~\cite{Bean:2007hc}.

\section{Theoretical restrictions on the parameter space}
\label{sec:restrictions}

We are ready to conclude our analytical investigation of the KKLMMT
inflationary model by deriving restrictions on its free parameters. For
our numerical calculations and for comparison with the WMAP3 data, we
use the observable parameter set $(M,\mu,\Nflux)$, where $M$ and $\mu$
are related by equation~(\ref{eq:vdef}). There are essentially three
theoretical consistency relations to be satisfied.

\subsection{Constraints from the size of the throat}
\label{subsec:throatsize}

\begin{figure}
\begin{center}
\includegraphics[width=7.7cm]{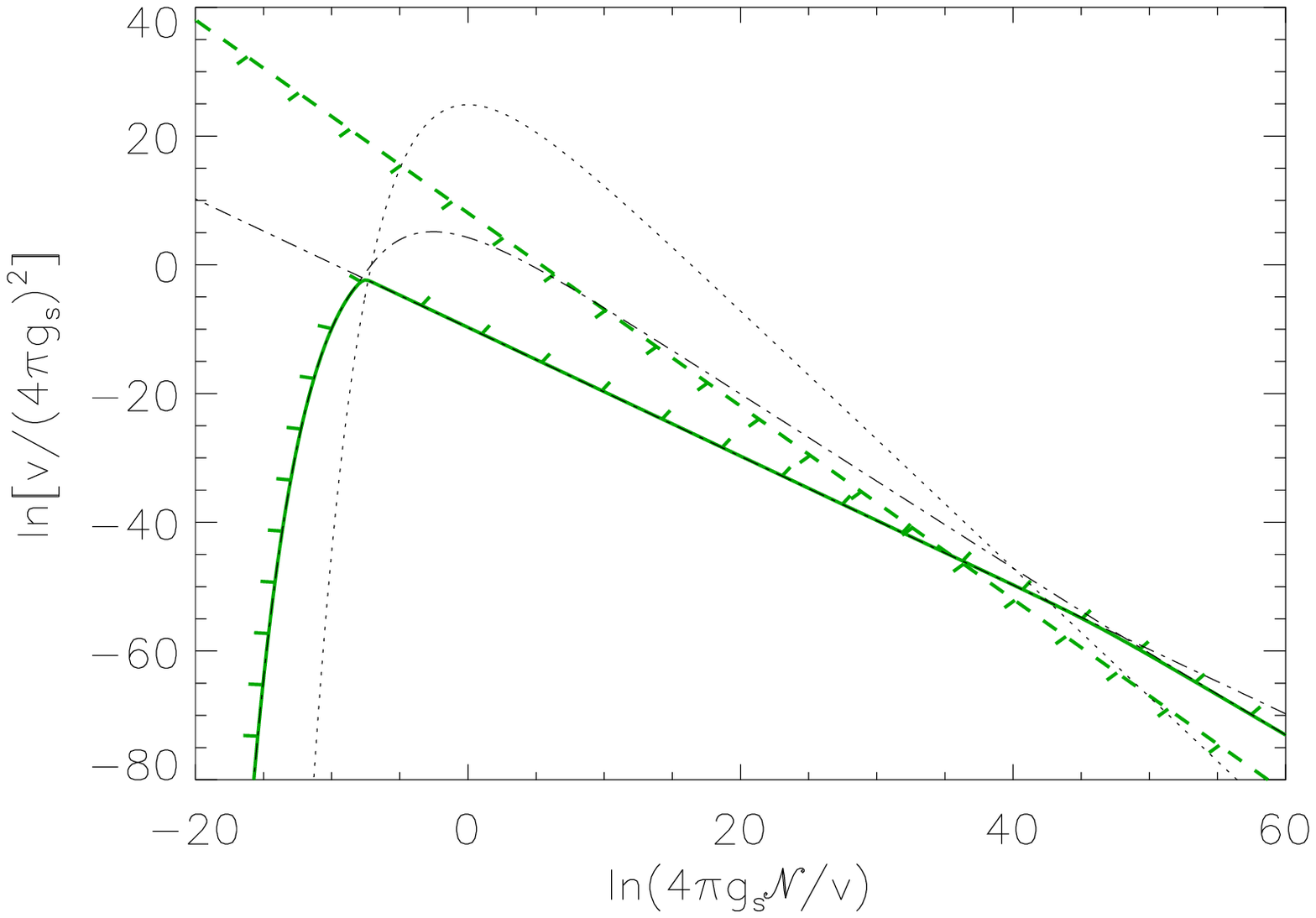}
\includegraphics[width=7.7cm]{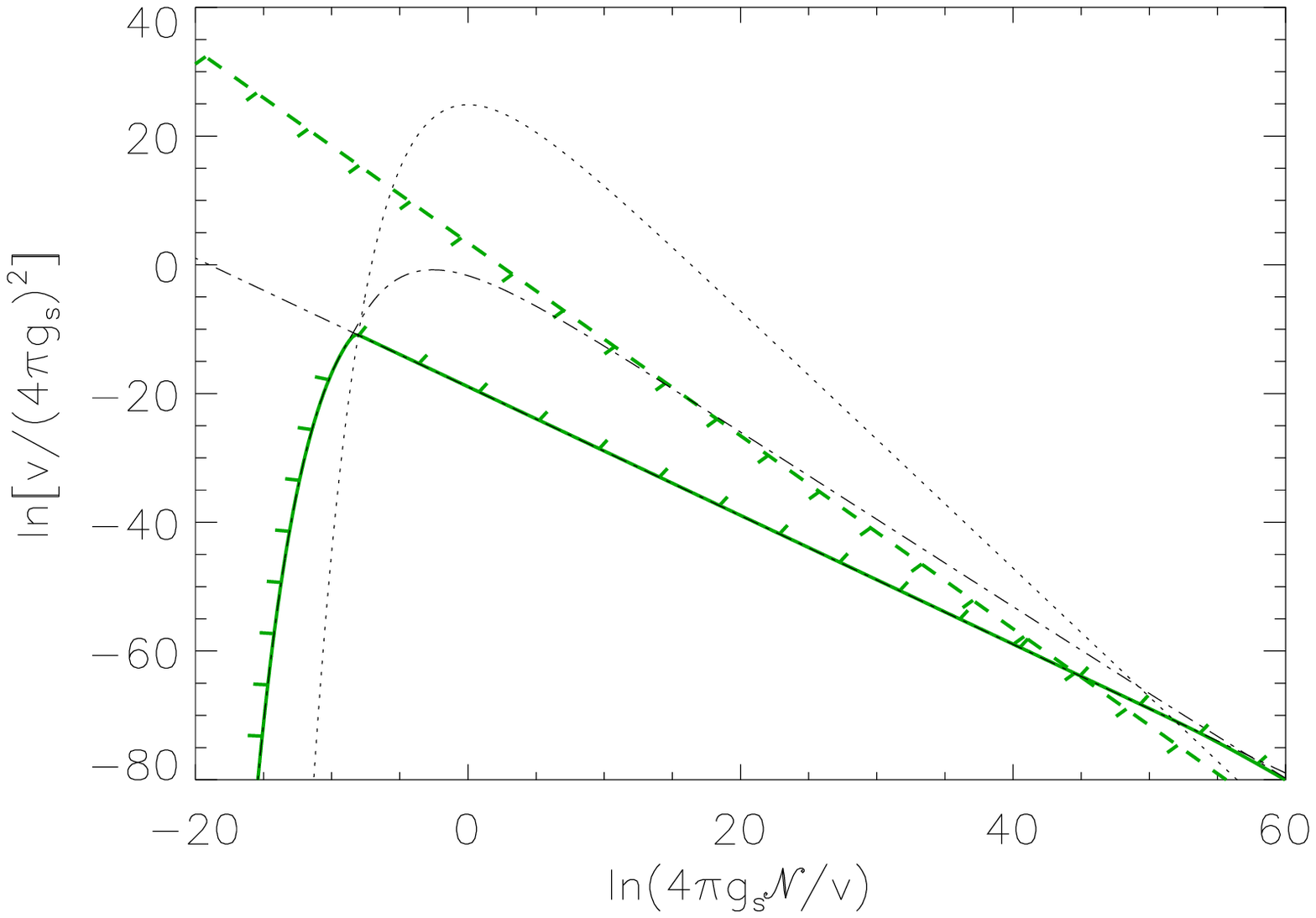} \caption{Allowed
regions (ticks in) in the rescaled parameter plane \((\ln
x,\ln\bar{v})\) for \(\alphas \mpl^{2}=1000\) (left panel) and \(\alphas
\mpl^{2}=10\) (right panel) with the fiducial values $N_*=50$ and
$N_{_{\rm T}}=60$. The black dotted curve is the contour
$\phitwo=\phistrg$ of figure~\ref{fig:rapp}. For all points located
inside this contour, the slow-roll conditions are violated before brane
annihilation. The green dashed line (with ticks down) represents the
volume ratio constraint (\ref{eq:throatsize-xvbar}), whose slope is
universal but whose offset depends on \(\alpha'\mpl^{2}\). Regions above
this line are therefore excluded. The solid green curve (with ticks up)
represents the condition that all of the brane evolution occurs within
one throat, and has been obtained through a numerical integration. All
points below that curve would not satisfy this condition. Its shape can
be piecewise analysed. In the region $\phistrg < \phitwo$ (inside the
black dotted contour), this is a straight line given by
equation~(\ref{eq:onethroat-phieps2}). The slope of this line is
universal, but the offset again depends on \(\alphas\mpl^{2}\). Outside
the dotted contour, namely for $\phistrg>\phitwo$, and in the limit
$\phistrg \gg \phitwo$, this boundary is described by
equation~(\ref{eq:onethroat-phistrings}). Combined, these pieces lead to
the solid green curve with ticks up; to make the shape of both pieces
visible individually, they have been extended outside their respective
domains of validity (dotted-dashed black curve).}
\label{fig:restrictions}
\end{center}
\end{figure}

For the model to be consistent, the warped throat in which the D3 and
anti-D3 branes are located should be smaller than the total size of the
compactified sub-manifold spanned by the extra dimensions. Earlier, we
derived the restriction (\ref{eq:throatsize}) from this condition. It
turns out that this condition can be conveniently re-written in terms of
our rescaled parameters ($x$, $\bar{v}$) defined in
equation~(\ref{eq:rescaledparam}) as
\begin{equation}
\label{eq:throatsize-xvbar}
\ln\bar{v}<\ln\left(\alpha'\mpl^{2}\pi\right)-\frac{3}{2}\ln x .
\end{equation}
This is a straight line in the plane $(\ln x,\,\ln \bar{v})$ whose
offset depends on $\alphas$ (see figure~\ref{fig:restrictions}). A
significant fraction of the parameter space is cut-out by this
requirement. Notice that, as announced above, the rescaling
(\ref{eq:rescaledparam}) has removed any dependence of this bound on
\(\gs\) and it is universal in the sense that it does not involve the
precise inflationary trajectory, nor the mechanism that ends inflation.

\subsection{Constraints from starting inflation inside the throat}
\label{subsec:onethroat}

For all the e-folds to occur inside one throat, we enforce
\(\phi<\phiuv\). Since \(\phi\) decreases during inflation, this
restriction applies to the initial field value \(\phi_{\rm in}\). On the
other hand, there is a lower bound on $\phiin$ such that at least
$\Ntot$ e-folds are produced. This bound in turn depends on the
mechanism ending the inflationary expansion.

\par

Let us start with the case in which the slow-roll approximation is
violated before brane annihilation. Making use of the classical
trajectory (\ref{eq:trajecapprox}), the initial field value necessary to
produce $\Ntot$ e-folds before $\phitwo$ can be derived from
equation~(\ref{eq:phieps2mu}) and reads
\begin{eqnarray}
\label{eq:phiineps2} 
\frac{\phi_{{\rm in},\epsilon_{2}}}{\mu}&\simeq  &\left\{
\frac{3}{\pi}\left(\frac{\mpl}{\mu}\right)^{2}
\left[\frac{5}{3}+N_{_{\rm T}}\right]\right\}^{1/6} .
\end{eqnarray} 
For the KKLMMT case, imposing the CMB normalisation
(\ref{eq:scalemuclas}) calculated earlier and using
equation~(\ref{eq:phiuv}), one can recast the bound $\phiintwo <\phiuv$
in terms of the rescaled variables
\begin{eqnarray}
\label{eq:onethroat-phieps2}
\ln\bar{v} > - \ln x - 4\ln \calD  ,
\end{eqnarray}
where the constant $\calD$ reads
\begin{eqnarray}
\calD = \left(\pi \alphas \mpl^{2}\right)^{-1/2}
\left(6N_{*}+10\right)^{5/12}\left(6\Ntot + 10 \right)^{-1/6} \left(45
\QQoverTT \right)^{-1/4} .
\end{eqnarray}
This is also a straight line in the plane $(x,\bar{v})$ and further
constrains from below the parameter space (see
figure~\ref{fig:restrictions}). As before, all the $\gs$ dependence
has been removed and only the $\alpha '$ one remains in the expression
of the offset. This offset also depends on $N_*$ and $N_{_{\rm
    T}}$ but only weakly because these two quantities appear in a
logarithm. In the following, the fiducial values $N_*=50$ and
$N_{_{\rm T}}=60$ will be used

\par

Notice that by using the normalisation~(\ref{eq:scalemuclas}) derived
from $\phiend=\phitwo$, the previous result is valid only inside the
contour~(\ref{eq:ratiocontour}). Therefore, we still have to treat the
case where inflation ends prematurely at $\phistrg>\phitwo$. Since the
entire field evolution occurs in the slow-roll regime, the
trajectory~(\ref{eq:trajecapprox}) remains valid all the time. The
minimal initial field value \(\phiinstrg\) leading to $\Ntot$ e-folds of
inflation is therefore obtained by setting \(\phiend=\phistrg\). This
leads to
\begin{equation}
\label{eq:phiinstrings} 
\frac{\phiinstrg}{\mu}=\left[\frac{3
    \Ntot}{\pi}\left(\frac{\mpl}{\mu}\right)^{2} +
  \left(\frac{\phistrg}{\mu}\right)^{6}\right]^{1/6} ,
\end{equation}
where \(\phistrg\) is given by equation~(\ref{eq:phistrg}). In order to
use the CMB normalisation, the new value of $\phistar$ has to be derived
from equation~(\ref{eq:phistarapprox}) where, now,
$\phiend=\phistrg$. One obtains
\begin{equation}
\frac{\phi_{*}}{\mu}=
\left[\left(\frac{\phistrg}{\mu}\right)^{6}
+\frac{3}{\pi}
\left(\frac{\mu}{\mpl}\right)^{-2}N_{*}\right]^{1/6} .
\end{equation}
The slow-roll parameter associated with this field value reads
\begin{eqnarray}
\epsilon_{1} &\simeq& \frac{1}{\pi}\left(\frac{\mu}{\mpl}\right)^{-2}
\left(\frac{\phi_{*}}{\mu}\right)^{-10} ,
\end{eqnarray}
and can be plugged into equation~(\ref{eq:WMAPnormalization}) to give
the CMB normalisation condition
\begin{eqnarray}
\left(\frac{M}{\mpl}\right)^6 &=& 45 \QQoverTT
v^{1/2} \left[\left(\frac{\phistrg}{\mu}\right)^{6}+6N_{*}
  \left(\frac{M}{\mpl}\right)^{-2} v^{1/2}\right]^{-5/3} .
\end{eqnarray}
Unlike the case $\phiend=\phitwo$, this equation for \(M/\mpl\) cannot
be made explicit but one can use equation~(\ref{eq:phieps2mu}) to
replace the last term in the square brackets and obtain
\begin{equation}
\label{eq:cfnorm}
\left(\frac{M}{\mpl}\right)^6=45 \QQoverTT v^{1/2}
\left[\left(\frac{\phistrg}{\mu}\right)^{6} +\frac{3N_{*}}{5}
  \left(\frac{\phitwo}{\mu}\right)^{6}\right]^{-5/3}.
\end{equation}
An explicit analytical solution can be derived in the limit \(\phistrg
\gg \phitwo\). Due to the presence of the $N_*$ term, this approximation
may be violated only for large $\Nstar$ values. But if we are not in
this extreme situation, in the limit $\phistrg \gg \phitwo$, one gets
\begin{eqnarray}
\frac{M}{\mpl} & \simeq & \left(45 \QQoverTT \right)^{1/6} v^{1/12}
\left(\frac{\phistrg}{\mu}\right)^{-5/3},
\label{eq:scaleMstring}\\
\frac{\mu}{\mpl} & \simeq & \left(45 \QQoverTT \right)^{1/6}
\left(2\pi\right)^{-1/2} v^{-1/6} \left( \frac{\phistrg}{\mu}
\right)^{-5/3}.
\label{eq:scalemustring}
\end{eqnarray}
Enforcing $\phiinstrg<\phiuv$ from equations~(\ref{eq:phiuv}) and
(\ref{eq:phiinstrings}) gives, in terms of the rescaled parameters,
\begin{eqnarray}
\ln\Biggl[1 &+ & 6\Ntot
\left(45 \QQoverTT \right)^{-1/3} x^{-2/3}\bar{v}^{-1/3}
\exp \left(-\frac{8}{3}x^{-1/4}\right)\Biggr]
< 4 x^{-1/4} \nonumber\\ & + & 2 \ln\bar{v}
+\frac{5}{2}\ln x - 3 \ln\left(\pi\alpha'\mpl^{2}\right)
- \ln \left(45 \QQoverTT \right).
\label{eq:onethroat-phistrings}
\end{eqnarray}
Again, there is no trace of \(\gs\) left after rescaling, but instead of
a straight line as was the case for
equations~(\ref{eq:throatsize-xvbar}) and (\ref{eq:onethroat-phieps2}),
this is an implicit contour in the $(\ln x,\,\ln\bar{v})$ plane. It is
represented in figure~\ref{fig:restrictions}. Let us recall that
equation~(\ref{eq:onethroat-phistrings}) is only valid far enough from
the contour $\phistrg=\phitwo$ in such a way that $\phistrg \gg
\phitwo$. In the intermediate regime where inflation does end by
instability at \(\phistrg\), but the two terms in the square brackets of
equation~(\ref{eq:cfnorm}) are of comparable value, the correct
normalisation of \(M/\mpl\) to the CMB quadrupole is only accessible
numerically and has been also represented in
figure~\ref{fig:restrictions} for convenience. It is remarkable that
when one normalises with either (\ref{eq:scaleMclas}) (obtained from
$\phiend=\phitwo$) or with (\ref{eq:scaleMstring}) (assuming $\phistrg
\gg \phitwo$), the $\gs$ dependence can be absorbed into the unique
rescaling given in equation~(\ref{eq:rescaledparam}).

\begin{figure}
\begin{center}
\includegraphics[width=7.7cm]{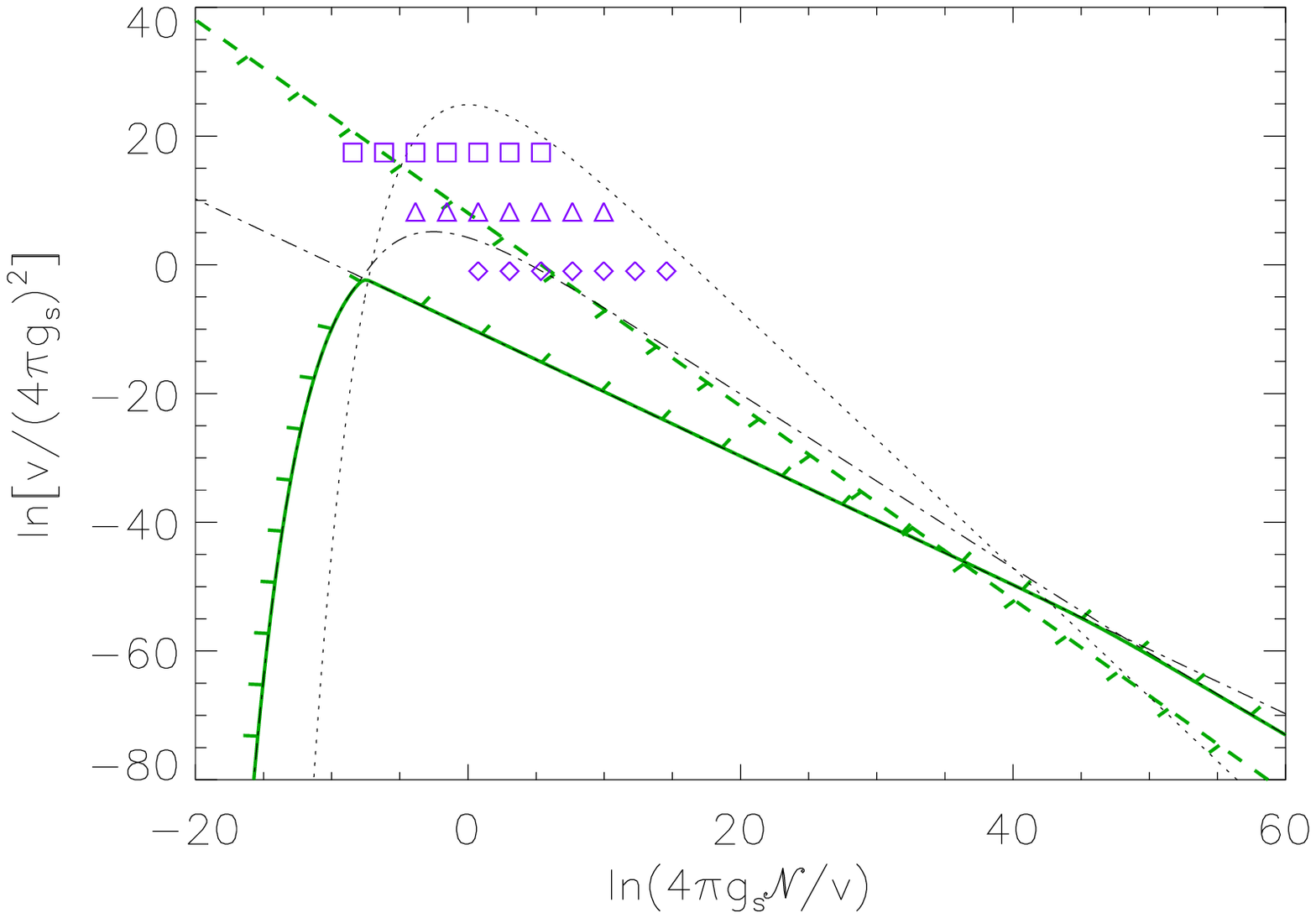}
\includegraphics[width=7.7cm]{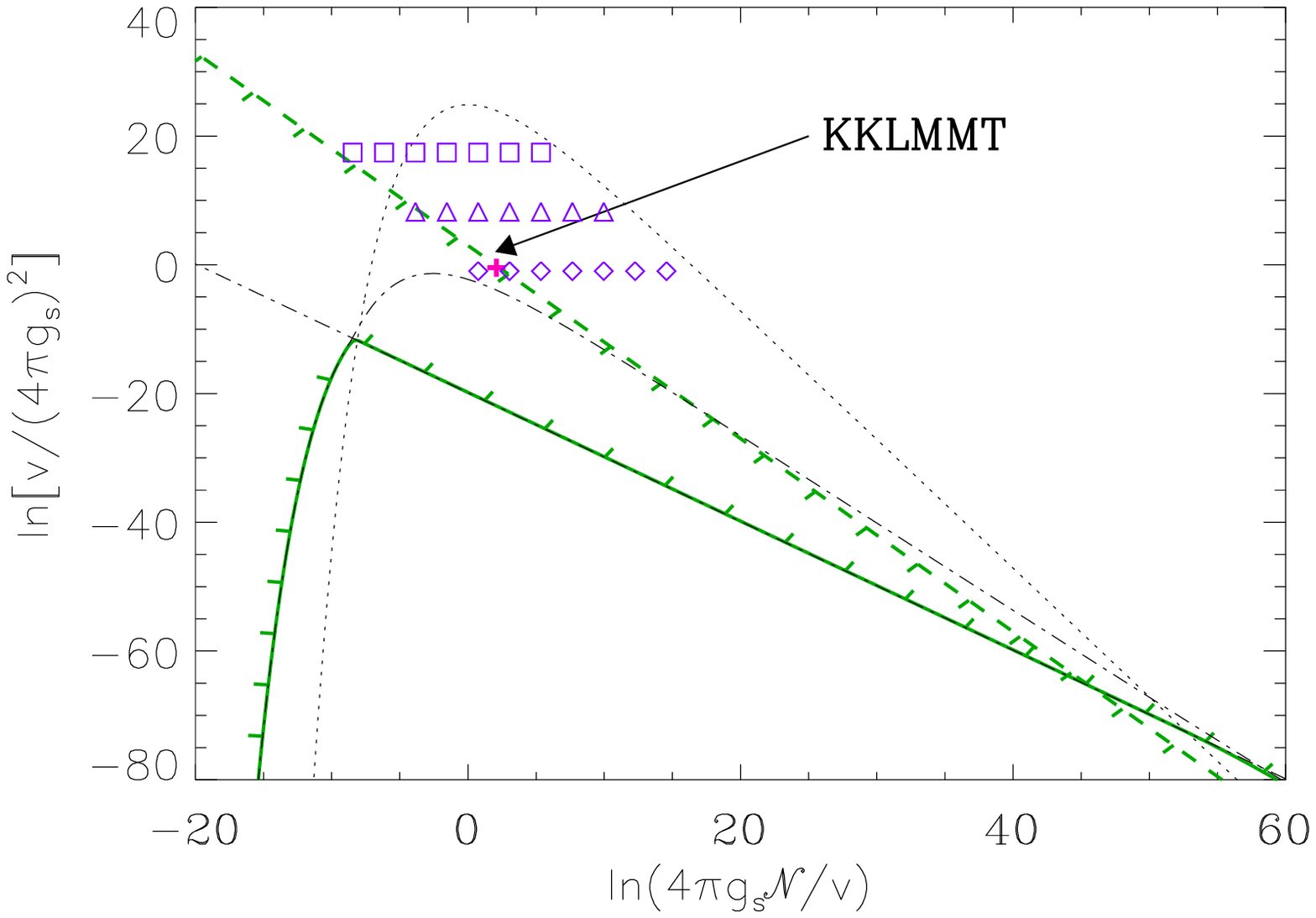} \caption{Plots similar
to those in figure~\ref{fig:restrictions}, for $\alphas\mpl^{2}=1000$ on
the left and for $\alphas\mpl^{2}=6.4$ (as chosen in
\cite{Kachru:2003sx}) on the right. The quantities $N_*$ and $N_{_{\rm
T}}$ are fixed to their fiducial values $N_*=50$ and $N_{_{\rm
T}}=60$. The additional symbols correspond to various models selected
according to the values of $v$, $\Nflux$ and $\gs$. All purple symbols
correspond to the KS throat value of $v=16/27$, and respectively have
$\Nflux=1$, $10$, $100$, $10^3$, $10^{4}$, $10^{5}$ and $10^{6}$ from
left to right. The diamond-shaped points are for $\gs=0.1$, the
triangles for $\gs=10^{-3}$, and the squares for $\gs=10^{-5}$. The
original KKLMMT model \cite{Kachru:2003sx} is indicated by a red cross
in the right panel; it corresponds to $v=1$, $\Nflux=160$ and $\gs=0.1$
at $\alphas\mpl^{2}=6.4$.} \label{fig:restrictions-kklt}
\end{center}
\end{figure}

Finally, in figure~\ref{fig:restrictions-kklt}, various concrete
models for different values of $v$, $\Nflux$ and $\gs$ are compared
with the consistency conditions derived above. In particular, one
notices that the volume ratio constraint is quite restrictive and that
many possible models are already ruled out by this condition.

\subsection{Constraints from stochastic inflation}
\label{subsec:constraintstocha}

If the field \(\phi\) starts out at a value comparable to $\phifluct$
given in equation~(\ref{eq:phifluct}), then stochastic effects will be
important and the classical field trajectory does no longer suffice to
describe the field evolution. As a minimum requirement, in the case
where brane annihilation occurs after \(\phitwo\), one has to impose
\begin{equation}
\label{eq:stochaeps2}
\phi_{{\rm in},\epsilon_{2}}\ll \phi_{\rm fluct} ,
\end{equation}
where $\phiintwo$ is given by equation~(\ref{eq:phiineps2}). Comparing
equation~(\ref{eq:phiineps2}) with equation~(\ref{eq:phifluct}), and
using the correct normalisation~(\ref{eq:scaleMclas}), we notice that
the dependence on $\Nflux$ and $v$ disappears. As a result, the
restriction coming from (\ref{eq:stochaeps2}) concerns the maximal total
number of e-folds $\Ntot$ achievable,
\begin{equation*}
\Ntot <- \frac{5}{3}+\left(3^{2}\,2^{11}\right)^{-1/5}\,
\left(6N_{*}+10\right) \left(45\QQoverTT \right)^{-3/5},
\end{equation*}
which gives $\max(\Ntot) \simeq 10^{7}$ for $\Nstar \simeq
50$. Similarly, if inflation ends at $\phistrg>\phitwo$, one imposes
\begin{equation}
\label{eq:stochastring}
\phiinstrg \ll \phi_{\rm fluct},
\end{equation}
to avoid quantum fluctuation dominance. In this case, using
equations~(\ref{eq:phifluct}) and (\ref{eq:phiinstrings}), together with
the normalisation given in equation~(\ref{eq:scaleMstring}), gives
\begin{eqnarray}
\ln\Biggl[1 & + & 6 \Ntot \left(45 \QQoverTT \right)^{-1/3} x^{-2/3}
    \bar{v}^{-1/3} \exp\left(-\frac{8}{3} x^{-1/4} \right) \Biggr]
\nonumber \\
& &  < \frac35\ln \frac34-\frac35\ln \left(45\QQoverTT\right)\, .
\end{eqnarray}
As usual, the rescaling allows us to get rid of any explicit dependence
in $\gs$. We do not pursue this issue further here. For a detailed study
of the presence of quantum fluctuation effects in the throat, see
\ref{app:stocha}.

\section{Implications of the WMAP slow-roll bounds}
\label{sec:srwmap}

\begin{figure}
\begin{center}
\includegraphics[width=7.7cm]{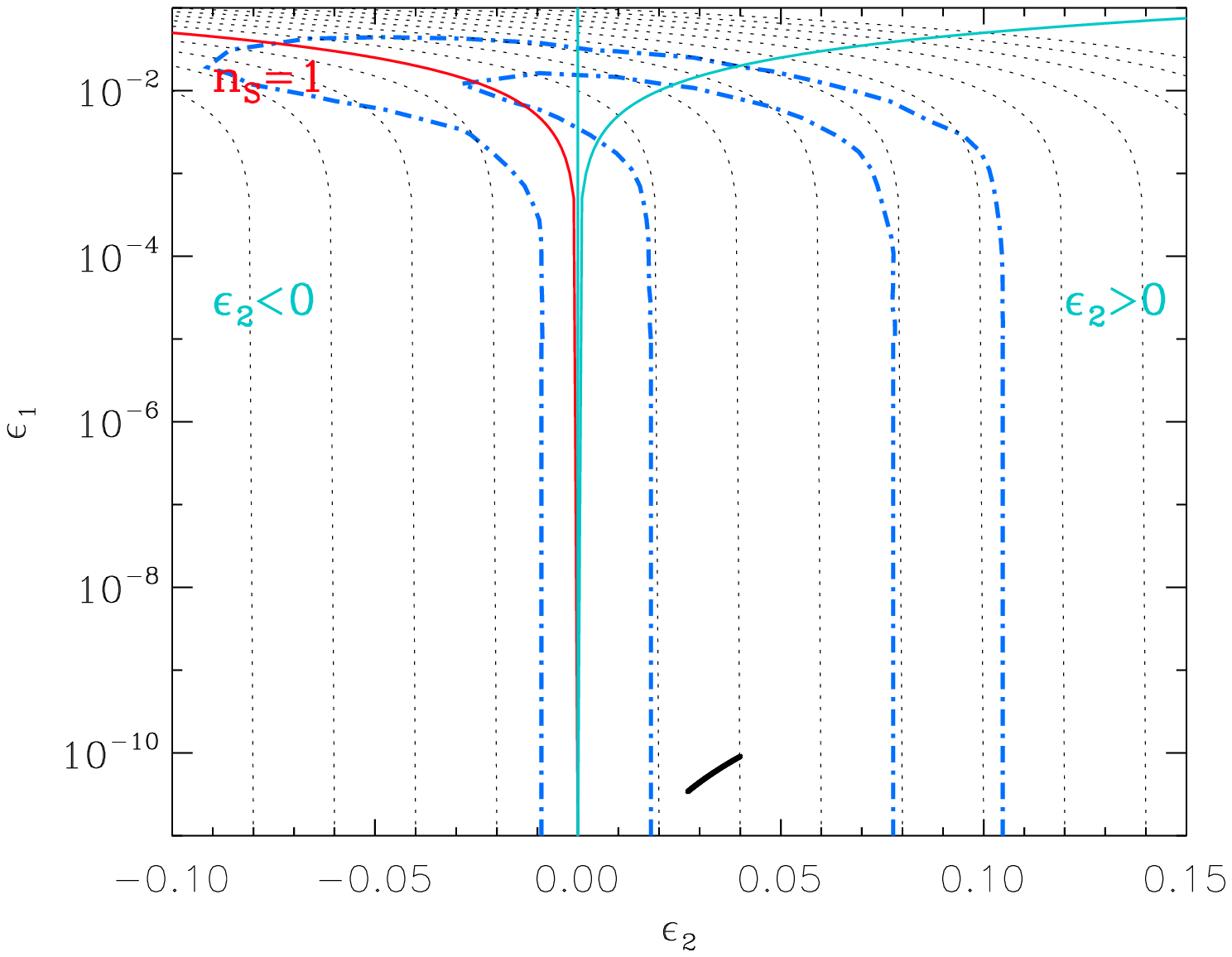}
\includegraphics[width=7.7cm]{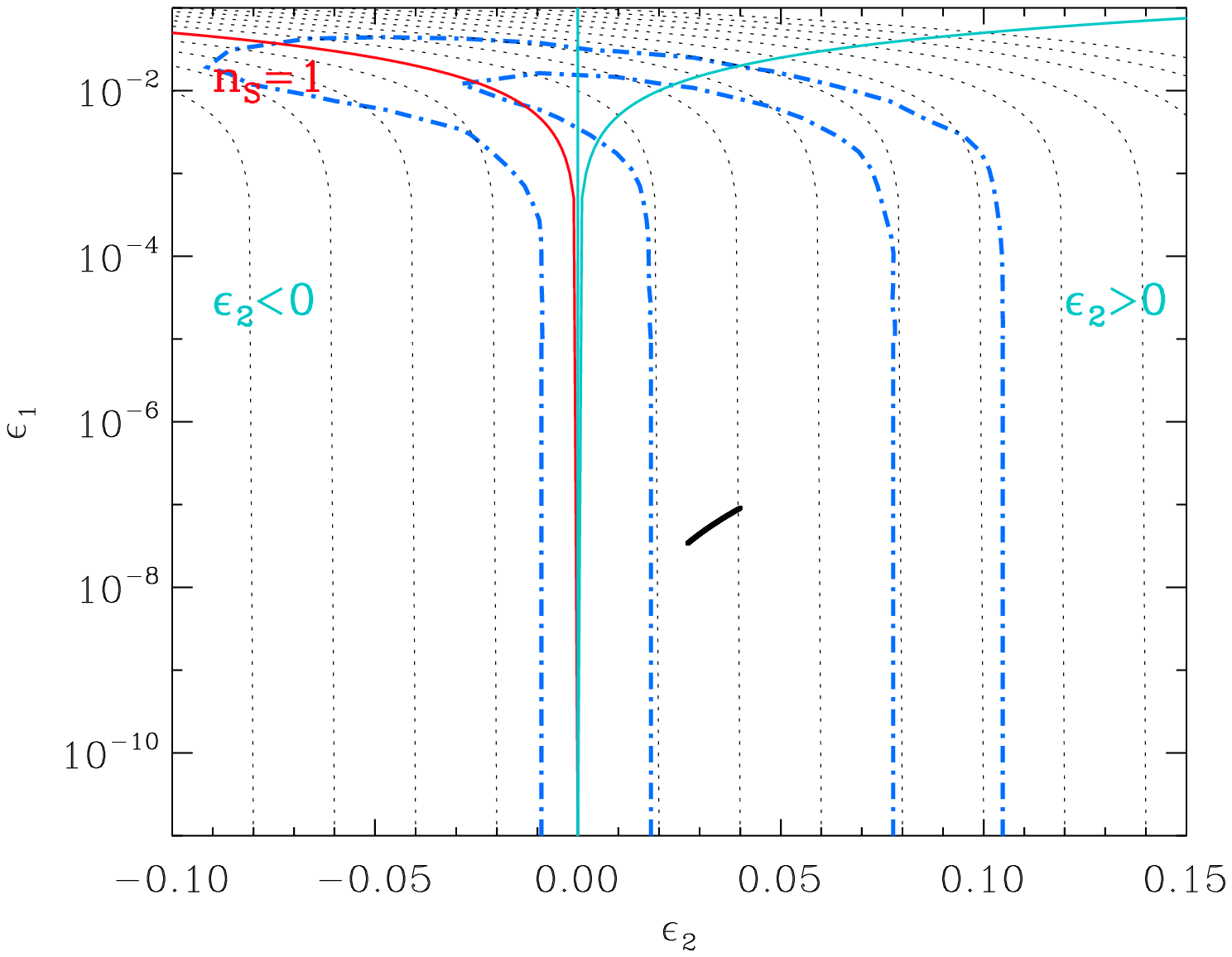} \caption{Left panel:
Predictions for the KKLMMT model with \(v=1\) in the case where one
considers that inflation stops at $\epstwo=1$. Right panel: Same as left
panel but with $v=10^{-6}$.}  \label{fig:kkltsr}
\end{center}
\end{figure}

In this section, the theoretically predicted values of the KKLMMT
slow-roll parameters are confronted with the third year WMAP data
(WMAP3). The WMAP3 bounds on the Hubble-flow parameters $\epsone$ and
$\epstwo$ have been discussed in reference~\cite{Martin:2006rs,
Kinney:2006qm, Finelli:2006fi} under various prior choices and
hypotheses. For the sake of robustness, in the following we use the
constraints derived in reference~\cite{Martin:2006rs} on the first order
Hubble-flow parameters obtained by marginalising over the second order
slow-roll parameter $\epsthree$ and under an uniform prior choice for
$\log(\epsilon_1)$ in $[-5,0]$ and for $\epstwo$ in $[-0.2,0.2]$. The
resulting one and two-sigma contour intervals of the two-dimensional
marginalised posteriors are represented in figure~\ref{fig:kkltsr}.

\par

Let us first consider the situation where violation of the slow-roll
conditions occurs before brane annihilation. In this case, if one
chooses for instance $N_*=50$, the CMB normalisation from
equations~(\ref{eq:scaleMclas}) and (\ref{eq:scalemuclas}) yields
\begin{equation}
\label{eq:wmapmuM}
\frac{\mu }{\mpl}\simeq 5.6\times
10^{-6} v^{-3/8},\qquad \frac{M}{\mpl} \simeq 1.4 \times
10^{-5} v^{-1/8}.
\end{equation}
Hence these two parameters become functions of $v$ only. The resulting
numerical values for $\epsone$ and $\epstwo$ are given by
equations~(\ref{eq:sr1approxclas}) and (\ref{eq:sr2approxclas}) and read
\begin{equation}
\epsilon_{1}\simeq4.8\times10^{-11}\,v^{-1/2},\qquad\epsilon_{2}\simeq
0.03\, .
\end{equation}
This has two very important consequences: it is clear that except for
very small values of the volume ration $v<10^{-18}$, $\epsone$ remains
negligible compared to $\epstwo$. In particular, since the spectral
indices of the scalar and tensor primordial power spectra, at first
order in the Hubble-flow parameters, read
\begin{eqnarray}
\label{eq:spectralindices}
\nS-1 &=& -2\epsilon_{1}-\epsilon_{2},\qquad \nT =-2\epsilon_{1},
\end{eqnarray}
we have unobservably small tensor modes and a scalar spectral index
\begin{equation}
\nS\simeq 0.97.
\end{equation}

In figure~\ref{fig:kkltsr}, we have studied the slow-roll predictions in
more detail. The slow-roll parameters have been calculated exactly in
the sense that equation~(\ref{eq:defphieps2}) has been solved
numerically, that is to say we have obtained the exact value of
$\phitwo$. The same has been done for $\phistar$. Then, the
corresponding values of the Hubble-flow parameters have been derived
with the help of equations~(\ref{eq:sr1approxclas}) and
(\ref{eq:sr2approxclas}) for different $N_*$ such that $N_*\in
[40,60]$. For each value of $N_*$, this gives a point in the space
$(\epsilon _1, \epsilon _2)$ and for the range under consideration, a
segment line spanning twenty e-folds (see figure~\ref{fig:kkltsr}). The
``left'' end of this segment line corresponds to the largest value of
$N_*$, namely $N_*=60$ and the ``right'' end corresponds to
$N_*=40$. The left panel assumes $v=1$, while the right one was plotted
for $v=10^{-6}$. The dotted-dashed blue contours are the WMAP3
constraints on the slow-roll parameters and corresponds to the $68\%$
and $95\%$ confidence intervals of the two-dimensional marginalised
posteriors. The dotted black lines are the lines of constant $\nS$, the
red one tracing the scale invariant case $\nS=1$. The figure confirms
the previous analysis. The gravitational waves level is generically very
low and the spectral index is slightly red in full agreement with the
WMAP3 data. Moreover, changing the size of the extra dimension changes
$\epsilon _1 $ without affecting $\epsilon _2$. However, for reasonable
values of $v$, $\epsilon _1 $ remains so small that there is no hope to
detect primordial gravitational waves in the future.

\par

Let us now turn to the situation where the parameters of the model are
such that inflation ends by instability at $\phistrg$. In this case, the
WMAP normalisation leads to the scale $M/\mpl$ given by
equation~(\ref{eq:scaleMstring}). Then, straightforward calculations
lead to the following expressions for the Hubble-flow parameters
\begin{eqnarray}
\epsilon _1 &=& \frac{v^{1/3}}{2\pi ^2}\left(45 \QQoverTT
\right)^{-1/3}\left(\frac{\phistrg}{\mu}\right)^{-20/3} ,\\ \epsilon
_2 &=& \frac{5v^{1/3}}{2\pi ^2}\left(45 \QQoverTT
\right)^{-1/3} \left( \frac{\phistrg}{\mu}\right)^{-8/3} .
\end{eqnarray}
In the limit $\phistrg/\mu \gg 1$, one has $\epsilon _1\ll \epsilon
_2\ll 1$ except maybe for extreme values of $v$. Therefore, $\nS$ is now
pushed towards one, while $\nT$ becomes even smaller for all values of
$N_*$. As a consequence, one expects this scenario to be disfavoured by
the data. Indeed, as can be noticed in figure~\ref{fig:kkltsr}, a scale
invariant power spectrum is not likely if the gravitational waves level
is low (this is no longer true if $\epsilon _1>10^{-2}$).

To conclude this section, let us recap what can be deduced from the
slow-roll analysis. Firstly, the situation where inflation ends by
violation of the slow-roll condition is favoured by the data as compared
to the case where inflation stops by brane annihilation. Secondly, one
expects the level of gravitational waves to be extremely low. The
stringy interpretation of this result is linked to the so-called Lyth
bound~\cite{Lyth:1996im,Baumann:2006cd} which relates the tensor to
scalar ratio to the total variation of the inflaton field value. In our
case, this gives $\Delta\phi=\sqrt{T_3}\left(r_{_{\rm
UV}}-r_{0}\right)\simeq\sqrt{T_3}r_{_{\rm UV}}=\phiuv$. Therefore, given
equation (\ref{eq:phiuvmax}), the volume bound on the throat limits the
gravitational wave contribution, as recovered here. Thirdly, the
constraint $\epsilon_1 \lesssim 3\times 10^{-2}$ implies the limit
\begin{equation}
\label{eq:limv}
\log\left(4\pi ^2v \right)\gtrsim -16\, .
\end{equation}
This is clearly not a very stringent constraint. In the next section, we
will see that, combined with the constraint on the size of the throat,
the data can lead to a better limit.

\section{WMAP constraints in the exact numerical approach}
\label{sec:mcmc}

In this section, the cosmological consequences of the KKLMMT model are
investigated using a numerical integration of the brane motion up to its
linear perturbations to extract the exact scalar and tensor primordial
power spectra. The tachyonic instability occurring at the bottom of the
throat triggering a reheating era is considered through a simple
phenomenological model. The predicted CMB power spectra can then be
computed by integrating the seeded cosmological perturbations through
the radiation and matter era and we have used a modified version of the
\CAMB code for this purpose~\cite{Lewis:1999bs}. This method allows a
Markov-Chain Monte-Carlo (MCMC) analysis of the WMAP third year data
involving the usual cosmological parameters together with the KKLMMT
parameters. A modified version of the \COSMOMC code~\cite{Lewis:2002ah}
has been used to derive the probability density distributions of the
cosmological, KKLMMT and reheating parameter values given the data.

\subsection{Method and hypotheses}
\label{subsec:mcmcmethod}

The numerical method used has been introduced in
references~\cite{Martin:2006rs,Ringeval:2007am} and consists in the mode
by mode integration of both the background and the perturbed quantities
till the end of inflation (see also Refs.~\cite{Salopek:1988qh,
Grivell:1999wc, Adams:2001vc, Tsujikawa:2002qx, Parkinson:2004yx,
Makarov:2005uh, Chen:2006xj}). As shown in section~\ref{sec:dbi}, the
DBI regime is not relevant for the model under scrutiny and we have
chosen to integrate the background equations stemming from the
action~(\ref{eq:generalaction2}) in the standard kinetic-term limit. The
resulting equations of motion are obtained from
equations~(\ref{eq:kglike}), (\ref{eq:epsonedef}) and
(\ref{eq:hubbledbi}) in the $\gamma=1$ limit and for the KKLMMT
potential (\ref{eq:Vofphifull}).

\par

In the longitudinal gauge, we consider the scalar and tensor linear
perturbations to the flat FLRW metric~(\ref{eq:flrwmetric}),
\begin{equation}
\ud s^2 = a^2\left\{ - (1+2\Phi) \ud \eta^2 + \left[(1-2\Psi)
  \delta_{ij} + \tens_{ij} \right] \ud x^i \ud x^j \right\}\,,
\end{equation}
where $\Phi$ and $\Psi$ are the Bardeen potentials~\cite{Bardeen:1980kt}
and $\tens_{ij}$ is a transverse and traceless tensor. From the
perturbed Einstein and Klein-Gordon equations, the dynamics of the
scalar and tensor modes, in Fourier space, reduces to the equations of
motion of two uncoupled parametric oscillators~\cite{Mukhanov:1981xt,
Grishchuk:1974ny, Grishchuk:1975uf, Martin:1997zd}
\begin{equation}
\label{eq:paramoscillator}
\dfrac{\ud^2\muST}{\ud \eta^2}+\omegaST^2(k,\eta ) \muST=0 \, ,
\end{equation}
where the frequencies are respectively
\begin{equation}
\label{eq:frequencies}
\omegaS^2\left(k,\eta \right)=k^2 -\frac{(a\sqrt{\epsilon
_1})''}{a\sqrt{\epsilon _1}}\, , \qquad \omegaT^2\left(k,\eta
\right)=k^2 -\frac{a''}{a} \,,
\end{equation}
the prime denoting differentiation with respect to conformal time. The
scalar and tensor mode functions are gauge invariant and
read~\cite{Mukhanov:1981xt}
\begin{equation}
\muS = a \sqrt{2\kappa} \left( \delta\phi + \dfrac{\ud \phi}{\ud \eta}
  \dfrac{\Phi}{\calH} \right), \qquad \muT = a \tens_{ij}\,,
\end{equation}
where $\delta \phi$ stands for the linear perturbations, in the
longitudinal gauge, of the inflaton field $\phi$ [see
equation~(\ref{eq:phir})]. The initial conditions for the perturbed
quantities $\muST$ are set in the decoupling limit $k\gg\calH$ where the
deep sub-Hubble modes behave as free quantum fields. Assuming their
initial state to be the usual Bunch-Davies vacuum, one
gets~\cite{Mukhanov:1990me}
\begin{equation}
\label{eq:quantumic}
\lim_{k/\calH \rightarrow \infty} \muST = \sqrt{\kappa}
 \, \dfrac{\ue^{-ik\left(\eta - \eta_\ui \right)}}{\sqrt{k}}\, ,
\end{equation}
with $\eta_\ui$ some (arbitrary) initial conformal time. These initial
conditions uniquely determine the solutions of
equation~(\ref{eq:paramoscillator}) which can be numerically integrated
along the lines described in reference~\cite{Ringeval:2007am}. From such a
numerical integration, the primordial scalar and tensor power spectra
can be evaluated at the end of inflation for all the relevant observable
wavenumbers
\begin{equation}
\label{eq:spec}
\calP_\zeta \equiv k^3P_{\zeta }(k)=\frac{k^3}{8\pi ^2}\left \vert
\frac{\muS}{a\sqrt{\epsilon _1}}\right \vert ^2 , \qquad \calP_h
\equiv k^3P_h(k) = \frac{2k^3}{\pi ^2}\left \vert \frac{\muT}{a}
\right \vert ^2 .
\end{equation}

Concerning the background quantities, the attractor mechanism during
inflation ensures that the initial field value $\phiini$ has no direct
observable effects since it determines only the total number of e-folds
$\Ntot$. Its value can therefore be, a priori, arbitrarily chosen
provided $\Ntot$ is big enough to solve the problems of the standard hot
Big-Bang phase and allows the observable perturbations to be sub-Hubble
initially. As thoroughly discussed in sections~\ref{sec:stringeffects}
and \ref{sec:restrictions}, the brane motion ends by the appearance of a
tachyon from its coalescence with the anti-brane at the bottom of the
throat, and this determines the field value at which the above classical
evolution cannot longer be used, namely $\phi \simeq \phistrg$. The
initial field value $\phiini$ can therefore be chosen in such a way that
the brane motion generates $\Ntot$ e-folds of inflation before reaching
$\phistrg$. However, one still has to impose that the brane motion
starts inside the throat for the model to be consistent, \ie $\phiini <
\phiuv$.

\par

According to the previous discussion, it is natural to assume that the
reheating era starts when the brane alights on the
anti-brane. Although, strictly speaking, brane annihilation occurs at
the bottom of the throat $\rzero$, the String Theory details of the
evolution for $\phi<\phistrg$ are out of the scope of our simple
approach~\cite{Sen:1998sm,Brodie:2003qv, Barnaby:2004gg}. As a result,
we have adopted the phenomenological reheating model discussed in
references~\cite{Martin:2006rs,Ringeval:2007am}. It assumes that the
reheating can only influence the observed perturbations through its
effects on the cosmological redshift $\zstrg$, which will be
identified with the beginning of the reheating era. Such an assumption
is motivated by the fact that adiabatic super-Hubble perturbation
modes should not be significantly modified during reheating, at least
in the absence of entropy modes (here, we deal with a single field
model). The influence of $\zstrg$ can be seen by rewriting
equation~(\ref{eq:paramoscillator}) in terms of the number of e-folds
as a time variable. The wavenumbers to be considered during inflation
always appear in the ratio $k/\calH$ which should be determined from
the observable wavenumbers $k/a_0$ measured today, typically three
decades around $\kstar = 0.05\,\Mpc^{-1}$. At a given e-fold $N$
during inflation,
\begin{equation}
\label{eq:kinfknow}
\dfrac{k}{\calH} = \dfrac{k}{a_0} \left(1+\zstrg \right)
\dfrac{\ue^{\Ntot-N}}{H(N)}\,.
\end{equation}
Assuming instantaneous transitions between the successive expansion
eras, the redshift $\zstrg$ is given by
\begin{equation}
\label{eq:lnzstrg}
\ln(1+\zstrg) = \dfrac{1}{4} \ln\left(\kappa^2 \rho_\ureh \right) -
\ln \dfrac{a_\ustrg}{a_\ureh} - \dfrac{1}{2}\ln\left(\sqrt{3 \OmegaR
  \kappa} H_0 \right),
\end{equation}
where $\rho_\ureh$ is the total energy density at the end of the
reheating era, and $\OmegaR$, $H_0$ are the density parameter of
radiation and the Hubble parameter today. The first two terms in
equation~(\ref{eq:lnzstrg}) are clearly reheating dependent and in
absence of a microscopic model we can define a phenomenological
reheating parameter $\Rrad$ such that
\begin{equation}
\label{eq:lnRrad}
\ln \Rrad \equiv - \dfrac{1}{4} \ln
\left(\dfrac{\rho_\ureh}{\rho_\ustrg} \right)
+\ln \dfrac{a_\ustrg}{a_\ureh}\, .
\end{equation}
This parameter has a simple physical interpretation: it encodes the
global deviation the dynamics of the reheating era may have with respect
to a pure radiation-like era (for which $\Rrad$ vanishes). Notice that,
from the numerical integration, the energy density $\rho_\ustrg$ when
$\phi=\phistrg$ is known and only depends on $\phi_\ustrg$ and the
potential parameters.

\par

Under the previous assumptions, the KKLMMT inflation era requires the
knowledge of five primordial parameters: the KKLMMT potential
parameters, namely $M$ and $\mu$, the reheating parameter $\Rrad$ and
the field values $\phiuv$ and $\phistrg$ encoding the observable
properties of the throat and the branes. The resulting CMB anisotropies
are uniquely determined once the ``low energy'' cosmological model is
fixed. We are considering the $\Lambda$CDM flat universe model which
adds four cosmological parameters: the number density of baryons
$\OmegaB$, of cold dark matter $\OmegaCDM$, the reduced Hubble parameter
today $h$, and the redshift of reionisation $\zre$.

\subsection{MCMC parameters and priors}
\label{subsec:mcmcprior}

Due to parameter degeneracies with respect to the CMB temperature and
polarisation angular power spectra, some parameter combinations are more
appropriate for an efficient MCMC exploration. Concerning the
$\Lambda$CDM parameters, instead of directly sampling models according
to the values of the set $\left(\OmegaB,\OmegaCDM,\zre,H_0\right)$, it
is more convenient to use the equivalent set $\left(\OmegaB
h^2,\OmegaCDM h^2,\optdepth, \theta\right)$, where $\optdepth$ is the
optical depth and $\theta$ measures the ratio of the sound horizon to
the angular diameter distance~\cite{Lewis:2002ah}.

\par

Similarly, it is more convenient to use an optimal derived set for the
primordial parameters. The amplitude of the scalar primordial power
spectrum at a given observable wavenumber
$\calP_*=\calP_\zeta(\kstar)$ is a well measured quantity [see
  equation~(\ref{eq:WMAPnormalization})]. It is therefore more
convenient to directly sample the models according to the values of
$\calP_*$ rather than the potential energy scale $M/\mpl$. This can be
done by integrating the perturbations with the artificial value
$M/\mpl=1$ and then perform a rescaling of $M/\mpl$ from unity to its
physical value that would be associated with the wanted $\calP_*$. As
shown in reference~\cite{Martin:2006rs}, under the rescaling $M/\mpl
\rightarrow s M/\mpl$, the power spectra become $\calP(k) \rightarrow
s \calP\left(s^{1/2}k\right)$. As a result, the value of $s$ required
is the ratio $\calP_*/\calPnum_\diamond$, where $\calPnum_\diamond$ is
the amplitude of the scalar power spectrum obtained with $M/\mpl=1$
and evaluated at $k_\diamond = k_* s^{-1/2}$. In fact, to circumvent
such a rescaling on the wavenumbers, it is more convenient to
introduce the rescaled reheating parameter $\Rreh$ defined by
\begin{equation}
\label{eq:lnRreh}
\ln \Rreh \equiv \ln \Rrad + \dfrac{1}{4}\ln \left(\kappa^2
\rho_\ustrg\right).
\end{equation}
The quantity $\Rreh$ represents an effective energy, in Planck units, at
the time of brane merging, which is exactly equal to $\sqrt{\kappa}
\rho_\ustrg^{1/4}$ for a radiation-like reheating. The advantage of
$\Rreh$ with respect to $\Rrad$ is that, at fixed $\Rreh$, one has
$k_\diamond=k_*$~\cite{Martin:2006rs}. Finally, as shown in
section~\ref{sec:stringeffects}, one may expect some observable effects
coming from the situations in which there is violation of the slow-roll
conditions before brane annihilation, \ie according to the value of
$\phistrg/\phitwo$. Since $\phitwo$ is determined by $\mu/\mpl$ only
[see equation~(\ref{eq:phieps2mu})], we have chosen to perform the MCMC
exploration on the parameter $\phistrg/\mu$ instead of $\phistrg$.

\par

Finally, the nine MCMC parameters are: $\OmegaB h^2$, $\OmegaCDM h^2$,
$\optdepth$, $\theta$ on the cosmological side, $\calP_*$, $\mu/\mpl$,
$\phiuv$, $\phistrg/\mu$ for the primordial parameters and $\Rreh$ for
the reheating era. We still have to specify their prior probability
distributions according to the theoretical constraints and our best
knowledge on their value. For the base cosmological parameters $\OmegaB
h^2$, $\OmegaCDM h^2$, $\tau$ and $\theta$, wide top hat uniform
distributions have been chosen centred over their preferred value from
the previous analysis of the CMB
data~\cite{Spergel:2006hy,Lewis:2002ah}. For the primordial parameters,
$\calP_*$ fixes the amplitude of the cosmological perturbations and we
have chosen a uniform prior on the logarithm compatible with the
amplitude of the CMB fluctuations:
\begin{equation}
2.7 \le\ln \left(10^{10}\calP_*\right)\le 4.0\,.
\end{equation}
The parameter $\mu/\mpl$ is related to the underlying String Theory
model by equation~(\ref{eq:mandmu}) and we will assume that the order of
magnitude of the volume ratio $v$ is not known. As a result, it is
natural to also use a uniform prior probability distribution on $\log
\left(\mu/\mpl\right)$ to ensure its conformal invariance. The same
considerations hold for the two other primordial parameters, $\phiuv$
and $\phistrg$, whose dependency with respect to the fundamental string
theory parameters is given in equations~(\ref{eq:phiuv}) and
(\ref{eq:phistrg}). Our ignorance on the values of $\Nflux$, $\gs$ and
$\alphas$ motivates the use of uninformative uniform logarithmic
priors. However, the analysis of section~\ref{sec:restrictions} imposes
various bounds. The maximal size of the throat in
equation~(\ref{eq:phiuvmax}) and the requirement $\Nflux\ge1$ give the
upper limit
\begin{equation}
\log(\sqrt{\kappa}\phiuv) \le \log 2 \,.
\end{equation}
Similarly, the definition~(\ref{eq:phistrg}) of $\phistrg$ ensures
that $\phistrg \ge \mu$ implying the lower limit
\begin{equation}
\log(\phistrg/\mu)\ge 0\,.
\end{equation}
Since $\phistrg<\phiuv$, these conditions impose
\begin{equation}
\log(\sqrt{\kappa}\mu) < \log 2\,,
\end{equation}
which will be our upper prior limit for the parameter
$\mu/\mpl$. However, we should ensure that there are enough e-folds of
inflation to solve the flatness problem and to set the sub-Hubble
initial conditions~(\ref{eq:quantumic}) for the observable
perturbations. As shown in reference~\cite{Liddle:2003as}, only for
some extreme reheating models the values of $\Ntot$ may exceed
$10^2$. In order to include all the models, we have implemented a
``hard prior'' which, during the MCMC exploration, rejects any model
that does not support at least $110$ e-folds of inflation inside the
throat. This is implemented in the following way. Once $\phistrg$ is
known [from $\log(\phistrg/\mu)$ and $\log(\sqrt{\kappa}\mu$)], one
can determine $\phiini$ such that there are $110$ e-folds of inflation
in between: the model is accepted if $\phiini<\phiuv$ and rejected
otherwise. The prior limits on $\ln \Rreh$ are determined by requiring
that the end of reheating occurs before nucleosynthesis, characterised
by $\rho_{\unuc}$, and that the instantaneous equation of state
parameter $\wstate_\ureh$ satisfies the strong and dominant energy
conditions $-1/3<\wstate_\ureh<1$. Under these assumptions, one
gets~\cite{Martin:2006rs}
\begin{equation}
\label{eq:rehprior}
\frac{1}{4} \ln \left(\kappa ^2 \rho _{\unuc}\right) < \ln \Rreh <
-\frac{1}{12} \ln \left(\kappa ^2 \rho _{\unuc}\right) +\frac{1}{3} \ln
\left(\kappa ^2 \rho _{\ustrg}\right).
\end{equation}
This equation clearly involves $\rho_\ustrg$ and the upper prior limit
on $\ln \Rreh$ depends on the other primordial parameters. We have
therefore coded another ``hard prior'' checked during the MCMC
exploration to dismiss any model violating this limit. The energy
density $\rho_\unuc$ at nucleosynthesis time has been quite extremely
set around the $\MeV$ scale: $\ln \Rreh > -46$. Finally, we have to set
a lower limit on the $\log(\sqrt{\kappa} \mu)$ prior. According to the
values of $v$, $\mu/\mpl$ may be extremely small. Our choice has been
motivated by numerical convenience and we have chosen
$\log(\sqrt{\kappa}\mu) > -3$. Indeed, it turns out that, for smaller
values, one runs into tricky numerical difficulties. The dependency of
the results with respect to this choice will be carefully discussed in
the following. Notice that the lower limit of the
$\log(\sqrt{\kappa}\phiuv)$ prior is implicitly set by the others since
$\phiuv>\phiini$ and we ensure that there are at least $110$ e-folds of
inflation in between $\phiini$ and $\phistrg$.

\subsection{Data used}
\label{subsec:dataused} 

As already mentioned, the CMB measurements used are the WMAP third
year data~\cite{Jarosik:2006ib, Spergel:2006hy, Hinshaw:2006ia,
  Page:2006hz}. The degeneracies between some cosmological parameters
have been reduced by adding the Hubble Space Telescope (HST)
constraint $H_0=72\pm 8 \,\km/\scnd/\Mpc$~\cite{Freedman:2000cf} and a
uniform top hat prior on the age of the universe between $10\,\Gy$ and
$20\,\Gy$~\cite{Lewis:2002ah}. The likelihood estimator is provided by
the WMAP team\footnote{\texttt{http://lambda.gsfc.nasa.gov}} and we
have used the current 2.2.2 version. Use has been made of the
eigenvalue compression option at low multipoles to decrease the
computing time. The convergence of the Markov chain simulations has
been monitored from the Gelman and Rubin $R$--test implemented in
\COSMOMC~\cite{Gelman:1992}. The MCMC exploration has been stopped
once the ratio of the variance of the means to the mean of the
variances between the different chains was less than $R-1 <
0.2\%$. The total number of samples obtained from this convergence
criterion and drawing the posterior distributions is around $1.5$
millions. In the following, we present and interpret the posterior
marginalised probability distributions for the nine model parameters
obtained from the previous priors and data.

\subsection{Cosmological parameters}
\label{subsec:cosmopara}

The marginalised posteriors and mean likelihoods of the cosmological
parameters and power spectra amplitude are the same than those found in
previous analysis of the same data but that were using other primordial
power spectra, such as the first and second order slow-roll spectra in
reference~\cite{Martin:2006rs} or the phenomenological power law models
in reference~\cite{Spergel:2006hy}. This is expected since these
parameters are rather well constrained and supported by the slow-roll
analysis of section~\ref{sec:srwmap} showing that the KKLMMT model can
indeed produce the preferred value of the power spectra amplitude and
spectral indices. The best fit model is found with a $\chi^2\simeq
3538.1$ for nine parameters (or $3517$ degrees of freedom after
eigenvalue compression). This may be compared with the WMAP power law
best fit model having $\chi^2\simeq 3540.8$ with three parameters
less. Although it shows that the KKLMMT model can provide a good fit to
the data, it is not currently favoured because of its intrinsic numbers
of parameters.

\subsection{Primordial parameters}
\label{subsec:primpara}

\begin{figure}
\begin{center}
\includegraphics[width=13cm]{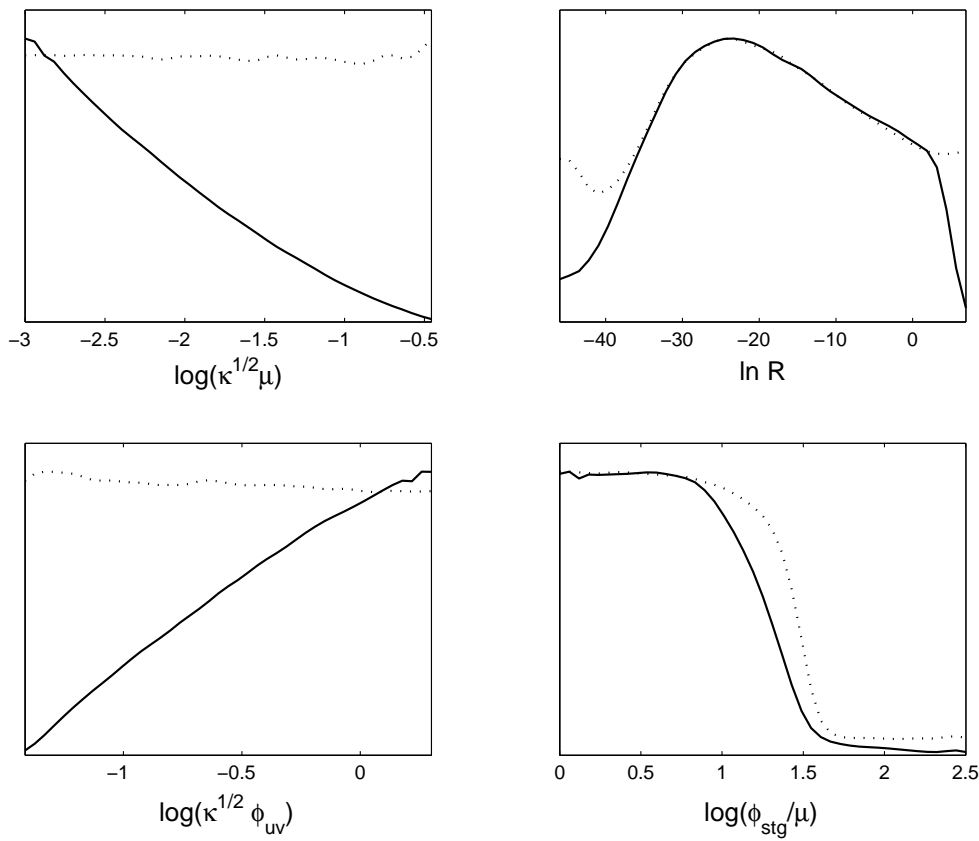}
\caption{Marginalised posterior probability distributions (solid
  lines) and mean likelihoods (dotted lines) for the sampled
  primordial parameters of the $\Lambda$CDM--KKLMMT model.}
\label{fig:infbase}
\end{center}
\end{figure}

The marginalised posteriors and mean likelihoods for the sampled
primordial parameters $\log(\sqrt{\kappa}\mu)$,
$\log(\sqrt{\kappa}\phiuv)$, $\log(\phistrg/\mu)$ and $\ln \Rreh$ are
represented in figure~\ref{fig:infbase}. The fact that the mean
likelihood (dotted curves) remains uniform for $\log(\sqrt{\kappa}\mu)$
and $\log(\sqrt{\kappa}\phiuv)$ shows that, on the prior range explored,
these parameters do not help to improve the fit to the data. On the
other hand, their probability distributions are not flat and show that
large values of $\mu/\mpl$ and small values of $\phiuv$ are strongly
disfavoured: $\log(\sqrt{\kappa} \mu) < -1.1$ at $95\%$ confidence
level. As discussed in reference~\cite{Martin:2004yi}, discrepancies
between the marginalised probability and the mean likelihood occur when
``volume effects'' are induced by strong correlations between the
parameters. The full likelihood in the multi-dimensional parameter space
can be uniform along peculiar regions whose volume may be parameter
dependent. The shape of the $\mu/\mpl$ and $\phiuv$ probabilities come
from the amount of fine tuning required to have a successful model of
inflation when $\mu/\mpl$ is close to the throat edge. In that case,
there is indeed not too much parameter space available for the brane
motion since $\phiuv$ has to saturate its maximal value to allow both
$\phistrg>\mu$ and $\phistrg<\phiini<\phiuv$. As a result of these
correlations, the marginalised probabilities for $\mu/\mpl$ and $\phiuv$
penalise the values for which strong fine-tunings are required to
satisfy the theoretical priors. Let us mention again that these priors
are imposed by the consistency of the model.

\par

On the other hand, the marginalised posteriors and mean likelihoods
associated with $\Rreh$ and $\phistrg/\mu$ show that these parameters
are directly constrained by the data. Clearly, $\phistrg \gg \mu$ is
disfavoured and at $95\%$ confidence level one has $\log(\phistrg/\mu)
< 1.4$. This result can be understood from
section~\ref{sec:srwmap}. Large values of $\phistrg/\mu$ correspond to
models in which brane annihilation occurs well inside the slow-roll
regime, \ie $\phistrg > \phitwo$. As a result, the observable
perturbation modes are generated during the brane motion in the flat
part of the potential (see figure~\ref{fig:potkklt_s}) and both the
slow-roll parameters are small: the spectral index is close to $\nS
\simeq 1$ while the amplitude of the tensor modes is negligible with
respect to the scalar modes. As already mentioned in
section~\ref{sec:srwmap}, this situation is disfavoured by the WMAP
data. The upper bound on $\phistrg/\mu$ is in fact slightly
dependant on the lower limit of the $\mu$ prior due to the presence of
some correlations with the other parameters. As discussed in the
following, a $\mu$-prior independent upper limit can be more
conveniently given for the rescaled parameter $\phistrg/\mu^{2/3}$ and
one gets, at $95\%$ confidence level,
\begin{equation}
\label{eq:phistrgmax}
\log \left(\dfrac{\kappa^{1/6}\phistrg}{\mu^{2/3}}\right) < 0.52\,.
\end{equation}

\par

The marginalised probability distribution for the reheating parameter
is peaked around $\ln \Rreh \simeq -22$. Its behaviour at large
values directly comes from the prior~(\ref{eq:rehprior}) which is a
function of $\rho_\ustrg$. As will be discussed in the following,
$\rho_\ustrg$ is related to the energy scale of inflation which is
bounded from above by the amplitude of the cosmological
perturbations. Here again, the differences between the marginalised
probability and the mean likelihood trace the correlations induced by
the prior hypothesis between these two parameters, namely that
reheating occurs after inflation. The lower tail of the distribution
is driven by the data and falls off till the prior lower bound is
saturated at $\ln \Rreh=-46$ (nucleosynthesis limit). This posterior
is very similar to the one derived for the small field models in
reference~\cite{Martin:2006rs} where it has been shown that the CMB data
were disfavouring a low energy scale reheating for these models. In
the present case, the $\ln \Rreh$ posterior of
figure~\ref{fig:infbase} does not fall-off to zero in its lower part
but still yields a limit
\begin{equation}
\label{eq:lnRrehmin}
\ln \Rreh > -38 \, , 
\end{equation}
at a $95\%$ confidence level.

\par

\begin{figure}
\begin{center}
\includegraphics[height=7.1cm]{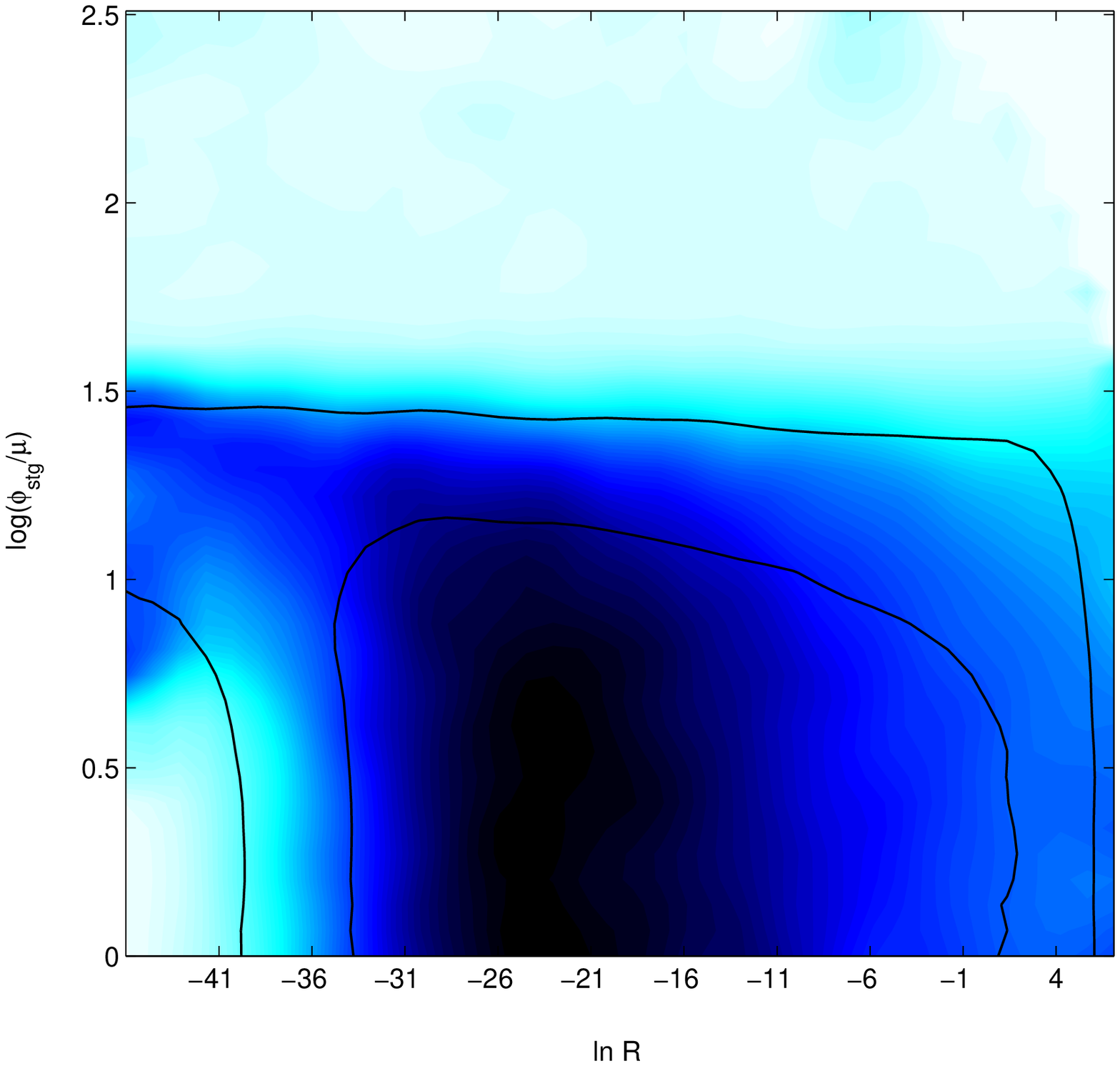}
\includegraphics[height=7.4cm]{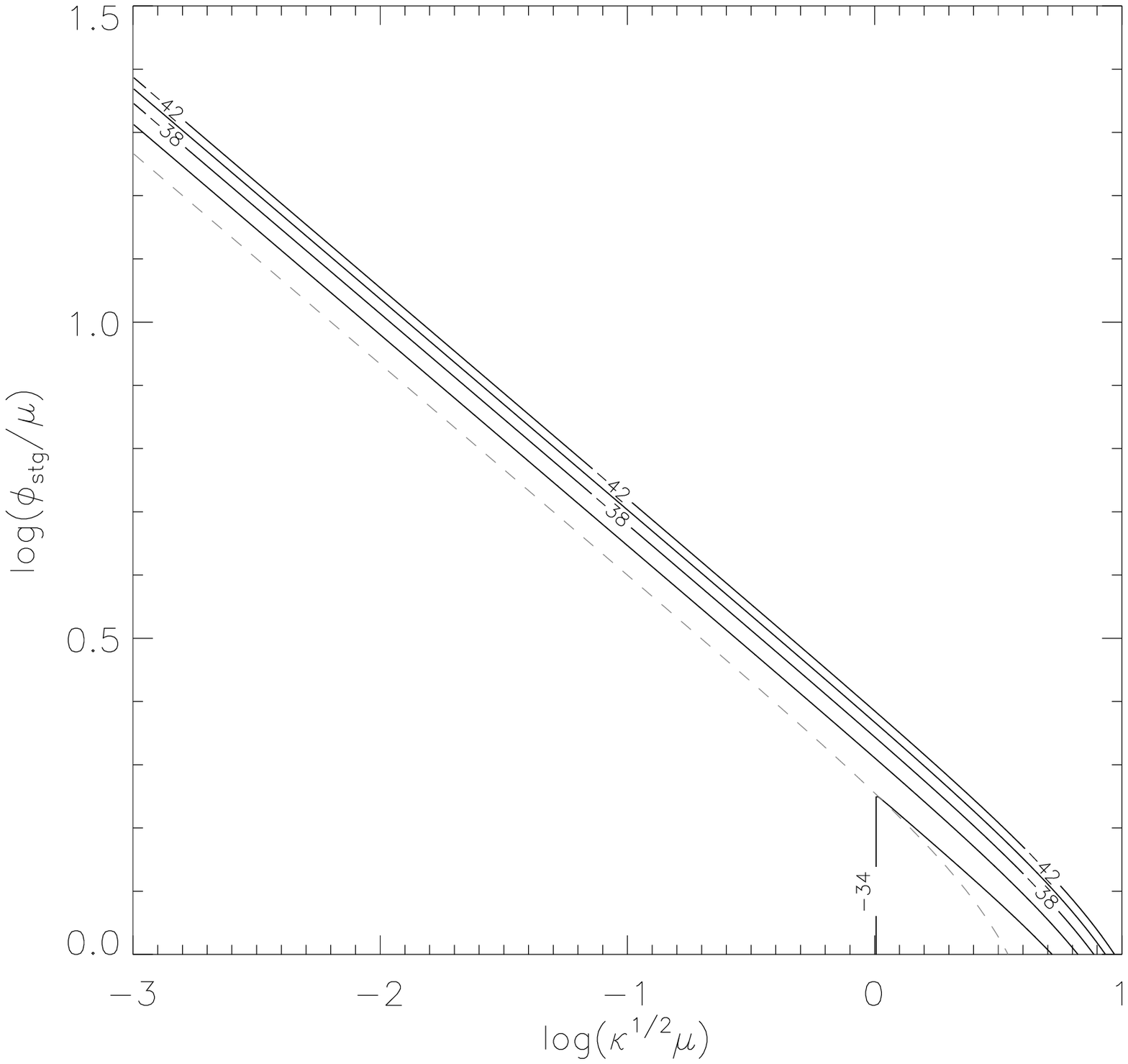} \caption{The left
panel shows the one- and two-sigma confidence intervals of the
two-dimensional marginalised posterior as a function of the parameters
$\ln \Rreh$ and $\log(\phistrg/\mu)$. The shading traces the
corresponding two-dimensional mean likelihood. The right panel is a
contour plot of $\ln \Rreh$ as a function of $\log (\sqrt{\kappa} \mu)$
and $\log(\phistrg/\mu)$ obtained from the slow-roll approximation by
assuming that $\epstwo = \epstwoobs \simeq 0.05$. The dashed line marks
the limit $\phistrg=\phitwo$. This plot shows that, although $\phistrg >
\mu$ is generically disfavoured (see figure~\ref{fig:infbase}), it is
still possible to obtain a good fit to the data provided some fine
tuning is performed between $\phistrg/\mu$, $\mu/\mpl$ and $\ln \Rreh$:
a low energy reheating allows for some models where brane annihilation
occurs before slow-roll violation to have red-tilted power spectra.}
\label{fig:lnRreh2D}
\end{center}
\end{figure}

The physical interpretation of this limit comes from the influence of
$\Rreh$ on $\Nstar$, the e-fold at which an observable perturbation
mode with wavenumber $\kstar$ crossed the Hubble radius during
inflation. Indeed, one has $a_*/a_0=a_*/a_\ustrg \times a_\ustrg/a_0$,
with, by definition, $a_*/a_0\simeq k_*/(a_0 \kappa ^{1/2 }V_*^{1/2})$,
where we have expressed the Hubble parameter in terms of the
potential. The quantity $a_*/a_\ustrg$ can be expressed in terms of
$N_*$ and $a_\ustrg/a_0 =\left(1+z_\ustrg\right)^{-1}$ is given by
equations~(\ref{eq:lnzstrg}), (\ref{eq:lnRrad}) and (\ref{eq:lnRreh}),
expressing $\rho_\ustrg$ in terms of the potential evaluated at
$\phistrg$. One gets~\cite{Martin:2006rs}
\begin{equation}
\label{eq:lnRrehNstar}
N_*\simeq 58 -\ln
\left[\frac{\kstar}{a_0}\left(\mbox{Mpc}^{-1}\right)\right] +\ln \Rreh
+\dfrac{1}{2} \ln \dfrac{V_*}{V_\ustrg}\,.
\end{equation}
As shown in section~\ref{sec:srwmap}, since the WMAP data prefer a
slightly red-tilted spectral index in the absence of tensor modes, it is
not surprising that, for a given model of inflation, some values of
$\Nstar$, and therefore $\Rreh$, end up being favoured. This
interpretation can be further explored by using the slow-roll
approximation. As already mentioned, the KKLMMT model generically leads
to observable $\epsone$ values much smaller than $\epstwo$. Let us first
determine determine $N_*$ such that $\epstwo (\phi _*)=\epstwoobs$,
where $\phi_* = \phi(\Nstar)$ and $\epstwoobs\simeq 0.05$ is the
preferred observed value (see figure~\ref{fig:kkltsr}). Using the
classical trajectory given by equation~(\ref{eq:trajectory}), one finds
that
\begin{equation}
\Nstar\simeq \frac{\kappa \mu ^2}{24}
\left[\left(\frac{\phistar}{\mu }\right)^6
-\left(\frac{\phiend}{\mu }\right)^6\right],
\end{equation}
and using the expression of the second horizon flow
parameter~(\ref{eq:eps2kklt}), one arrives at
\begin{equation}
\epstwo\simeq \frac{40}{\kappa \mu ^2}
\left(\frac{\phistar}{\mu }\right)^{-6}\, .
\end{equation}
Combining these two last equations, one obtains
\begin{equation}
\label{eq:Nstarobs}
\Nstar =\frac{5}{3\epstwoobs}\left[1-\frac{\kappa \mu ^2\epstwoobs}{40}
\left(\frac{\phiend}{\mu}\right)^6\right]\, .
\end{equation}
As already discussed, if the slow-roll is violated before brane
annihilation, we can use the slow-roll trajectory with the approximation
$\phiend \simeq \phitwo=[40/(\kappa \mu ^2)]^{1/6}$, from which one
deduces that
\begin{equation}
\Nstar \simeq \dfrac{5}{3 \epstwoobs}\,.
\end{equation}
Inserting this value into formula~(\ref{eq:lnRrehNstar}) with
$\kstar/a_0\simeq 0.05 \, \Mpc^{-1}$ and neglecting the potential term
gives $\ln \Rreh\simeq -28$. This is the preferred value of $\ln \Rreh$
that would give an observable spectral index at $\kstar$ compatible with
the favoured value and in agreement with the shape of the marginalised
posterior of figure~\ref{fig:infbase}. Let us notice that the two-sigma
limit~(\ref{eq:lnRrehmin}) is the best that can be extracted from the
data. Indeed, the three-sigma lower limit matches the nucleosynthesis
prior bound. This differs from the results obtained in
reference~\cite{Martin:2006rs} for the small field models and is due to
the marginalised posterior which does not vanish for low value of $\ln
\Rreh$. Moreover, the mean likelihood appears to increase again in that
region. This effect comes from correlations between parameters and
suggests that some, but quite fine-tuned, models provide a good fit to
the data even for $\ln \Rreh$ small. To understand this property, we
have represented in figure~\ref{fig:lnRreh2D} the two-dimensional
marginalised posterior and mean likelihood in the plane $[\ln
\Rreh,\log(\phistrg/\mu)]$ (left panel).
One recovers the favoured value for $\ln \Rreh \simeq -22$ as the
central vertical broad region (one-sigma contour), and as shown above,
this domain is associated with the models in which slow-roll inflation
ends before the branes collide. The two-sigma contour extends towards
the left side of the plot but only for
$1<\log\left(\phistrg/\mu\right)<1.5$. This tail is associated with the
models in which brane annihilation occurs before the slow-roll
conditions are violated. To see why such models are not disfavoured we
can still use equations~(\ref{eq:lnRrehNstar}) and (\ref{eq:Nstarobs})
but, now, with $\phiend=\phistrg$ to get
\begin{equation}
\ln \Rreh \simeq -61 +\frac{5}{3\epstwoobs}\left[1-\frac{\kappa \mu
    ^2\epstwoobs}{40} \left(\frac{\phistrg}{\mu}\right)^6\right].
\end{equation}
The $\ln \Rreh$ contour plot coming from this function is represented on
the right panel of figure~\ref{fig:lnRreh2D} for different values of
$\phistrg $ and $\mu/\mpl$. One sees that it is indeed possible to get
$\epstwoobs \simeq 0.05$ for $\phistrg>\phitwo$ provided $\ln
\Rreh\simeq -40$, in agreement with the plot. The physical
interpretation is that a low energy reheating allows the observable
window of wavenumbers to be shifted towards the end of brane evolution
in such a way that, although brane annihilation occurs before violation
of the slow-roll conditions, it is possible to get not so small values
for $\epstwo$. Of course, this requires some amount of fine tuning
between $\ln \Rreh$, $\mu/\mpl$ and $\phistrg$, which is statistically
penalised and explains why these models are out of the one-sigma
contour.

\subsection{Derived primordial parameters}
\label{subsec:derivprim}

\begin{figure}
\begin{center}
\includegraphics[width=13cm]{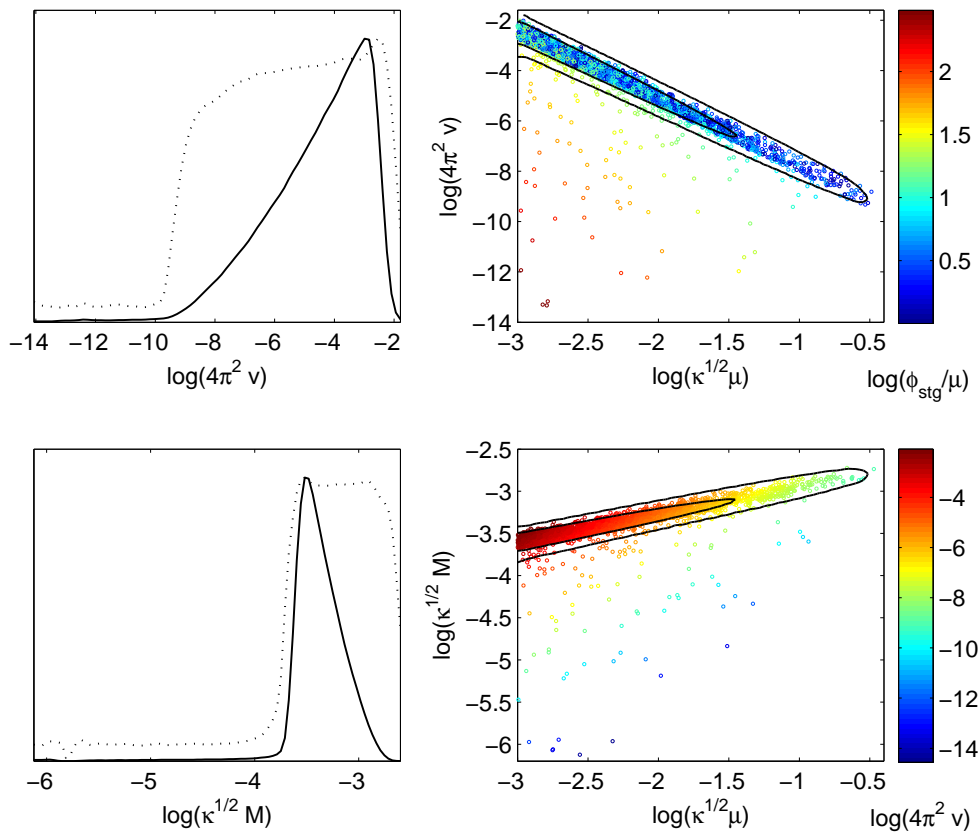} \caption{Marginalised
posterior probability distributions for the $M/\mpl$ and $v$ parameters
(solid curves) and their associated mean likelihoods (dotted
curves). The right panel shows the corresponding one- and two-sigma
contour of the two-dimensional posteriors obtained without marginalising
over $\log(\sqrt{\kappa}\mu)$. The two-dimensional probability is
proportional to the point density while the colormap traces correlations
with a third parameter.}  \label{fig:infderiv}
\end{center}
\end{figure}

As described in section~\ref{subsec:mcmcmethod}, the Markov chains have
been performed on the power spectra amplitude $\calP_*$ and its
posterior probability distribution is plotted in
figure~\ref{fig:infbase}. Since $\calP_*$ and $M/\mpl$ are in direct
relation (see section~\ref{subsec:mcmcprior}), it is possible to derive
the marginalised probability distribution of $M/\mpl$ from the one of
$\calP_*$ . The same considerations apply to the volume ratio $v$ which
is given by equation~(\ref{eq:vdef}): using importance sampling, one can
extract its posterior distribution from the one of $\mu/\mpl$ and
$M/\mpl$~\cite{Lewis:2002ah}. Both probability distributions are plotted
in figure~\ref{fig:infderiv} and seem to favour some peculiar values of
$M/\mpl$ and $v$. Once again, the discrepancies between the posteriors
and the mean likelihoods come from the correlations with other
parameters and may hide some effects coming from the priors. The right
panel of figure~\ref{fig:infderiv} represents the two-dimensional
probability distribution (point density), as well as its one- and
two-sigma confidence level regions, obtained without marginalising over
the parameter $\log(\sqrt{\kappa}\mu)$. As can be seen on these plots,
the narrow highly probable regions trace the strong correlations between
$M/\mpl$ and $\mu/\mpl$, and between $v$ and $\mu/\mpl$. Remembering
that the lower limit on the $\mu/\mpl$ parameter comes from our
numerically convenient prior $\sqrt{\kappa} \mu > 10^{-3}$, one
immediately sees that this choice has a direct influence on the upper
limit, and respectively the lower limit, of the $\log(4\pi^2 v)$ and
$\log(\sqrt{\kappa}M)$ probability distributions. If we had chosen a
smaller limit for $\mu/\mpl$, the $v$ posterior would have been shifted
towards larger values, whereas the $M/\mpl$ one towards the smaller. As
a result, we conclude that there is no upper, respectively lower,
constraints on $v$ and $M/\mpl$.

The same correlations are at work for the other tails of the probability
distribution associated with $v$ and $M/\mpl$. However, this time, we
are probing the highest allowed values for $\mu/\mpl$ which are
disfavoured due to the amount of fine-tuning required for a successful
inflation nearby the throat edge (see previous section). As a result, we
obtain the $95\%$ confidence levels
\begin{equation}
\label{eq:Mmaxvmin}
\log \left( \sqrt{\kappa} M \right) < -2.9\,, \qquad
\log \left(4 \pi^2 v \right) > -8.5\,.
\end{equation}
Let us stress that these bounds come from both the data and the
requirement that inflation proceeds inside the throat, as imposed by the
self-consistency of the model used. 

\begin{figure}
\begin{center}
\includegraphics[width=14cm]{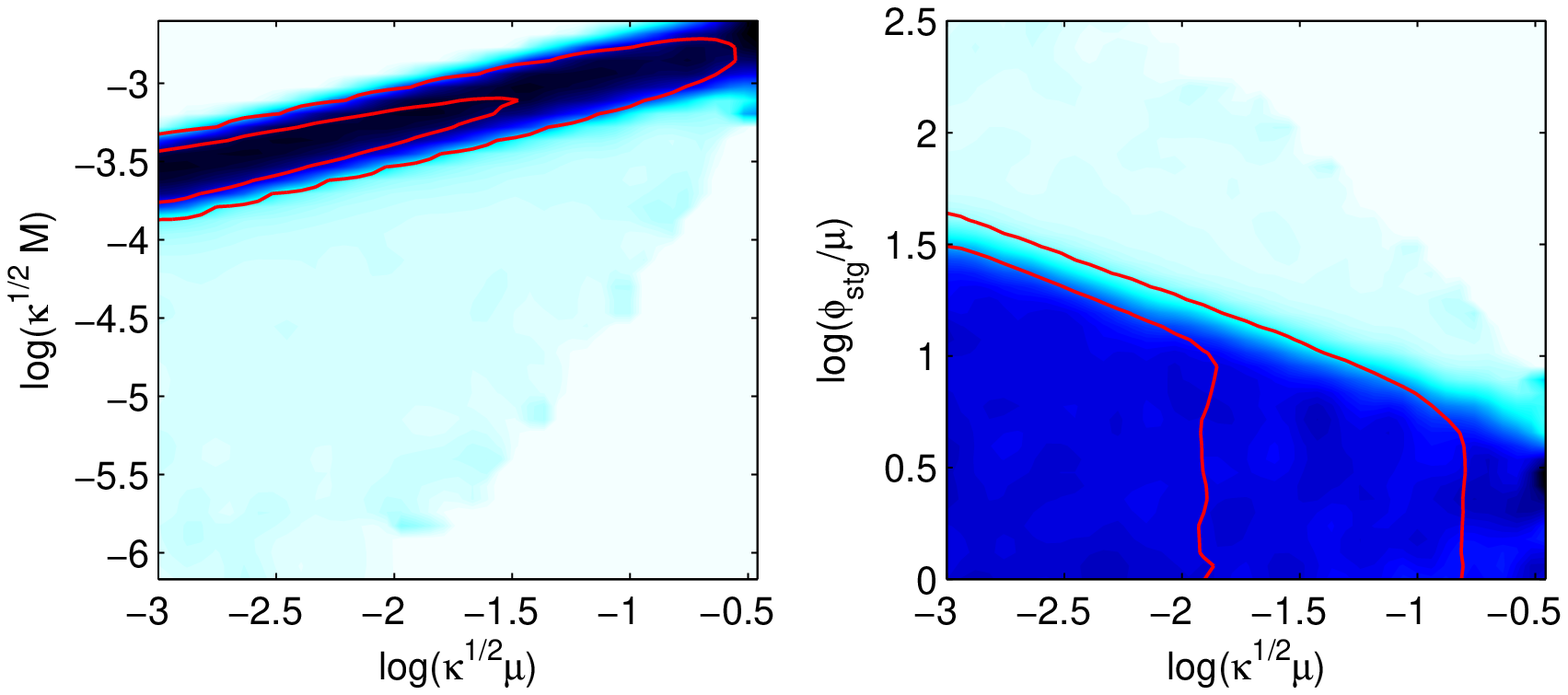}
\includegraphics[width=14cm]{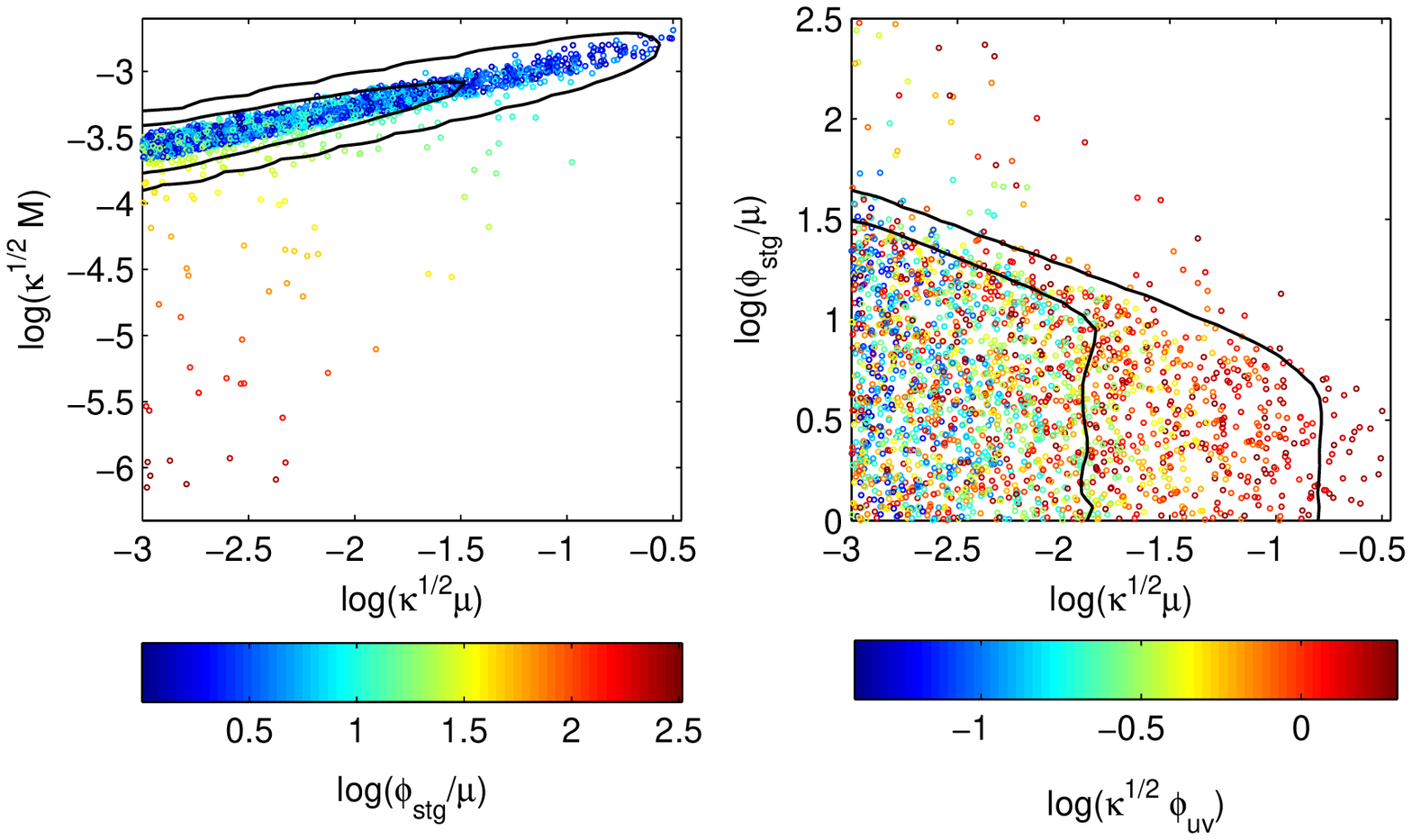} \caption{One- and
two-sigma contours (solid curves) of the two-dimensional marginalised
probability distribution for various pairs of the primordial
parameters. The two-dimensional mean likelihood is traced by the shading
intensity. The bottom panels show the two-dimensional posterior (point
density) and its correlation with a third parameter (colorbar).}
\label{fig:infderiv2D}
\end{center}
\end{figure}

To understand how the data lead to the correlations observed between the
parameters $\mu/\mpl$, $M/\mpl$ and $v$, we have plotted in
figure~\ref{fig:infderiv2D} the one- and two-sigma contours (solid
curves) of the two-dimensional posteriors associated with various pairs
of primordial parameters. The corresponding two-dimensional mean
likelihood is traced by the shaded areas. The highly probable region
spanning the plane $\left[\log(\sqrt{\kappa} M),
\log(\sqrt{\kappa}\mu)\right]$ directly comes from the constraints on
the power spectrum amplitude and spectral index. This can be shown by
using the slow-roll approximations. Indeed, if one assumes that the
slow-roll conditions are violated before brane annihilation, namely
$\phistrg < \phitwo$, then the WMAP normalisation through
equations~(\ref{eq:scaleMclas}) and~(\ref{eq:scalemuclas}) leads to
\begin{eqnarray}
\label{eq:Mmu}
\log\left(\sqrt{\kappa}M\right)&=&\log
\left[\left(45\QQoverTT\right)^{1/4}
  \left(6N_*+10\right)^{-5/12}\left(4\pi
  ^2\right)^{1/12}\left(8\pi\right)^{1/3} \right]\nonumber \\ & +
&\frac{1}{3} \log\left(\sqrt{\kappa}\mu\right) .
\end{eqnarray}
This is a straight line with slope equal to $1/3$, exactly as observed
for the high probable regions in figure~\ref{fig:infderiv2D}. Using a
fiducial value for the quadrupole and $\Nstar=50$, the offset is around
$-2.6$. Consequently, for $\log\left(\sqrt{\kappa}\mu\right)\simeq -3$
one gets $\log\left(\sqrt{\kappa }M\right)\simeq -3.6$, as confirmed by
the plot. The effect of the spectral index is indirect and, as
previously discussed, tends to favour the models in which $\phistrg <
\phitwo$. This suggests that the less probable and wider shaded region
in the plane $\left[\log(\sqrt{\kappa} M),
\log(\sqrt{\kappa}\mu)\right]$ corresponds to the cases where $\phistrg
> \phitwo$. Again, the slow-roll approximation in this limit gives the
relation
\begin{equation}
\label{eq:massmu} \left(\frac{M}{\mpl}\right)^6 \simeq 45 \QQoverTT
v^{1/2} \left(\frac{\phistrg}{\mu}\right)^{-10}.
\end{equation}
Since $v$ is also a function of $M/\mpl$ and $\mu/\mpl$, this is a three
parameters relation, or a surface in the plane $\left[\log(\sqrt{\kappa}
M), \log(\sqrt{\kappa}\mu)\right]$. We can, however, derive the slope
associated with the lower boundary of the blurred region (beyond which
there is no acceptable models). This boundary is reached for the largest
$\phistrg$ compatible with the throat size, \ie for $\phiinstrg =
\max(\phiuv)=\mpl/\sqrt{2\pi}$ [see equation~(\ref{eq:phiuvmax})]. This
limit is reached when the ratio of throat to bulk volume is
maximal. Using the expression of $\phiinstrg$ given by
equation~(\ref{eq:phiinstrings}) and solving for $\phistrg$ leads to
\begin{equation}
\label{eq:phistrguv}
\frac{\phistrg}{\mu}=\left[\left(\frac{\mpl}{\sqrt{2\pi}\mu}\right)^6-
\frac{3N_{_{\rm T}}}{\pi}\left(\frac{\mpl}{\mu}\right)^2\right]^{1/6} .
\end{equation}
This expression can now be inserted into into equation~(\ref{eq:massmu})
and one gets
\begin{eqnarray}
\log\left(\sqrt{\kappa}M\right)&\simeq &
\log\left[\left(45\QQoverTT\right)^{1/4} 4\sqrt{\pi}\right] -\frac12
\log\left(\sqrt{\kappa}\mu\right)\nonumber \\ & -&
\frac{5}{12}\log\left[\frac{64}{\left(\sqrt{\kappa}\mu\right)^6}
  -\frac{24\Ntot}{\sqrt{\kappa \mu}}\right] \, .
\end{eqnarray}
For not too large values of $\sqrt{\kappa }\mu$ (in practice
$\sqrt{\kappa}\mu \lesssim 0.1$), the term proportional to $N_{_{\rm
T}}$ can be neglected and one obtains
\begin{eqnarray}
\log\left(\sqrt{\kappa}M\right)&\simeq & 
\log\left[\left(45\QQoverTT\right)^{1/4}
4\sqrt{\pi}\right]-\frac{5}{12}\log\left(64\right) \nonumber
\\ & + &2\log\left(\sqrt{\kappa}\mu\right).
\end{eqnarray}
This is a straight line in the plane $\left[\log(\sqrt{\kappa} M),
\log(\sqrt{\kappa}\mu)\right]$ the slope of which is $+2$: this matches
with the observed lower boundary of the blurred region in
figure~\ref{fig:infderiv2D}. The offset is around $-2.1$ giving
$\log\left(\sqrt{\kappa}M\right)\simeq -4.1$ for
$\log\left(\sqrt{\kappa}\mu\right)\simeq -1$, again in agreement with
the figure. One might also notice that the boundary curve in
figure~\ref{fig:infderiv2D} slightly bends over for small values of
$\mu/\mpl$ as the effects of the terms proportional to $\Ntot$ start to
appear. The previous interpretation is confirmed by the left bottom
panel of figure~\ref{fig:infderiv2D}. The point density clearly
decreases in the domain of low likelihood which appears blurred in the
upper left panel. As traced by the colormap, this region indeed
corresponds to large value of $\phistrg/\mu$. Let us notice that,
although this region is out of the $95\%$ confidence contour, it remains
inside the three-sigma one.

\par

The previous discussion also applies to the right plots of
figure~\ref{fig:infderiv}. The above-described relations between
$M/\mpl$, $\mu/\mpl$ and $\phistrg$ are directly converted into
correlations between $v$ and the other parameters through
equation~(\ref{eq:vdef}). As a result, the main degeneracy seen in the
plane $[\log(\sqrt{\kappa}\mu),\log(4 \pi^2 v)]$ is also a consequence
of both the power spectra normalisation and spectral index
constraints. The bottom right panel illustrates the degeneracy between
$M/\mpl$ and $\mu/\mpl$ coming from the data and differs from the one
associated with equation~(\ref{eq:vdef}): this difference explains the
two-sigma lower limit on $\log(4 \pi^2 v)$. Let us elaborate on this
point.  The limit on $v$ can be understood from the slow-roll result
even if, at first sight, there is a mismatch between the constraint
obtained above and equation~(\ref{eq:limv}). In fact,
equation~(\ref{eq:wmapmuM}) implies
\begin{equation} 
\label{eq:vmu}
\log \left(4\pi ^2v\right)\simeq -\frac{8}{3}
\log \left(\sqrt{\kappa}\mu\right)-9\, .
\end{equation}
This relation is consistent with the top right panel in
figure~\ref{fig:infderiv}. Let us notice that, in this context, it is
relevant to use equation~(\ref{eq:wmapmuM}), which is derived under the
assumption that inflation stops by violation of the slow-roll
conditions, since one observes in figure~\ref{fig:infderiv} that the
two-sigma contour corresponds to ``small'' values of
$\phistrg$. Moreover, if one inserts the limit~(\ref{eq:mumax}), namely
$\sqrt{\kappa }\mu <2$, in the above relation then the
constraint~(\ref{eq:Mmaxvmin}) is reproduced. This constraint is
different from equation~(\ref{eq:limv}) because it has a different
origin. The limit~(\ref{eq:limv}) assumes that the maximum allowed
contribution of tensor modes to the observed CMB data can indeed be
generated during KKLMMT inflation. However, the requirement that the
brane motion proceeds inside the throat strongly limits the generation
of tensor modes and leads to the stronger bound of
equation~(\ref{eq:Mmaxvmin}). Therefore, the two limits are consistent
and the stronger bound is given by equation~(\ref{eq:Mmaxvmin}).

\par

Finally, we have plotted in the right hand panels of
figure~\ref{fig:infderiv2D} the two-dimensional probability distribution
and mean likelihood in the plane
$[\log(\sqrt{\kappa}\mu),\log(\phistrg/\mu)]$, as well as the effect of
$\phiuv$. These plots exhibit the volume effects associated with the
large values of the parameter $\mu/\mpl$. Clearly, the allowed range of
both $\phistrg/\mu$ and $\phiuv$ are all the more reduced as $\mu/\mpl$
increases, thereby decreasing the statistical weight of these domains in
the marginalised posterior of $\mu/\mpl$ (see
figure~\ref{fig:infbase}). The two-sigma contours are found to follow
the high likelihood region, and as before, it corresponds to models in
which brane annihilation occurs after violation of the slow-roll
conditions ($\phistrg < \phitwo$). Indeed, the edge between the dark and
the blurred regions is just given by the equation
$\phistrg=\phitwo$. Using equation~(\ref{eq:phieps2mu}), which gives
$\phitwo$ when $\sqrt{\kappa}\mu$ is small, one gets
\begin{equation}
\label{eq:edgestrg2D}
\log \left(\frac{\phistrg}{\mu}\right)\simeq \frac{1}{6} \log (40)
-\frac{1}{3} \log\left(\sqrt{\kappa }\mu\right).
\end{equation}
This is a straight line in the plane $[\log(\sqrt{\kappa}
  \mu),\log(\phistrg/\mu)]$ with a $-1/3$ slope, as observed in the
figure. The offset is close to $0.3$ implying that for
$\log(\sqrt{\kappa }\mu)=-3$ one would have $\log(\phistrg/\mu) \sim
1.3$. There is a slight difference since the contour seems to
intercept the vertical axis at $\log \left(\phistrg/\mu\right)\sim
1.5$. Such a difference may be due to the fact that $\phistrg $ must
significantly deviate from $\phitwo$ for the effect to be
visible. Moreover, the reheating effects already mentioned are not
considered in the derivation of equation~(\ref{eq:edgestrg2D}).
Equation~(\ref{eq:edgestrg2D}) renders explicit the $\mu$-prior
dependence previously noted on the upper two-sigma limit associated
with the marginalised probability distribution of $\phistrg/\mu$ (see
figure~\ref{fig:infbase}). The $-1/3$ slope explains why the prior
independent limit of equation~(\ref{eq:phistrgmax}) is more
conveniently express in terms of $\phi/\mu^{2/3}$. Finally, the
upper edge of the blurred region is, as before, given by the condition
$\phiinstrg=\max(\phiuv)$ (see lower right panel of
figure~\ref{fig:infderiv2D}). Using equation~(\ref{eq:phistrguv}), one
finds
\begin{equation}
\log \left(\frac{\phistrg}{\mu}\right)\simeq \log\left(2\right)
-\log\left(\sqrt{\kappa }\mu\right) .
\end{equation}
Again, one can check that the slope and the offset are consistent with
what is observed in figure~\ref{fig:infderiv2D}. Beyond that limit,
inflation would not take place in the throat and those models are
rejected.

\subsection{Fundamental parameters}
\label{subsec:fundpara}

Our choice of MCMC parameters is minimal as far as observable effects
are concerned. The cosmological consequences of the KKLMMT model are
described by the set of four parameters $(M, \mu, \phistrg, \phiuv)$
whereas, as discussed in section~\ref{sec:params}, the underlying string
theory model involves five parameters: $(\gs,\alphas,M,v,\Nflux)$. As a
result, it is impossible without additional assumption to extract more
information for the string parameters. As found in the previous section,
the CMB data, however, allow some constraints to be derived for the
$M/\mpl$ and $v$ parameters. Their probability distributions do not
assume anything on the value of $\gs$, $\alphas$ or $\Nflux$ (apart from
consistency requirements) and are also robust against the reheating as
long as it can be described by our phenomenological model. If one wants
to go further on the fundamental theory parameters, additional
assumptions have to be made. In the following, such a step is performed
first on the reheating model by assuming that it proceeds with a
constant equation of state $P=\wstate_\ureh \rho$. Then, a similar
approach is adopted to derive some posterior probability distributions
on the remaining string parameters by assuming that the value of
$\alphas$ is known. As a working example, $\alphas=10^3 \mpl^{-2}$ and
$\alphas=10^5 \mpl^{-2}$ are considered.

\begin{figure}
\begin{center}
\includegraphics[width=13cm]{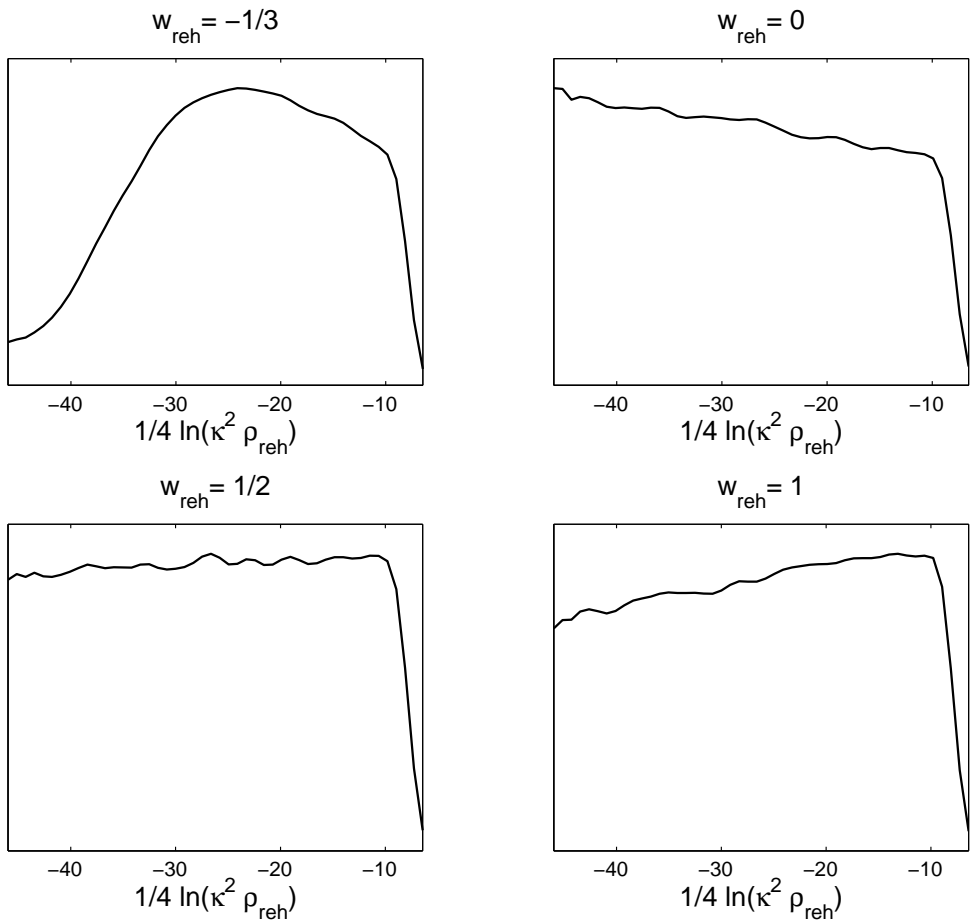} \caption{Marginalised
probability distributions on the reheating energy scale derived from the
WMAP data under the assumption that the reheating proceeds with a
constant equation of state: $P=\wstate_\ureh \rho$. For an extreme
equation of state $\wstate_\ureh\gtrsim -1/3$, one finds a weak
(non-trivial) lower bound $\rho_\ureh^{1/4} > 20\,\GeV $
at two-sigma level. For all the other cases, the reheating can occur at
any energy scale higher than nucleosynthesis (from the CMB point of
view).}  \label{fig:wreh}
\end{center}
\end{figure}

For a constant equation of state parameter during reheating, $\rho
\propto a^{-3(1+\wstate_\ureh)}$ and equation (\ref{eq:lnRreh}) can be
further simplified into
\begin{equation}
\label{eq:lnRrehw}
\ln \Rreh = \dfrac{1-3 \wstate_\ureh}{12 + 12 \wstate_\ureh} \ln
\left( \kappa^2 \rho_\ureh \right) + \dfrac{1 + 3 \wstate_\ureh}{6 + 6
  \wstate_\ureh} \ln \left(\kappa^2 \rho_\ustrg \right).
\end{equation}
{}From the MCMC analysis performed in the previous section, samples on
$\Rreh$ and $\rho_\ustrg$ can be used to extract by importance sampling
the probability distribution associated with $\rho_\ureh$ (the prior
choice remains that the reheating occurs before nucleosynthesis). The
resulting posteriors are plotted in figure~\ref{fig:wreh} for four
values of $\wstate_\ureh$ spanning the range allowed by the dominant and
strong energy conditions in General Relativity. For an extreme equation
of state with $\wstate_\ureh\gtrsim -1/3$, \ie which is on the verge of
an accelerated expansion, one has the $95\%$ confidence limit
\begin{equation}
\label{eq:Trehmin}
\rho_\ureh^{1/4} > 20 \, \GeV\,.
\end{equation}
Although this kind of models are certainly already ruled out by particle
physics experiments, let us notice that this limit comes from CMB data
only. As can be seen in figure~\ref{fig:wreh}, all the other cases
associated with more reasonable values of the equation of state
parameter are not constrained.

\begin{figure}
\begin{center}
\includegraphics[width=15cm]{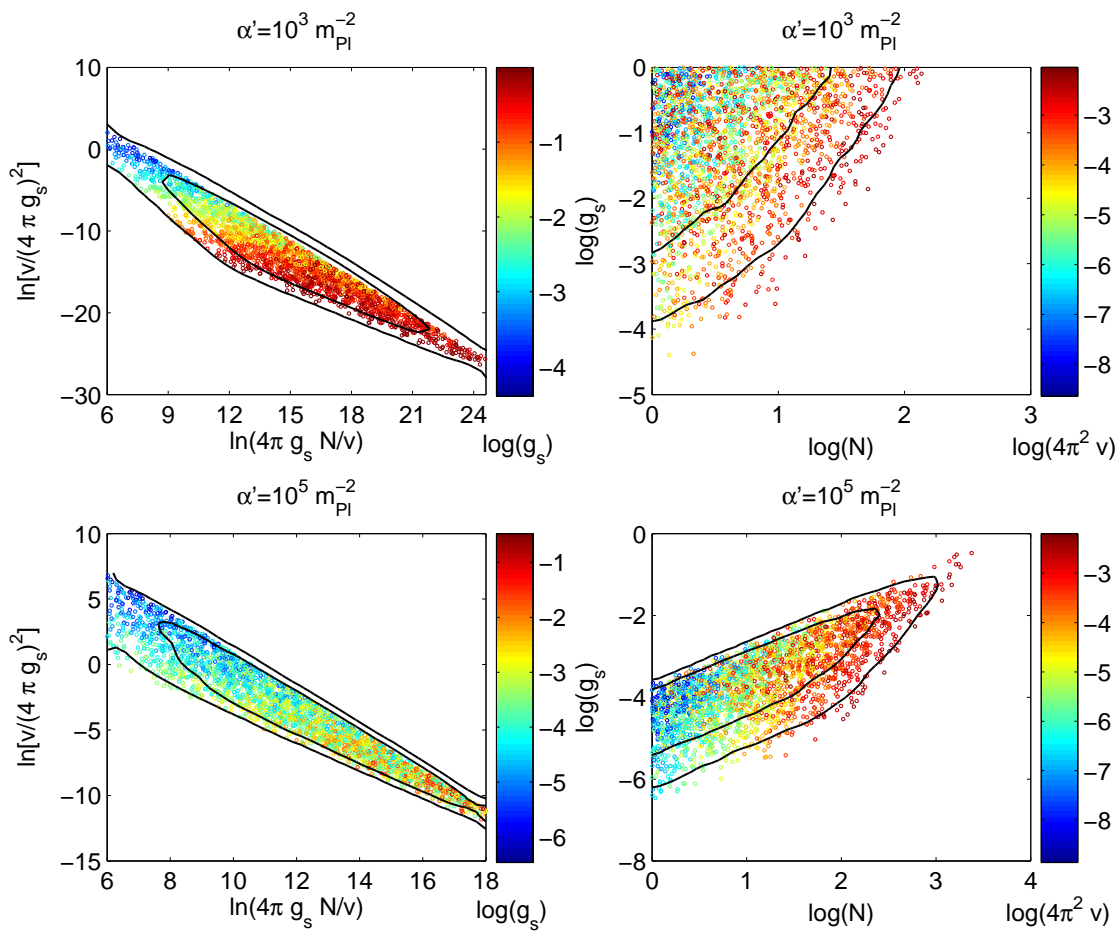} \caption{Two-dimensional
posteriors of the string parameters and their one- and two-sigma
contours obtained by importance sampling for two fiducial values of
$\alphas$: $\alphas=10^3 \mpl^{-2}$ and $\alphas=10^5
\mpl^{-2}$. Correlations with a third parameter are traced by the
colormap. The CMB data give some non-trivial constraints only for large
values of $\alphas$. This appears as the degeneracy observed between
$\Nflux$ and $\gs$ in the bottom right panel. The left panels show that
the models lie in the regions predicted by the slow-roll analysis in
section~\ref{sec:restrictions}.}  \label{fig:strgparams}
\end{center}
\end{figure}

A similar method has been applied to the parameters $\gs$ and $\Nflux$
by assuming that the value of $\alphas$ is known. In
figure~\ref{fig:strgparams}, we have plotted the two-dimensional
marginalised probability distributions, as well as their one- and
two-sigma contours, in the plane $(\log \Nflux, \log \gs)$ and for the
two fiducial values of $\alphas$. As can be seen on these plots (right
panels), there is basically no constraint coming from the data for the
case $\alphas =10^3 \mpl^{-2}$. Indeed, on one hand, the two-dimensional
posterior is limited by the consistency conditions $\Nflux \ge 1$ and
$\gs < 1$ to satisfy flux quantisation and perturbative treatment. On
the other hand, the colorbar shows that the two-sigma contour simply
traces the high values of $v$ which have been shown in
section~\ref{subsec:derivprim} to be essentially related to our lower
prior on $\mu/\mpl$. The case $\alphas = 10^5 \mpl^{-2}$ is slightly
different. As can be seen in the lower right panel of
figure~\ref{fig:strgparams}, the two-sigma contour is detached from the
axes and therefore the data give slightly more information than the
prior consistency limit. Again, the right end of the contour is not
physical since it comes only from our $\mu/\mpl$ prior, as
before. However, the observed degeneracy between $\log \Nflux$ and $\log
\gs$ is a non-trivial information. It means that for a given flux
number, a value of $\gs$ is favoured by the data. We have not
represented any one-dimensional posterior for $\gs$ or $\Nflux$ since
there is typically no more information than the above-mentioned
degeneracy. To do so, another assumption would have to be made, such as
some prior knowledge on the flux number or on the coupling constant. On
the left panels of figure~\ref{fig:strgparams}, we have plotted the
corresponding two-dimensional posteriors in the plane
$\{\ln(4\pi\gs\Nflux/v),\ln[v/(4\pi\gs)^2]\}$ to allow a comparison with
the slow-roll results of section~\ref{sec:restrictions} (see also
figure~\ref{fig:restrictions}). As expected, all the models lie in the
range predicted by the slow-roll analysis, up to the prior limit fixed
by the numerically convenient lower limit on $\mu/\mpl$. The associated
$\gs$ values are traced by the colormap.

\section{Conclusions}
\label{sec:conclusions}

To conclude, let us summarise and discuss the main results obtained in
this article. Our goal was to compare a typical model of string-inspired
brane inflation to the CMB data. In this scenario, the inflaton field is
interpreted as the distance between two branes moving in six dimensions
along a warped throat. The end of inflation either occurs by violation
of the slow-roll conditions or by tachyonic instability at brane
annihilation.  Our MCMC analysis was carried out by using an exact
numerical approach in which the only approximation used is the linear
theory of cosmological perturbations. Moreover, various effects have
been considered as, for instance, the presence of a DBI kinetic term or
the possible quantum effects. The MCMC exploration has also been
performed for arbitrary values of $\gs $ and $\alpha '$.

\par

{}From this analysis, we have obtained the following results. Firstly,
the WMAP data favour a scenario where inflation ends by violation of the
slow-roll conditions before brane annihilation rather than by the
tachyonic instability brought about by the annihilation. In other words,
$\phistrg$ cannot be too close to the edge of the throat and one finds
\begin{equation}
\log\left(\frac{\kappa^{1/6}\phi _\ustrg}{\mu^{2/3}}\right)  < 0.52\, ,
\end{equation}
at $95\%$ confidence level. This constraint originates from the fact
that ending inflation by instability pushes the spectral index towards
one while preserving a low level of gravitational waves, a situation
which is disfavoured by the data.

\par

Secondly, one has obtained a limit on $v$, the volume ratio of the
five-dimensional sub-manifold forming the basis of the six-dimensional
conifold to the volume of the five-sphere. This limit reads
\begin{equation}
\log v > -10\, ,
\end{equation}
at $95\%$ confidence level. This constraint comes both from the data
and a requirement of self-consistency of the model, namely that
inflation proceeds inside the throat.

\par

Thirdly, we find that the reheating period is slightly
constrained. Although we have not considered the detailed process of
brane annihilation, the total number of e-folds for which the universe
reheats may change the part of the inflaton potential probed by the
observations. Combined with the power spectra generated during brane
inflation, the WMAP data provide the $95\% $ confidence limit
\begin{equation}
\ln \Rreh > -38 \, .
\end{equation}
In the case where the reheating proceeds with a constant equation of
state parameter, we find that if $\wstate_\ureh \gtrsim -1/3$ then $\rho
_\ureh^{1/4} > 20 \GeV$ (at two-sigma). This limit is certainly already
ruled out by particle physics experiments, but it is worth recalling
that it has been obtained from the CMB data only. With more accurate
data, it will certainly be possible to improve this bound, and hopefully
for other values of the equation of state parameter.

\par

Fourthly, on the theoretical side, we have obtained approximate
solutions in the case where the DBI kinetic term does not reduce to the
standard one. The regime where the initial conditions are such that the
quantum effects are important has also been considered in detail, see
\ref{app:stocha}. In this case, using a perturbative treatment, we have
computed how the trajectory (with or without volume effects) can deviate
from the classical motion. We have also shown that this approximative
scheme breaks down when the brane approaches the bottom of the
throat. In this regime, only a numerical integration of the Langevin
equation could allow us to go further and, for instance, to see whether
the field really starts to climb the potential as indicated by the
calculation presented in the appendix. Clearly, this discussion is
relevant in order to know whether a regime of eternal inflation can be
established in this model.

\par

Finally, let us discuss how the present work could be
improved. Recently, various works have been devoted to the type of model
studied here. In particular, special attention has been paid to the
general set up and exactly calculable corrections to the
potential~(\ref{eq:Vofphifull}) have been proposed. It was also show
that, in some situations, these corrections play a crucial role
\cite{Baumann:2006cd,Baumann:2006th,Baumann:2007ah,
Baumann:2007np,Krause:2007jk}. The next step would therefore be to
include them in our analysis. Since they introduce new parameters into
the problem, and given the weakness of the constraints obtained here,
this would probably be meaningful only at the expense of including other
more accurate data sets to break the degeneracies. Time dependent
KS-like compactifications have also been discussed in the literature to
provide a late-time acceleration of the universe~\cite{Neupane:2005nb,
Neupane:2006in}. These models may provide an alternative to the standard
$\Lambda$CDM universe we have considered at low energy. Concerning the
approach of the present paper, more precise data would certainly help to
disentangle the correlations between $\Nflux $ and $\gs$, and might
therefore allow to use astrophysical data to constrain String Theory, at
least under the strong theoretical prejudice that the relevant model of
inflation is indeed the one described here.

\acknowledgments

It is a pleasure to thank Philippe Brax and Dan Isra\"el for helpful
discussions and careful reading of the manuscript. We wish to thank
Andrew Frey for useful clarifications. The computations have been
performed at the Institut du D\'eveloppement des Ressources en
Informatique Scientifique\footnote{\texttt{http://www.idris.fr}} and at
the French Data Processing Center for
Planck-HFI\footnote{\texttt{http://www.planck.fr}} hosted in the
Institut d'Astrophysique de Paris. This work is partially supported by
the Belgian Federal Office for Scientific, Technical and Cultural
Affairs, under the Inter-University Attraction Pole grant P6/11. LL
acknowledges financial support through a DAAD PhD scholarship.

\begin{appendix}

\section{Stochastic inflation}
\label{app:stocha}

\subsection{Regime of quantum fluctuations within the throat}

\begin{figure}
\begin{center}
\includegraphics[width=7.7cm]{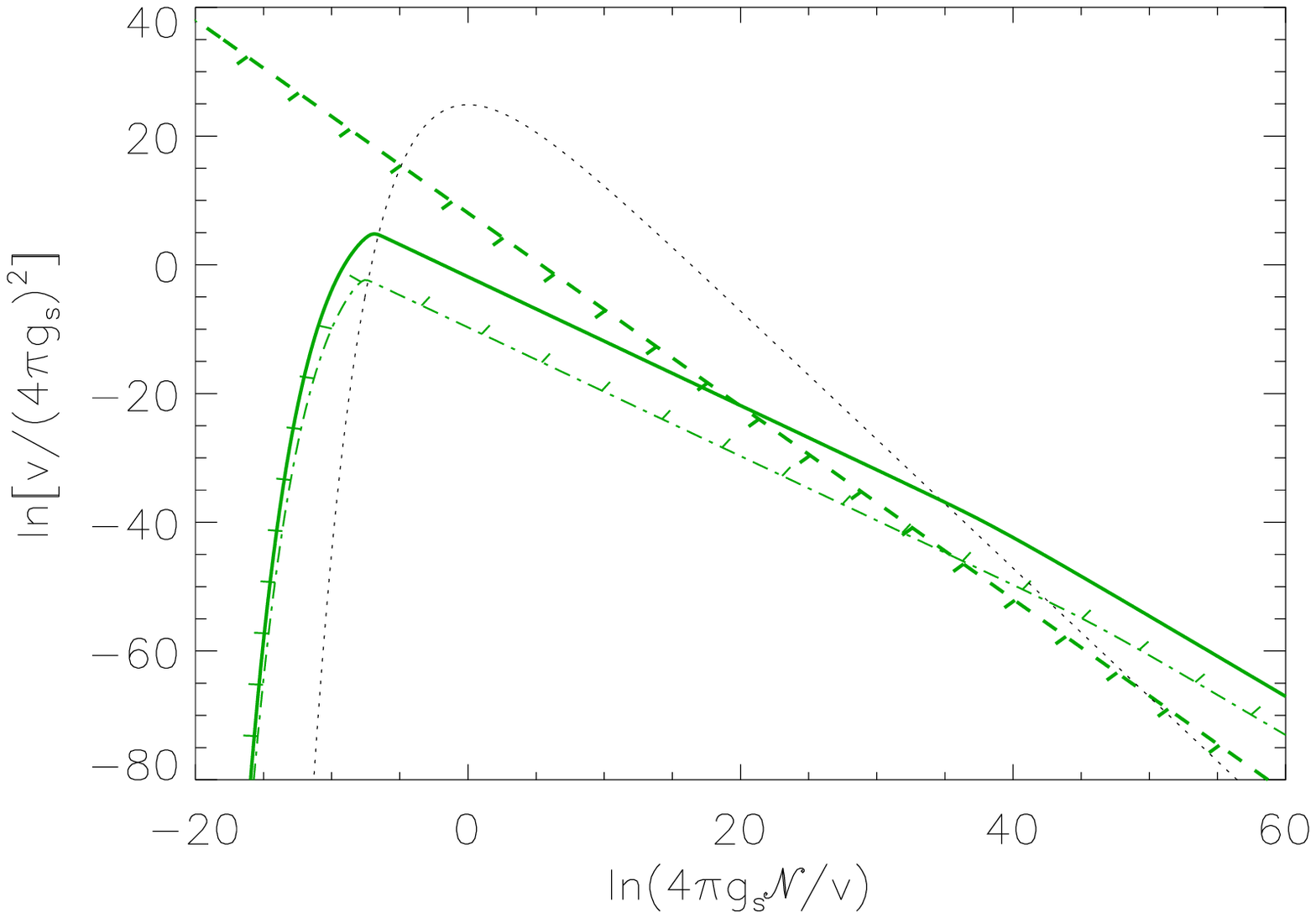}
\includegraphics[width=7.7cm]{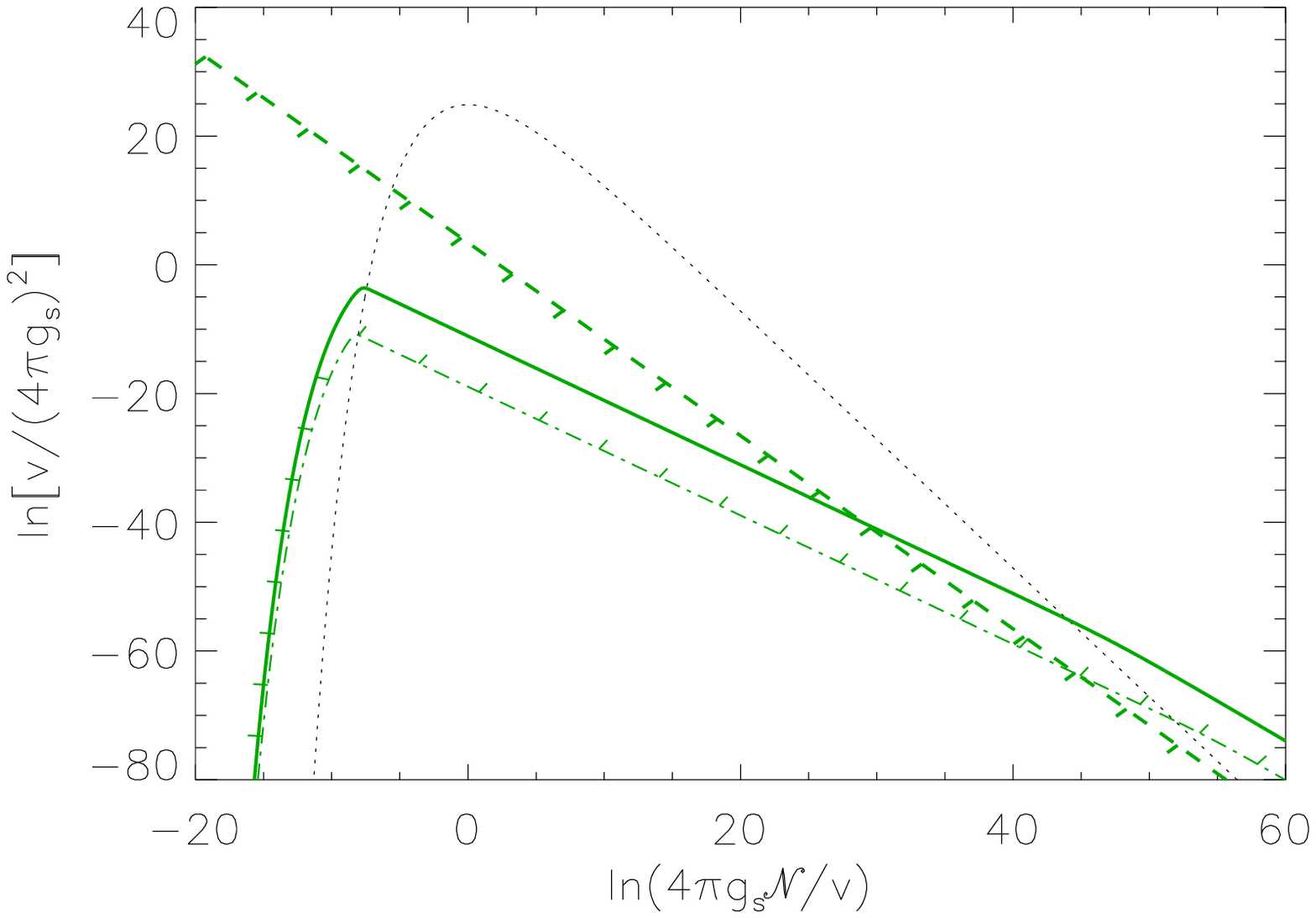}

\caption{The panels illustrate the parameter regions allowing the
  stochastic regime to take place inside the throat, for $\alpha' \mpl^2
  =1000$ (left panel) and $\alpha' \mpl^2 = 10$ (right panel). All
  curves are the same as in figure~\ref{fig:restrictions}, but in
  addition a solid green curve illustrates the condition
  $\phifluct=\phiuv$. Stochastic inflation may occur inside the throat
  only for the parameter values lying above the solid green curve curve,
  and below the volume constraint (dashed green curve with ticks down).}
  \label{fig:restrictionsUV}
\end{center}
\end{figure}

We previously stated (section~\ref{subsec:validity}) that all e-folds of
inflation should occur within a single throat and hence
\(\phi<\phi_{_{\rm UV}}\) always, \(\phi_{_{\rm UV}}\) being the field
value at the edge of the throat. In section~\ref{sec:stochastic}, we
further calculated the field value \(\phi_{\rm fluct}\) above which
quantum fluctuations have an impact on the classical trajectory. It
seems reasonable to ask under which conditions, if any, the field value
\(\phi_{\rm fluct}\) is located within the throat and hence stochastic
inflation effects can, at least in principle, affect the field
evolution. Let us try to derive the parameter restriction resulting from
\begin{equation}\label{eq:fluctUV}
\phi_{\rm fluct}<\phi_{_{\rm UV}}\,.
\end{equation}
Again, it will be important to decide which normalisation is used to
express \(\phi_{\rm fluct}\) given by equation~(\ref{eq:phifluct}). With
\(\phi_{_{\rm UV}}\) from equation~(\ref{eq:phiuv}), consider first the
normalisation (\ref{eq:scaleMclas}) valid for \(\phi_{\rm
end}=\phi_{\epsilon_{2}}\). In this case, the rescaled parameters
\((x,\,\bar{v})\) of equation~(\ref{eq:rescaledparam}) are again a
useful set to express (\ref{eq:fluctUV}) in a simple form,
\begin{eqnarray}
\label{eq:fluctUV-eps2}
\ln\bar{v}&>&4\ln{\cal F}-\ln x\, ,
\end{eqnarray}
where the constant ${\cal F}$ can be expressed as
\begin{eqnarray}
{\cal F}&=&3^{1/10}2^{-1/5}\left(45\,\frac{Q^{2}_{\rm
rms-PS}}{T^{2}}\right)^{3/20}\left(6N_{*}+10\right)^{-1/4}
\left(\pi\alpha'\mpl^{2}\right)^{1/2}\, .
\end{eqnarray}
This is a condition of the same type as the one given by
equation~(\ref{eq:onethroat-phieps2}) (from \(\phi_{{\rm
in},\epsilon_{2}}<\phi_{_{\rm UV}}\)), but more
restrictive. The parameter region in the \((\ln x,\,\ln\bar{v})\)
plane for which stochastic effects can occur within the throat is
therefore smaller than the one obtained from
equation~(\ref{eq:onethroat-phieps2}) (see
figure~\ref{fig:restrictionsUV}).

Now let us investigate the case of equations~(\ref{eq:scaleMstring})
and~(\ref{eq:scalemustring}) where the WMAP normalisation is done using
\(\phi_{\rm end}=\phistrg\gg\phi_{\epsilon_{2}}\). Inserting this
normalisation into equation~(\ref{eq:phifluct}) for \(\phi_{\rm
fluct}\), we find the following inequality in terms of \((\ln x,\ln
\bar{v})\):
\begin{eqnarray}
\label{eq:fluctUV-strings}
\ln{\calG}&<&\frac{5}{12}\,\ln x+\frac{1}{3}\,
\ln\bar{v}+\frac{2}{3}\,x^{-1/4}\, ,
\end{eqnarray}
where 
\begin{eqnarray}
{\cal G} &=& 3^{1/10}2^{-1/5}
\left(\pi\alpha'\mpl^{2}\right)^{1/2}\left(45\,\frac{Q^{2}_{\rm
rms-PS}}{T^{2}}\right)^{1/15}\, .
\end{eqnarray}
This is another curve in the plane \((\ln x,\ln\bar{v})\), analogous to
(\ref{eq:onethroat-phistrings}) though slightly simpler. Again,
equation~(\ref{eq:fluctUV-strings}) constrains the possible parameter
space outside the contour $\phi_{\epsilon _2}=\phistrg$ in the same
manner, but somewhat tighter than
equation~(\ref{eq:onethroat-phistrings}). Note that inside and outside
of the contour $\phi_{\epsilon _2}=\phistrg$, the requirements
$\phi_{{\rm in},\epsilon_{2}}<\phi_{_{\rm UV}}$ and $\phi_{\rm
fluct}<\phi_{_{\rm UV}}$ are consistent with each other. Both are
represented in figure~\ref{fig:restrictionsUV}. We have therefore shown
that the condition (\ref{eq:fluctUV}) can be met for sufficiently
generic parameter combinations, and hence stochastic effects may occur
inside the throat.

\subsection{Detailed analysis of the stochastic regime}

In the following, we apply the treatment established in
reference~\cite{Martin:2005ir} to the case of the brane inflationary
potential (\ref{eq:Vofphifull}). The underlying approach of stochastic
inflation consists in describing the evolution of a coarse-grained field
\(\varphi\) corresponding to the original scalar field \(\phi\) averaged
over \eg Hubble patch. The quantum fluctuations correspond in this
picture to stochastic noise due to small-scale Fourier modes. Therefore,
the dynamics of the coarse-grained field are controlled by a Langevin
equation, whose solution enables one to calculate the probability
density function of the field \(\varphi\). In the slow-roll
approximation, this coarse-grained field obeys the equation
\begin{equation}
\label{langevin}
\dot{\varphi}+\frac{V_{\varphi}}{3H}=\frac{H^{3/2}}{2\pi}\,\xi(t),
\end{equation}
where the Hubble parameter \(H\) is entirely described by the
coarse-grained field as well, i.e. one has (in the slow-roll
approximation)
\begin{equation}
H^{2}\simeq\frac{\kappa}{3}V(\varphi).
\end{equation}
In equation~(\ref{langevin}), $\xi (t)$ is the noise field and its mean
and two-point correlation function satisfy
\begin{equation}
\mean{\xi(t)}=0\, , \qquad \mean{\xi(t)\xi(t')}=\delta (t-t')\, ,
\end{equation}
where $\delta$ is the Dirac delta function.  The approach in
\cite{Martin:2005ir} has been to solve this system of equations using a
perturbative expansion in the noise, that is, one writes \(\varphi\) as
\begin{equation}
\varphi(t)=\phi_{\cl}(t)+\delta\varphi_{1}(t)
+\delta\varphi_{2}(t)+\dots,
\end{equation}
where \(\phi_{\cl}(t)\) is the solution of the Langevin equation without
the noise, and \(\delta\varphi_{1}(t)\) is linear in \(\xi(t)\),
\(\delta\varphi_{2}(t)\) quadratic and so on. To avoid notational
confusion, the classical field value, which is denoted by $\phi$ in the
rest of this article, is called $\phi_{\cl}$ in this Appendix. As
discussed in reference~\cite{Martin:2005ir}, there are, up to second
order, two resulting probability distributions for \(\varphi\). The
first of these, $P_{\rm c}(\varphi,t)$, is the probability of the
stochastic process to assume a given field value at a given time in a
single coarse-grained domain. The second distribution, $P_{\rm
v}(\varphi,t)$, takes into account volume effects, namely the dependence
of the volume size on the field value within this domain. These two
distributions are given by
\begin{eqnarray}
\label{Pc}
P_{\rm c}(\varphi,t)&\equiv &\mean{\delta(\varphi-\varphi[\xi])}
=\frac{1}{\sqrt{2\pi\mean{\delta\varphi_{1}^{2}}}}
\exp\left[-\frac{(\varphi-\mean{\varphi})^{2}}
{2\mean{\delta\varphi_{1}^{2}}}\right],
\end{eqnarray}
and
\begin{eqnarray}
\label{Pv} 
P_{\rm v}(\varphi,t)&=&\frac{\mean{\delta(\varphi-\varphi[\xi])
\exp\left[3{\int\rm
d}\tau\,H(\varphi[\xi])\right]}}{\mean{\exp\left[3{\int\rm
d}\tau\,H(\varphi[\xi])\right]}}\nonumber\\
&=&\frac{1}{\sqrt{2\pi\mean{\delta\varphi_{1}^{2}}}}
\exp\left[-\frac{(\varphi-\mean{\varphi}-3I^{T}J)^{2}}
{2\mean{\delta\varphi_{1}^{2}}}\right] ,
\end{eqnarray}
where the mean field value is given by
\(\mean{\varphi}\approx\phi_{\cl}+\mean{\delta\varphi_{2}}\) and where
$I$ and $J$ are two continuous vectors~\cite{Martin:2005ir}.  The
influence of stochastic effects therefore requires three quantities to
be determined for any given inflationary potential, namely
\(\mean{\delta\varphi_{1}^{2}}\), \(\mean{\delta\varphi_{2}}\), and
\(I^{T}J\). These quantities have been calculated in
reference~\cite{Martin:2005ir} as functions of the Hubble parameter (in
the slow-roll approximation) and read
\begin{eqnarray}
\mean{\delta\varphi_{1}^{2}}&=&\frac{\kappa}{2}
\left(\frac{H'}{2\pi}\right)^{2}\int_{\phi_{\cl}}^{\phi_{\ini}}
\left(\frac{H}{H'}\right)^{3}{\rm
d}\phi\, ,\label{deltaphi1sqrd}\\
\mean{\delta\varphi_{2}}&=&\frac{H''}{2H'}
\mean{\delta\varphi_{1}^{2}}+\frac{H'}{4\pi
\mpl^{2}}\left[\frac{H_{\ini}^{3}}{H_{\ini}'{}^{2}}-\frac{H^{3}}{H'{}^{2}}
\right] ,\label{deltaphi2}
\end{eqnarray}
where, in the present context, a prime denotes a derivative with
respect to $\varphi $. In the slow-roll approximation, the Hubble
parameter can be directly replaced by $\sqrt{\kappa V/3}$ and the
correction due to volume effects can be expressed as
\begin{eqnarray}
\label{voleffect}
3I^{T}J&=&\frac{12H'}{\mpl^{4}}\int_{\phi_{\cl}}^{\phi_{\ini}}
\frac{H^{4}}{H'{}^{3}}{\rm d}\phi-12\pi\frac{H}{H'}
\frac{\mean{\delta\varphi_{1}^{2}}}{\mpl^{2}}\, .
\label{3ItJ}
\end{eqnarray}

\par

In the stochastic inflation regime, we are dealing with very large field
values $\phi \gg \mu$ and one can use the expansion
(\ref{eq:Vofphiapprox}) of the potential. Equation~(\ref{deltaphi1sqrd})
then yields
\begin{eqnarray}
\label{phi1sqrdmu}
\frac{\mean{\delta\varphi_{1}^{2}}}{\mu^{2}}
&=&\frac{\kappa^{2}M^{4}}{48\pi^{2}}
\left(\frac{\mu}{\phi_{\cl}}\right)^{6}
\left[\left(\frac{\phi_{\cl}}{\mu}\right)^{4}-1\right]^{-1}
\left\{ \frac{1}{16}
  \left[ \left(\frac{\phi_{\ini}}{\mu}\right)^{16}  -
    \left(\frac{\phi_{\cl}}{\mu}
    \right)^{16}\right] \right. \nonumber \\
& - &
\left. \frac{3}{12}\left[\left(\frac{\phi_{\ini}}{\mu}\right)^{12} 
    - \left(\frac{\phi_{\cl}}{\mu}\right)^{12}\right] + \frac{3}{8} 
  \left[\left(\frac{\phi_{\ini}}{\mu}\right)^{8}
    -\left(\frac{\phi_{\cl}}{\mu}\right)^{8}\right] \right. \nonumber \\
&  - & \left.\frac{1}{4}\left[\left(\frac{\phi_{\ini}}{\mu}\right)^{4}
    -\left(\frac{\phi_{\cl}}{\mu}\right)^{4}\right]\right\},
\end{eqnarray}
Since the brane motion in the throat occurs only for $\phi_{\cl}>\mu$
(see section~\ref{sec:slowroll}), one gets in the regime $\phi_{\cl }\ll
\phi_{\ini}$
\begin{eqnarray}
\frac{\mean{\delta\varphi_{1}^{2}}}{\mu^{2}} &\simeq
&\frac{\kappa^{2}M^{4}}{768\pi^{2}}\left(\frac{\phi_{\cl }}{\mu
}\right)^{-10}\left(\frac{\phi_{\ini}}{\mu}\right)^{16}\, .
\end{eqnarray}
Since $\phi_{\cl }/\mu $ is decreasing, the variance is, as
expected, increasing as inflation proceeds.

\par

Let us now determine the correction to the mean value. As can be seen
from (\ref{deltaphi2}), no additional integration is necessary to
calculate \(\mean{\delta\varphi_{2}}\), and after some algebra one finds
\begin{eqnarray}
\frac{\mean{\delta\varphi_{2}}}{\mu} &=& \frac{1}{2} \left[
3-5 \left( \frac{\phi_{\cl}}{\mu}\right)^{4} \right]
\left(\frac{\phi_{\cl}}{\mu}\right)^{-1}
\left[ \left( \frac{\phi_{\cl}}{\mu}\right)^{4}-1\right]^{-1}
\frac{\mean{\delta\varphi_{1}^{2}}}{\mu^{2}} \nonumber\\
&+&\frac{\kappa^{2}M^4}{192\pi^{2}}
\left(\frac{\mu}{\phi_{\cl}}\right)^{5}
\left[1-\left(\frac{\mu}{\phi_{\cl}}\right)^{4}\right]^{-1/2}
\left\{
\left(\frac{\phi_{\ini}}{\mu}\right)^{10}
\nonumber \right. \\ &\times& \left.
\left[1-\left(\frac{\mu}{\phi_{\ini}}\right)^{4}\right]^{5/2}
-\left(\frac{\phi_{\cl}}{\mu}\right)^{10}
\left[1-\left(\frac{\mu}{\phi_{\cl}}\right)^{4}\right]^{5/2}
\right\} .
\end{eqnarray}

\par

\begin{figure}
\begin{center}
\includegraphics[width=7.7cm]{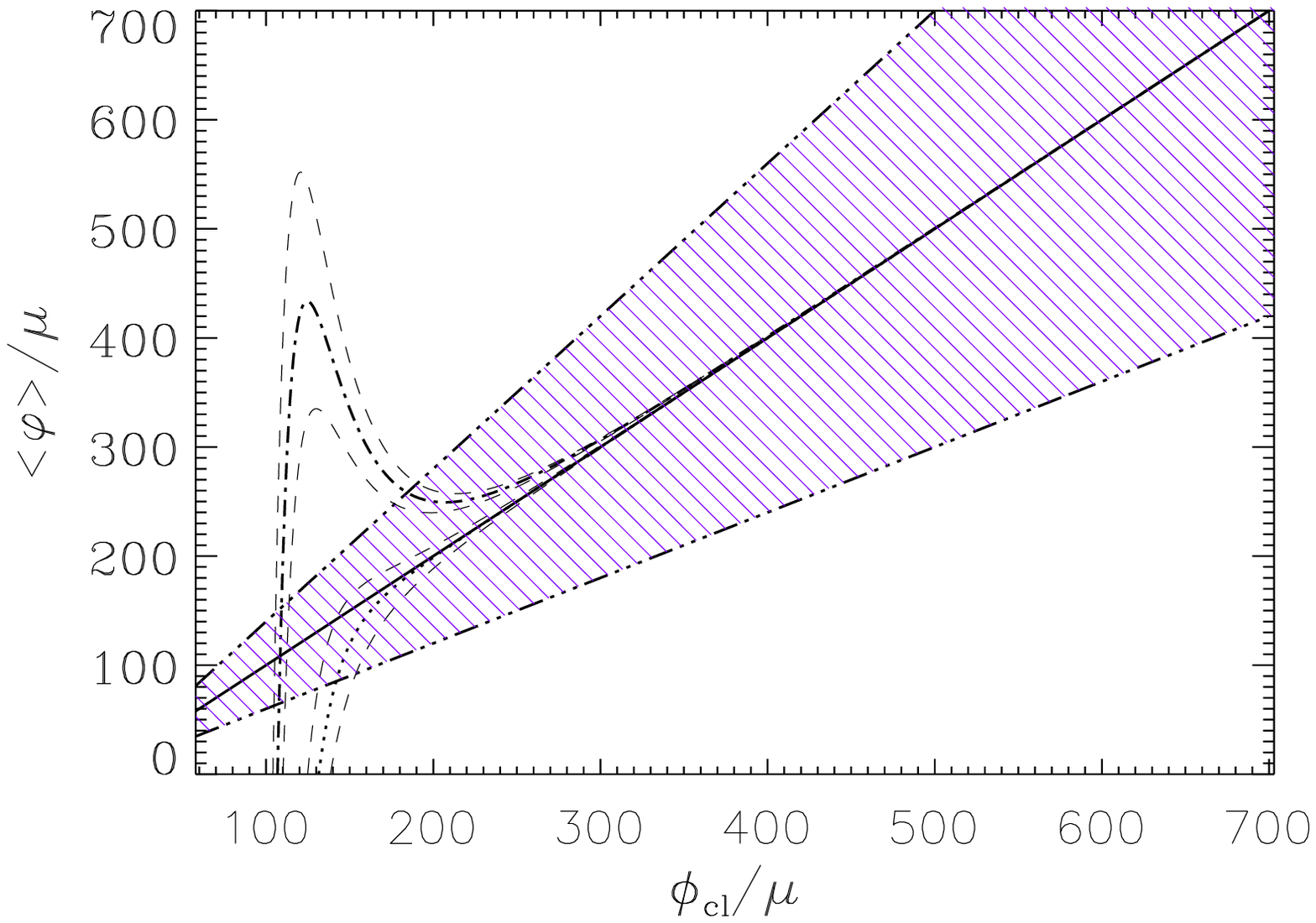}
\includegraphics[width=7.7cm]{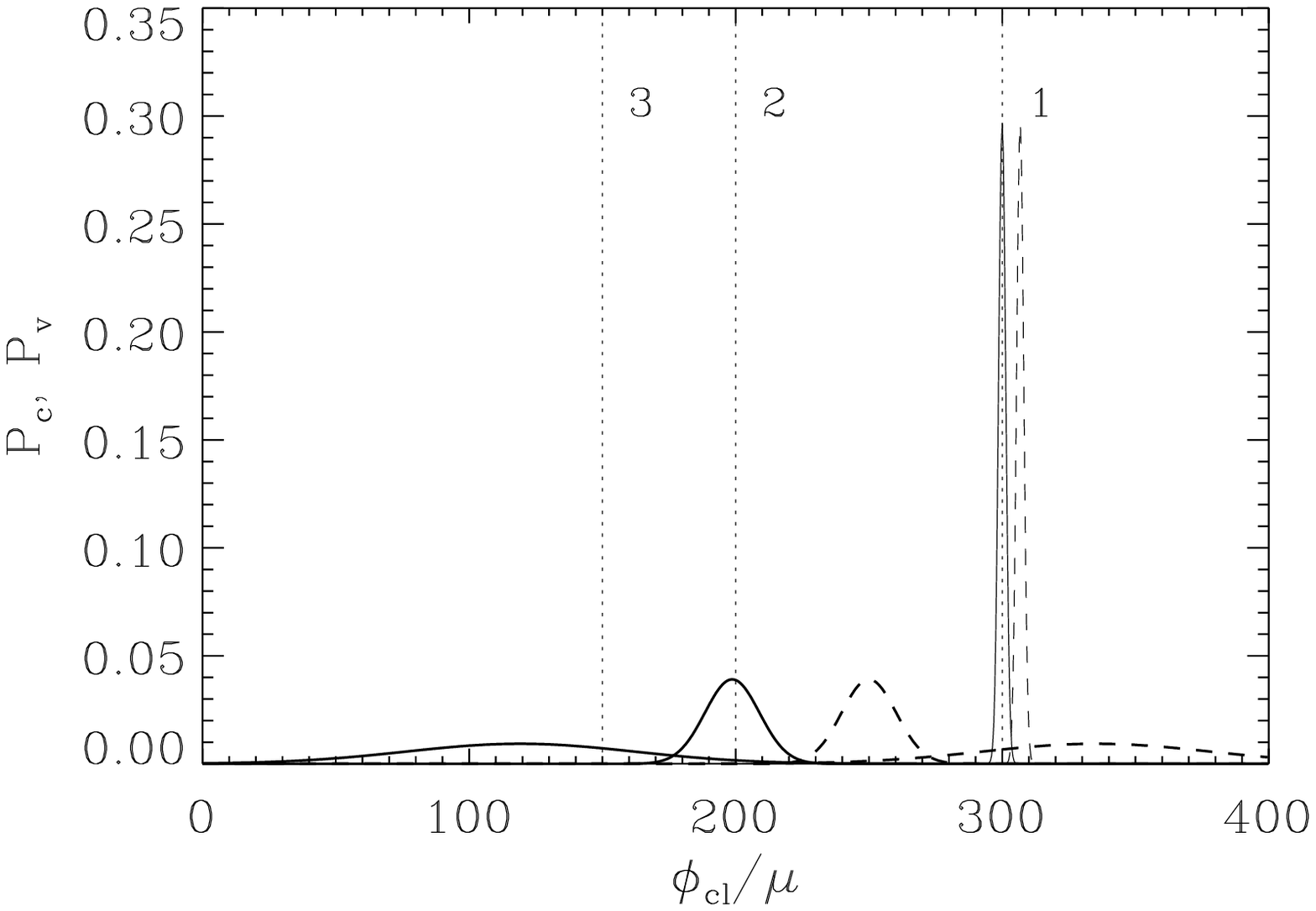} 
\caption{Left panel: Evolution of the
inflaton field when the quantum effects (with or without volume effects)
are taken into account. The underlying model is such that ``classical''
inflation stops by violation of the slow-roll conditions. Hence, the
WMAP normalisation is calculated with the help of
equation~(\ref{eq:scaleMclas}). With $v=1$ and $N_*=50$, this leads to
$\phi_{\rm fluct}/\mu \sim 793.5$. The initial condition is chosen to be
$\phi _{\ini}/\mu =700$. The solid line represents the classical
trajectory, the dotted line the mean value without the volume effects
and the dotted-dashed line the mean value with the volume effects. Right panel: Evolution of the
probability distribution $P_{\rm c}(\varphi)$ (solid lines) and $P_{\rm
v}(\varphi)$ (dashed lines). As in figure~\ref{fig:stochatrajec}, the case
displayed here corresponds to $v=1$ and $N_*=50$, leading to $\phi_{\rm
fluct}/\mu \sim 793.5$. The initial shape of the probability density
function has been chosen to be $\delta(\phi-\phi_{\rm in})$ with $\phi
_{\ini}/\mu =700$. The vertical dashed lines represent the location
of the field at successive points of the classical trajectory (the field
moves ``from right to left'').}
\label{fig:stochatrajec}
\end{center}
\end{figure}

Finally, in order to evaluate the volume effects characterised by the
term $3I^{T}J$, one needs to calculate the integral in
equation~(\ref{voleffect}). For the potential~(\ref{eq:Vofphiapprox}),
this can still be solved exactly. One obtains
\begin{eqnarray}
\frac{3I^{T}J}{\mu}&=&\frac{\kappa^{3}M^4\mu^2}{128\pi ^2}
\left(\frac{\mu}{\phi_{\cl}}\right)^{5}
\left[1-\left(\frac{\mu}{\phi_{\cl}}
\right)^{4}\right]^{-\frac{1}{2}}\times 
\left[S\left(\frac{\phi_{\ini}}{\mu}\right)
-S\left(\frac{\phi_{\cl}}{\mu }\right)\right]
\nonumber \\ & & 
-\frac{3\kappa\mu^{2}}{4}
\left[1 -\left(\frac{\mu}{\phi_{\cl}}\right)^{4}\right] 
\left(\frac{\phi_{\cl}}{\mu}\right)^{5}
\frac{\mean{\delta\varphi_{1}^{2}}}{\mu^{2}}\, ,
\end{eqnarray}
where the function $S$ is defined by
\begin{eqnarray}
S(x)&\equiv & \frac{1}{383}\left[x^2\sqrt{x^4-1}
\left(-279+326x^{4}-200x^{8}+48x^{12}\right) \right. \nonumber \\ & 
+& \left. 105\ln\left(x^{2}+\sqrt{x^{4}-1}\right)\right] .
\end{eqnarray}
The corresponding trajectory is plotted in
figure~\ref{fig:stochatrajec} (left panel). The solid line represents the classical
trajectory without the quantum effects. Since we are using $\phi _{\cl}$
as a clock, this line is just the straight line ``$y=x$''. The dotted
line gives $\langle \varphi \rangle =\phi_{\cl} +\langle \delta
\varphi_2 \rangle $, \ie the mean value of the field in the situation
where the volume effects are ignored. The two dashed lines above and
below this line are $\phi_{\cl} +\langle \delta \varphi_2 \rangle \pm
\sqrt{\langle \delta \varphi_1^2 \rangle}$. As discussed at length in
reference~\cite{Martin:2005ir}, the quantum effects accumulate and
manifest themselves only once the field is approaching the end of
inflation. As clearly shown in figure~\ref{fig:stochatrajec}, the
quantum effects are such that the field rolls down the potential more
quickly than its classical counterpart. However, the story is different
when volume effects are taken into account. The corresponding trajectory
is represented by the dotted-dashed line, the two dashed lines above and
below being the limits of the one sigma interval. In this case, the
field is first slowed down and starts moving back to even climb up the
potential. Then, at some point, it stops and rolls down again the
potential. However, as we will see in the next subsection, this last
part of the evolution can not be tracked within our approximation since
it is outside its domain of validity (represented by the shaded
regions).

In figure~\ref{fig:stochatrajec} (right panel), we have represented the corresponding
distributions without and with volume effects as given by
equations~(\ref{Pc}) and~(\ref{Pv}) respectively. The vertical dotted
lines corresponds to the classical values of the field at three
successive times (denoted ``1'', ``2'' and ``3''), namely $\phi
_{\cl}^{(1)}/\mu =300$, $\phi _{\cl}^{(2)}/\mu =200$ and
$\phi_{\cl}^{(3)}/\mu =150$. The solid lines are the three distributions
$P_{\rm c}$ corresponding to the three classical field values while the
dashed lines are the three $P_{\rm v}$'s. This plot essentially confirms
the above analysis. In particular, one notices that the maximum of
$P_{\rm c}$ is ahead the classical vacuum expectation value as discussed
before. On the contrary, the volume effects ``retain'' the field and
this one can even change the direction of its motion (from right to left
to left to right). Finally, the two distributions strongly spread out
while inflation is proceeding.

\subsection{Reliability of the perturbative treatment}

In this section, we discuss in which range of field values
\(\mean{\varphi}\in[\phi_{\cl}-|\Delta\varphi_{\rm
min}(\phi_{\cl})|,\phi_{\cl}+\Delta\varphi_{\rm max}(\phi_{\cl})]\) the
previous analysis is valid. Indeed, as discussed above, in order to
estimate the mean value of the field with or without the volume effects,
a perturbative expansion in the noise has been used which may break down
at some point. It has been shown in \cite{Martin:2005hb} that
\(\Delta\varphi_{\rm min}\) and \(\Delta\varphi_{\rm max}\) can be found
from requesting that the two conditions
\begin{eqnarray}
\max_{x\in[\phi_{\cl},\phi_{\cl}
+\Delta\varphi_{\rm max}(\phi_{\cl})]}\left|\frac{H^{(4)}(x)}{6}
\Delta\varphi^{3}\right|&\ll&\left|\frac{H_{\rm
cl}'''}{2}\right|\Delta\varphi^{2},
\label{cond1}\\ 
\max_{x\in[\phi_{\cl},\phi_{\cl}
+\Delta\varphi_{\rm max}(\phi_{\cl})]}
\left|\frac{\left[H^{3/2}(x)\right]''}{2}\right|
\Delta\varphi^{2}&\ll&\left|\left(H_{\rm
cl}^{3/2}\right)''\Delta\varphi\right|,
\label{cond2}
\end{eqnarray}
are simultaneously satisfied. The strongest bounds on
\(\Delta\varphi_{\rm min}\) and \(\Delta\varphi_{\rm max}\) that follow
from (\ref{cond1}) and (\ref{cond2}) then give the reliability interval
for the distributions \(P_{\rm c}(\varphi,t)\) and \(P_{\rm
v}(\varphi,t)\). In the present case, a straightforward calculation
leads to
\begin{equation}
\frac{\Delta\varphi_{\rm min}}{\mu}\approx-\frac{2}{5}\,
\frac{\phi_{\cl}}{\mu},\qquad \frac{\Delta\varphi_{\rm
max}}{\mu}=\frac{2}{5}\, \frac{\phi_{\cl}}{\mu}\, .
\end{equation}
The resulting reliability region is plotted in
figure~\ref{fig:stochatrajec} and corresponds to the purple hatched
region. We see that the approximation breaks down more quickly for
\(P_{\rm v}(\varphi,t)\) than for \(P_{\rm c}(\varphi,t)\). In
particular, the conclusion that, due to volume effects, the field
reverses its motion comes from a regime which is at the border of the
reliability interval. We conclude that, in order to have access to this
phenomenon and, in particular, to discuss eternal inflation in this
model, a more powerful method is needed, as, for instance, a full
numerical integration of the Langevin equation (\ref{langevin}).

\section{Brane inflation from String Theory}
\label{app:kklt}

\subsection{The background action of type IIB Superstring Theory}
\label{appsec:background}

We begin by considering the background action only, and to that end
place ourselves in the context of ten-dimensional type IIB Superstring
Theory. In the string frame, the effective action reads (see \eg
references~\cite{Klebanov:2000nc,Polchinski:1998rr,Johnson:2000ch})
\begin{eqnarray}
\label{eq:10daction}
\mathscr{S}_{\mathrm{tot}} & = & -\dfrac{1}{2 \hat{\kappa}} \int
\ud^{10}x \sqrt{-g} \left\{ \ue^{-2 \Phi} \left[R + 4 \left(\nabla
  \Phi\right)^2 - \dfrac{1}{12} \left(\ud B_{(2)} \right)^2 \right] \right.
\nonumber \\ & - & \left.\dfrac{1}{2} \left(\ud C_{(0)} \right)^2 -
\dfrac{1}{12} \left[\ud C_{(2)} - C_{(0)} \ud B_{(2)} \right]^2 -
\dfrac{1}{480} \left(F_{(5)}\right)^2 \right\} \nonumber \\
&+& \dfrac{1}{4 \hat{\kappa}} \int\,C_{(4)}\wedge \ud B_{(2)}\wedge\ud C_{(2)}\,,
\end{eqnarray}
where $\hat{\kappa}$ stands for
\begin{equation}
\hat{\kappa} = \dfrac{1}{2} (2 \pi)^7 {\alpha'}^4\,.
\end{equation}
The field strength denoted by \(F_{(5)}\) in
equation~(\ref{eq:10daction}) is given by
\begin{equation}\label{defF5}
F_{(5)}={\rm d} C_{(4)}+\frac{1}{2}\left[B_{(2)}\wedge{\rm d}
C_{(2)}-C_{(2)}\wedge{\rm d} B_{(2)}\right].
\end{equation}
The brackets \((i)\) indicate the number of indices of the corresponding
field. The fields in this theory are therefore the scalars \(\Phi\) (the
dilaton) and \(C_{(0)}\) (the axion), \(g_{mn}\) (ten-dimensional
gravity) as well as the antisymmetric fields \(B_{(2)}\) (Kalb-Ramond),
\(C_{(2)}\), and \(C_{(4)}\). Note that the totally antisymmetric
\(\epsilon\) tensor refers to curved space, where we define
\begin{eqnarray}
\epsilon_{(10)}=\pm 1&&\textnormal{for pair/odd 
permutation of the indices,}
\\
\epsilon_{(10)}=0&&\textnormal{if two or more indices are equal,}\\
\epsilon^{(10)}=\frac{1}{g}\,\epsilon_{(10)}.&&
\end{eqnarray}
The first line of (\ref{eq:10daction}) describes the
Neveu-Schwarz--Neveu-Schwarz (NS-NS) sector, while the second line
results from the Ramond-Ramond (RR) sector. The third line is the
Chern-Simons (C-S) coupling between the sources of \(C_{(4)}\) (that
is, RR charged objects such as D3 branes) and background fluxes. From
the definition (\ref{defF5}) of \(F_{(5)}\) it is clear that if
\(\left({\rm d} B_{(2)}\right)\) and \(\left({\rm d} C_{(2)}\right)\)
have non-zero background values, they can produce a non-zero
\(F_{(5)}\) field strength even in the absence of localised
\(C_{(4)}\) sources. Turning this argument around, one can also
interpret such fluxes as intrinsic ``background'' \(C_{(4)}\) sources.

\subsection{The equations of motion}
\label{subsec:motion}

{}From equation~(\ref{eq:10daction}), it is straightforward to derive
the equations of motion for these fields. The resulting system of
equations has been studied in the references~\cite{Klebanov:1999rd,
  Klebanov:2000nc, Klebanov:2000hb, PandoZayas:2000sq} and we now
present the solutions found there. Note that in addition to finding
simultaneous solutions for \emph{all} equations of motion, one has to
impose the condition of self-duality of the 5-form field strength,
$F_{(5)}=\star F_{(5)}$, onto these solutions.

\subsubsection{Ansatz for the metric.}
\label{subsec:ansatz}

Let us focus on the Einstein equations first, which follow from
(\ref{eq:10daction}) by variation with respect to the ten-dimensional
metric $g_{mn}$\footnote{Small Latin indices $m,n,\dots$ run from 0 to 9
(with the exception of $i=1,2,3$), while small Greek indices
$\mu,\nu,\dots$ run from 0 to 3. Capital Latin indices $M,N,\dots$ run
over the extra dimensions coordinates only, \ie from 4 to 9.}. In this
section, the ten-dimensional set of coordinates we use is
\(\left(x^{0},\,x^{i},\,r,\,\psi,\,\theta_{1},\,\phi_{1},\,
\theta_{2},\,\phi_{2}\right),\,i=1,2,3\). A super-gravity motivated
ansatz for the metric reads
\begin{equation}
\label{ansatz}
{\rm d}s^{2}=\frac{1}{\sqrt{h(r)}}\,\eta_{\mu\nu}
{\rm d}x^{\mu}{\rm d}x^{\nu}+\sqrt{h(r)}\,{\rm d}s_{6}^{2}\,,
\end{equation}
which involves a natural splitting into a four-dimensional
extended space-time and the six compactified extra dimensions. Note
that in (\ref{ansatz}), the four-dimensional metric has been set to
Minkowski \(\eta_{\mu\nu}={\rm diag}(-1,+1,+1,+1)\), multiplied by a
factor depending on the fifth dimension \(r\). In principle one can
also choose \eg a FLRW universe for the four-dimensional section of
the metric by replacing $\eta_{\mu\nu}$ with $g_{\mu\nu}={\rm
  diag}\left(-1,a^{2},a^{2},a^{2}\right)$. For the six-dimensional
section, the choice of interest here is
\begin{eqnarray}
{\rm d}s_{6}^{2}&=&{\rm d}r^{2}+r^{2}{\rm d}s_{T_{1,1}}^{2},
\end{eqnarray}
where \({\rm d}s_{T_{1,1}}^{2}\) is the metric on the Einstein space
\(T_{1,1}\)~\cite{Candelas:1989js}. Note that the choice of geometry
and compactification for the extra dimensions is an input parameter
for the resulting inflationary model. In particular, one can enforce
``warping'' on the conifold \(T_{1,1}\) by giving non-vanishing
background values to \(\left({\rm d} B_{(2)}\right)\) and \(\left({\rm
  d} C_{(2)}\right)\). The main virtue of these fluxes is that in
general they stabilise the scalar fields corresponding to the complex
structure moduli of the extra dimensions. The warped conifold geometry
was studied in reference~\cite{Klebanov:2000hb} and has become known
as the Klebanov-Strassler (KS) throat. Interest in the KS geometry is
due to the fact that it represents an exactly solvable example of
field strengths background fluxes. Nevertheless, this scenario can be
viewed as generic because ``warped throats'' appear in many flux
compactifactions invoked to stabilise the moduli. The metric of the
\(T_{1,1}\) space reads
\begin{eqnarray}
{\rm d}s_{6,\,T_{1,1}}^{2} &=&{\rm d}r^{2}+r^{2}{\rm d}
s_{T_{1,1}}^{2}\nonumber\\
&=&{\rm d}r^{2}+\frac{r^{2}}{9}\left({\rm d}\psi+\cos\theta_{1}\,
{\rm d}\phi_{1}+\cos\theta_{2}\,{\rm d}\phi_{2}\right)^{2}\nonumber\\
&+&\frac{r^{2}}{6}\sum_{i=1}^{2}\left({\rm d}\theta_{i}^{2}
+\sin^{2}\theta_{i}\,{\rm d}\phi_{i}^{2}\right),
\label{conifold}
\end{eqnarray}
whereas the one on its deformed counterpart is
\begin{eqnarray}
{\rm d}s_{6,\,{\rm warped}}^{2}&=&\frac{1}{K(r)}\,{\rm d}r^{2}
+\frac{r^{2}}{9}K(r)\,\left({\rm d}\psi
+\cos\theta_{1}\,{\rm d}\phi_{1}+\cos\theta_{2}\,{\rm d}
\phi_{2}\right)^{2}\nonumber\\
&+&\frac{r^{2}}{6}\,\left({\rm d}\theta_{1}^{2}+\sin^{2}
\theta_{1}\,{\rm d}\phi_{1}^{2}\right)\nonumber\\
&+&\frac{1}{6}\left(r^{2}+6z^{2}\right)\left({\rm d}
\theta_{2}^{2}+\sin^{2}\theta_{2}\,{\rm d}\phi_{2}^{2}\right),
\label{warpedcone}
\end{eqnarray}
where the warping function \(K(r)\) is given by
\begin{equation}
K(r)=\frac{r^{2}+9z^{2}}{r^{2}+6z^{2}}\, .
\end{equation}
The quantity \(z\neq0\) (in general complex) is the warping number
introduced in the definition of the deformed conifold (see below).

\par

We begin our investigation with the simple conifold
(\ref{conifold}). Using our coordinates $(r,\,\psi,\,\theta_{1},\,
\phi_{1},\,\theta_{2},\,\phi_{2})$ for the six-dimensional section of
the metric, equation~(\ref{conifold}) shows that we are dealing with a
generalised ``cone'' whose base is given by two spheres,
\(\mathscr{S}_{3}\times\mathscr{S}_{2}\)
\cite{Candelas:1989js}. However, the defining equation of the conifold
is more readily written in terms of four complex coordinates \(w_{i},
\,i=1\dots4\) which obey~\cite{Klebanov:2007us}
\begin{equation}\label{defconifold}
\sum_{i=1}^{4} w_{i}^{2}=0\,.
\end{equation}
{}From equation~(\ref{defconifold}), one sees that the geometry becomes
singular at \(w_{i}=0\) for all \(i=1\dots4\). In this case,
\(\mathscr{S}_{2}\) and \(\mathscr{S}_{3}\) are shrinking to a point at
the tip of the conifold where \(r=0\). Using the six-dimensional metric
(\ref{conifold}) within the ten-dimensional ansatz (\ref{ansatz}), the
components of the Einstein equations can be combined to yield the
differential equation
\begin{equation}
\label{deth}
\left(\frac{\partial^{2}}{\partial r^{2}}h\right)
+\frac{5}{r}\left(\frac{\partial}{\partial r}h\right)=0\, ,
\end{equation}
whose solutions read
\begin{equation}
\label{h}
h(r)=C_{2}+\frac{C_{1}}{4r^{4}}\, .
\end{equation}
Note that it is possible to choose \(C_{2}=0\) and that the
ten-dimensional space is Ricci flat. Equation~(\ref{deth}) can indeed
be obtained by setting the Ricci scalar \(\Ricci=0\).

\par

For the time being, we have considered the case of a simple (and
singular) conifold, where a priori no background fluxes have to be
present, \ie it is consistent to choose \({\rm} B_{(2)}={\rm}
C_{(2)}=0\). One can now turn these fluxes on; the resulting effect is
to induce corrections in equation~(\ref{h}) which are proportional to
the ratio of the flux quanta ${\cal M},\,{\cal K}$ in the NS-NS and
the RR sector respectively, namely \(\left({\rm d} B_{(2)}\right)\)
and \(\left({\rm d} C_{(2)}\right)\). This was the route taken in
references~\cite{Klebanov:1999rd,Klebanov:2000nc}. In the presence of
fluxes, the singular conifold is replaced by its deformed counterpart
and this amounts to modify equation~(\ref{defconifold}) into
\begin{equation}\label{defwarped}
\sum_{i=1}^{4} w_{i}^{2}=z\,,
\end{equation}
where the warping number $z$ is a non-vanishing complex number. The
solution for \(h(r)\) in the deformed conifold case (again found from
the Einstein equations or from $\Ricci=0$) is~\cite{PandoZayas:2000sq}
\begin{equation}\label{hwarped}
h(r)=C_{3}+C_{4}\left[\frac{1}{18r^{2}z^{2}}
-\frac{1}{162z^{4}} \ln\left(1 + \dfrac{9z^{2}}{r^2}\right )  \right].
\end{equation}
The constants \(C_{3},\,C_{4}\) can again be determined from the
chosen compactified geometry.

\par
 
In ``cone'' terminology, equation~(\ref{defwarped}) implies that the
\(\mathscr{S}_{2}\) still shrinks to a point at the tip of the cone, but
the \(\mathscr{S}_{3}\) remains finite. As mentioned above, this is
achieved by turning on background fluxes for \(\left({\rm d}
B_{(2)}\right)\) and \(\left({\rm d} C_{(2)}\right)\), \ie aligning them
with two Poincar\'e dual 3-cycles. Two such cycles which verify
equation~(\ref{defconifold}) are described by
\begin{equation}
\mathscr{A:}\,\sum_{i=1}^{4} x_{i}^{2}=z,
\qquad\mathscr{B:}\,x_{4}^{2}-\sum_{i=1}^{3} y_{i}^{2}=z \,,
\end{equation} 
where the complex coordinates \(w_{i}\) have been decomposed into their
real and imaginary parts, $x_i$ and $y_i$ respectively. Alignment of the
3-forms leads to a Dirac quantisation of the fluxes along the 3-cycles
\begin{equation}
\frac{1}{2\pi\alpha'}\int_{\mathscr{A}}\left({\rm d}
C_{(2)}\right)=2\pi {\cal M}\,, \qquad\frac{1}{2\pi\alpha'}
\int_{\mathscr{B}}\left({\rm d} B_{(2)}\right)=-2\pi {\cal K}\,.
\end{equation}
These conditions impose restrictions on the possible ans\"atze for the
fields, and eventually on the constants $C_{3},\,C_{4}$ in
equation~(\ref{hwarped}).

\par

The difference between the simple and the deformed  conifold will
be most remarkable close to the tip of the cone at $r=r_{0}$. Let us
notice that \(r_{0}\neq0\) since by deforming the conifold, we have
replaced the ``sharp'' tip by a finite size sphere \(\mathscr{S}_{3}\).
However, at some distance from the tip, both metrics should essentially
give the same description of the geometry and in particular, we can
``glue'' the warped metric (\ref{hwarped}) to its simple version
(\ref{h}) at a certain radius \(r_{*}\).  From this matching, the
constants $C_{1}$ and $C_{2}$ become some function of $C_{3}$, $C_{4}$
and \(r_{*}\). Far from the singularity, the metric (\ref{h}) expressed
in terms of those constants provides an accurate enough description of
the warped conifold geometry. Let us now link this setup with the brane
inflation model we are interested in. We place two branes (D3 and
anti-D3) into the background, and \(r\) will denote the coordinate
distance between them. Now, one of these branes (the anti-D3 brane) is
placed at the tip \(r_{0}\) while ensuring \(r\) to be large. In that
case, the metric in the vicinity of the D3 brane located at \(r_{1}\gg
r_{0}\) can still be accurately described by equation~(\ref{h}). It is
for this reason that, while the solution (\ref{hwarped}) for the warped
metric is known, for all ``practical'' applications the simple solution
(\ref{h}) suffices.

\par

While we do not write down the exact solutions for the other fields in
this theory exactly, let us state that the solution for the 4-form
$C_{(4)}$, in the string frame, is found to
be~\cite{Kachru:2003sx,Klebanov:1999rd,Klebanov:2000nc,
Klebanov:2000hb,Klebanov:1998hh,Klebanov:2007us}
\begin{equation}
\label{C4}
\left[C_{(4)}\right]_{0123}(r)=\gs^{-1} \frac{1}{h(r)}\, .
\end{equation}
This form turns out to be crucial in the derivation of the effective
inflaton potential as it is demonstrated below.

\subsubsection{Interpretation of the background fluxes.}
\label{subsubsec:interpretation}

We conclude our investigation of the background with some comments on
an intuitive interpretation of background fluxes. If we consider a
simple (singular) conifold geometry with non-zero fluxes, it can be
viewed as a stack of \({\cal N}={\cal M}{\cal K}\) branes with RR
charge placed at the \(r=0\) singularity. The total RR charge $\Nflux$
is therefore the product of the fluxes. If one uses the deformed
conifold instead, this singularity is replaced by a three-sphere
\(\mathscr{S}_{3}\) with radius \(r_{0}\), and the charge \({\cal N}\)
of the branes is literally smeared out across the three-sphere. In the
following sections, we first add one more D3 brane (so that there are
\(({\cal N}+1)\) RR charged branes in total) to this background. This
additional brane can be treated as a perturbation of the metric
background (\ref{h}). We then probe the perturbed background using a
test anti-D3 brane and calculate the force it experiences.

\subsection{Perturbed geometry}
\label{subsec:perturbed background} 

In order to describe a D\(p\) brane inserted into this super-gravity
background, one has to consider two additional contributions to the
action (\ref{eq:10daction})~\cite{Polchinski:1998rr}. In the string
frame, they are the DBI action of a D\(p\) brane,
\begin{equation}\label{DBI}
\mathscr{S}_{{\rm D}p}=-\mu_{p}\int_{M_{p}}{\rm d}^{p+1}
\xi \ue^{-\Phi}\left[-\det\left(G_{\alpha\beta}
+B_{\alpha\beta}+2\pi\alpha'F_{\alpha\beta}\right)\right]^{1/2},
\end{equation}
plus its Chern-Simons coupling
\begin{equation}\label{ChernSimons}
\mathscr{S}_{\rm C-S, brane}=\pm q_{p}\int_{M_{p}} C_{(p+1)}\, ,
\end{equation}
since D\(p\) branes in type IIB Superstring Theory are localised sources
of \(C_{(p+1)}\) field strength\footnote{More generally, a D\(p\) brane
in type IIB theory can source \(C_{(l)}\) fields for all even \(l\le
p+1\)~\cite{Polchinski:1998rr}.}. The RR charge $q_p$ is given by
$q_p=\mu_p$. For stabilised dilaton $\gs=\mathrm{e}^{\mean{\Phi}}$,
the effective brane tension \(T_{p}\) is related to the fundamental
brane tension \(\mu_{p}\) by
\begin{equation}
T_p = \dfrac{\mu _p}{\gs} ,
\end{equation}
and can be calculated from~\cite{Polchinski:1998rr}
\begin{equation}
T_{p}^{2}=\frac{\pi}{\kappa_{10}}\,\left(4\pi^2\alpha'\right)^{3-p}\,.
\end{equation}
The gravitational coupling constant $\kappa_{10}$ was given in equation~(\ref{kappaten}).
The upper sign in equation~(\ref{ChernSimons}) refers to the case of a
D\(p\) brane, the lower to an anti-D\(p\) brane. The quantities
\(\xi^{i}\) are the coordinates on the brane, \(G_{\alpha\beta}\) is
the induced four-dimensional metric and \(B_{\alpha\beta}\) the
induced anti-symmetric tensor (Kalb-Ramond field) on the brane. These
two are given by
\begin{eqnarray}
G_{\alpha\beta}&=&g_{mn}\frac{\partial x^{m}}{\partial 
\xi^{\alpha}}\frac{\partial x^{n}}{\partial \xi^{\beta}}\,,
\label{inducedG}\\
B_{\alpha\beta}&=&B_{mn}\frac{\partial x^{m}}{\partial 
\xi^{\alpha}}\frac{\partial x^{n}}{\partial \xi^{\beta}}\,,
\label{inducedB2}
\end{eqnarray}
\(g_{mn}\) and \(B_{mn}\) being ten-dimensional. It is moreover
possible (notably for parallel branes) to choose a configuration in
which the gauge field on the brane \(F_{\alpha\beta}\) vanishes.

\par

In our case, the brane has three spatial dimensions, and we may choose
the \(\xi^{\alpha}\) to be parallel to the \((x^{0},\,x^{i}),\,i=1,2,3\)
coordinate axes. If the coordinates \(y_{1}^{A},\,A=4\dots9\) of the
brane are assumed to be static and independent of the location \emph{on}
the brane, it can be shown that there is no \(B_{(2)}\) ``pull-back''
onto the brane. Then equation~(\ref{inducedG}) implies
\begin{equation}
G_{00}=g_{00},\qquad G_{ii}=g_{ii},\qquad \textrm{for $i=1,2,3$}.
\end{equation}
The determinant of this induced metric, given that the ten-dimensional
metric is (\ref{ansatz}), reads
\begin{equation}\label{detG}
\det G_{\alpha\beta}=G=-\frac{1}{h^{2}(r_{1})}\,.
\end{equation}
Notice that we would pick up an additional \(a^{6}(t)\) factor if we
were to use a FLRW universe on the brane instead of \(\eta_{\mu\nu}\).

\par

D3 branes are sources of \(C_{(4)}\) fields, and thus of the
corresponding \(F_{(5)}\) field strength, as can be seen from the
Chern-Simons coupling (\ref{ChernSimons})
\begin{equation}
\mathscr{S}_{\rm C-S}=\pm\sum_{i}\mu^{(i)}_{3}\int_{M_{4}^{(i)}}
{\rm d}\tau \,{\rm d}\sigma^{(i)}_{1}{\rm d}\sigma^{(i)}_{2}\,
{\rm d}\sigma^{(i)}_{3}\,\frac{{\rm d}x^{m}}{{\rm d}\tau}
\frac{{\rm d}x^{n}}{{\rm d}\sigma_{1}^{(i)}}
\frac{{\rm d}x^{k}}{{\rm d}\sigma_{2}^{(i)}}
\frac{{\rm d}x^{l}}{{\rm d}\sigma_{3}^{(i)}}C_{mnkl}\,,
\end{equation}
the summation over \(i\) running over all the D3 branes
present~\cite{Cline:2006hu,Burgess:2007pz}. In this equation, the
integration is carried out over the world volume \(M_{4}^{(i)}\) of each
brane, and \(\mu^{(i)}_{3}\) denotes their respective RR charge. It then
follows from equations (\ref{DBI}) and (\ref{ChernSimons}) that the
presence of one D3 brane at \(r_{1}\) adds the following terms to the
action
\begin{equation}\label{firstbrane}
\mathscr{S}_{{\rm D}3}=-T_{3}\int_{M_{3}}
{\rm d}^{4}\xi\,\sqrt{-G}+ \mu_{3}\int_{M_{3}}
\,C_{(4)}\,,
\end{equation}
where \(G\) is given by equation~(\ref{detG}). The D3 brane will
manifest itself as a small perturbation of the background, i.e. the
function \(h(r)\) calculated in \ref{appsec:background} is replaced by
\begin{equation}
\tilde{h}(r,r_{0})=h(r)+\delta h(r_{0})\,.
\end{equation}
Let us consider the effect of replacing $h(r)$ by $\tilde{h}(r,r_{1})$
in the Einstein equations. Since \(h(r)\) already fulfils
equation~(\ref{deth}), the resulting equation for \(\delta h(r_{1})\)
is
\begin{equation}
\left(\frac{\partial^{2}}{\partial r_{1}^{2}}\,\delta h\right)
+\frac{5}{r_{1}}\left(\frac{\partial}{\partial r_{1}}\,\delta h\right)=0\,.
\end{equation}
which has the solution
\begin{equation}\label{deltah}
\delta h(r_{1})=\tilde{C}_{2}+\frac{\tilde{C}_{1}}{4r_{1}^{4}}\,.
\end{equation}
The new constants $\tilde{C}_{1}$ and $\tilde{C}_{2}$ can be related to
the former $C_{1}$ and $C_{2}$ in equation~(\ref{h}) by noting that the
difference between the previous scenario and the present case comes from
adding one D3 brane. One may therefore intuitively guess that
\cite{Kachru:2003sx}
\begin{equation}
\tilde{C}_{i}=\frac{C_{i}}{{\cal N}}\,,\qquad \textrm{for $i=1,2$}\,.
\end{equation}
The perturbed background metric \(\tilde{h}(r,r_{1})\) is obtained
from equations~(\ref{h}) and (\ref{deltah}) and, without loss of
generality, one can set $C_{2}=\tilde{C}_{2}=0$ to get
\begin{equation}\label{fullh}
\tilde{h}(r,r_{1})=\frac{C_{1}}{4r^{4}}\left(1+\frac{1}{{\cal N}}\,
\frac{r^{4}}{r_{1}^{4}}\right)\,.
\end{equation}
Note that as a consequence of the presence of the D3 brane (which is a
localised RR charge source), the $C_{(4)}$ field is also perturbed, as
can easily be seen from (\ref{C4}).

\subsection{Test anti-brane in the perturbed geometry}
\label{subsec:test anti-brane}

Let us denote by \(y_{0}^{4}=r_{0}\) the radial position at which we
insert the anti-D3 brane, while for all other extra coordinates we
take \(y_{0}^{A}=y_{1}^{A},\,A=5\dots9\). The anti-brane has the same
tension but opposite charge as the brane and its contribution to the
action in the string frame reads
\begin{equation}\label{secondbrane}
\mathscr{S}_{\bar{{\rm D}3}}=-T_{3}
\int_{M'_{3}}{\rm d}^{4}\xi\,\sqrt{-G'}-\mu_{3}
\int_{M'_{3}}\,C_{(4)}\,,
\end{equation}
\(G'\) being the induced four-dimensional metric on the anti-brane.  We
assume that the anti-brane is ``light'' in the sense that it does not
affect the already perturbed geometry. While we keep the $y_{0}^{i}$
static ($i$ varying from $0$ to $5$), we assume that the radial
anti-brane coordinate $y_{0}^{4}=r_{0}$ \emph{does} depend on the time
\(x^{0}\) (but not on the \(x^{i}\)).  Deriving \(G'\) explicitly leads
to the expression
\begin{equation}\label{detGprime}
\det G'_{\alpha\beta}= G'= -
\frac{1}{\tilde{h}^{2}(r_{0},r_{1})}
\left[1-\tilde{h}(r_{0},r_{1})
\left(\frac{\partial r_{0}}{\partial t}\right)^{2}\right].
\end{equation}
Inserted into the perturbed geometry of the previous subsection,
\ie where the warp factor is $\tilde{h}(r,r_{1})$ and the 4-form
${C}_{(4)}$ is also perturbed, the mobile anti-brane contributes to
the action the terms
\begin{eqnarray}
\mathscr{S}_{\bar{{\rm D}3}}
&=&-\int{\rm d}^{4}\xi\, \frac{T_{3}}{\tilde{h}(r_{0},r_{1})}
\left\{1+
\left[1-\tilde{h}(r_{0},r_{1})
\left(\frac{\partial r_{0}}{\partial t}\right)^{2}\right]^{1/2}
\right\},\label{twobranes}
\end{eqnarray}
where we have used equation~(\ref{C4}). Expanding this expression
for small velocities yields
\begin{equation}
\label{scalaraction}
\mathscr{S}_{\bar{{\rm D}3}} =\int{\rm d}^{4}\xi \left[
  \frac{T_{3}}{2} \left(\frac{\partial r_{0}}{\partial t}\right)^{2}
  -\frac{2T_{3}}{\tilde{h}(r)}\right].
\end{equation}
By applying the canonic renormalisation
\(\phi=\sqrt{T_{3}}\,r\), the first term in (\ref{scalaraction})
becomes a standard kinetic term for a scalar field \(\phi\). The
second term in equation~(\ref{scalaraction}) can then be understood as
the field potential and has to be evaluated at the position of
the anti-brane. By means of equation~(\ref{fullh}), one gets
\begin{equation}
\label{Vofr}
V(r_{0},r_{1})=\dfrac{2T_{3}}{\tilde{h}(r_{0},r_{1})}
=\dfrac{2T_{3}}{\dfrac{C_{1}}{4r_{0}^{4}}
\left(1+\dfrac{1}{\Nflux} \dfrac{r^{4}_{0}}{r^{4}_{1}}\right)}\,.
\end{equation}
In the limit where the branes are far apart, namely \(r_{1}\ll
r_{0}\), one has
\begin{equation}
\label{Vofrapprox}
V(r_{0},r_{1})\approx 2T_{3}\,\frac{4r_{0}^{4}}{C_{1}}
\left(1-\frac{1}{\Nflux}\frac{r^{4}_{0}}{r^{4}_{1}}\right).
\end{equation}
It can moreover be shown that \(C_{1}\) is related to \({\cal N}\) and
\(v\) through
\begin{equation}
C_{1}=16\pi \frac{\cal N}{v} g_{\rm s}\,\alpha'^{2}=4\ruv^{4},
\end{equation}
where \(\ruv\) is the ``edge'' of the KS throat (see section
\ref{sec:params}).

\par

In this Appendix, we have recalled the derivation of the interaction
potential between the brane and the anti-brane in a ten-dimensional
background warped by fluxes.  The resulting potential would be the
same if one chose to insert the anti-brane first into the background,
and then study the motion of the brane. The only difference is that a
D3 brane, inserted into the warped background, does not feel any force
because of the compensation between the Chern-Simons coupling and the
gravitation due to its tension [note the opposite signs of the two
terms in equation~(\ref{firstbrane})]. This is not the case for an
anti-brane inserted first, for which the two forces add up due the
same signs in (\ref{secondbrane}).  Seeking to minimise its energy,
the anti-brane therefore sinks to the bottom of the KS throat. Let us
recall that the above treatment remains valid only if \(r_{1}\gg
r_{0}\) and breaks down as soon as the two branes become too close
(see section \ref{subsec:validity}).

\end{appendix}

\section*{References}

\bibliography{referenceskklt}

\end{document}